\documentclass{aa}

\usepackage{graphicx}
\usepackage{txfonts}
\usepackage{xcolor}
\usepackage[normalem]{ulem}
\usepackage[version=3]{mhchem} 
\usepackage{multirow}
\usepackage{amsmath,amssymb}

\newcommand*\setPci{\ensuremath{\mathcal{P}}_{c, i}}
\newcommand*\setPc{\ensuremath{\mathcal{P}}_{c}}
\newcommand*\setPs{\ensuremath{\mathcal{P}}_{s}}
\renewcommand{\vec}[1]{\mathbf{#1}}
\DeclareMathOperator{\E}{\mathbb{E}}
\DeclareMathOperator*{\argmin}{argmin}

\begin{document}

   \title{Indications for very high metallicity and absence of methane in the eccentric exo-Saturn WASP-117b}

   \subtitle{}

   \author{Ludmila Carone
          \inst{1},
          Paul Molli\`ere,\inst{1}
          Yifan Zhou,\inst{2}
          Jeroen Bouwman,\inst{1}
          Fei Yan,\inst{3},
          Robin Baeyens,\inst{4}
          D\'aniel Apai,\inst{5}
          Nestor Espinoza,\inst{6}
          Benjamin V. Rackham, \inst{7, 8}
          Andr\'es Jord\'an,\inst{9, 10}
          Daniel Angerhausen,\inst{11, 12, 13}
          Leen Decin,\inst{4}
          Monika Lendl,\inst{14}
          Olivia Venot,\inst{15}
          \and
          Thomas Henning\inst{1}
          }

   \institute{Max-Planck-Institut f\"ur Astronomie, K\"onigstuhl 17, 69117 Heidelberg, Germany\\
              \email{carone@mpia.de}
         \and
             Department of Astronomy/McDonald Observatory, The University of Texas, 2515 Speedway, Austin, TX, 78712, USA
             \and
             Institut f\"ur Astrophysik, Georg-August-Universit\"at, Friedrich-Hund-Platz 1, 37077 G\"ottingen, Germany
             \and
             Instituut voor Sterrenkunde, KU Leuven, Celestijnenlaan 200D, B-3001 Leuven, Belgium
             \and
             Steward Observatory, The University of Arizona, 933 N. Cherry Avenue, Tucson, AZ 85721, USA, and Lunar and Planetary Laboratory, The University of Arizona, 1629 E. Univ. Blvd., Tucson, AZ 85721, USA
             \and
             Space Telescope Science Institute, 3700 San Martin Drive, Baltimore, MD, USA
             \and 
             Department of Earth, Atmospheric and Planetary Sciences, and Kavli Institute for Astrophysics and Space Research, Massachusetts Institute of Technology, Cambridge, MA 02139, USA
             \and
             51 Pegasi b Fellow
             \and
             Facultad de Ingeniería y Ciencias, Universidad Adolfo Ib\'a\~nez, Av.\ Diagonal las Torres 2640, Pe\~nalol\'en, Santiago, Chile
             \and
             Millennium Institute for Astrophysics, Chile
             \and
             ETH Z\"urich, Inst. f. Teilchen- und Astrophysik,  Z\"urich, Switzerland
             \and
             CSH Fellow, Center for Space and Habitability, University of Bern, Switzerland
             \and 
             Blue Marble Space Institute of Science, Seattle, United States
             \and
             Observatoire de Gen\'eve, chemin des maillettes 51, 1290 Sauverny, Switzerland
             \and
             Laboratoire Interuniversitaire des Syst\`{e}mes Atmosph\'{e}riques (LISA), UMR CNRS 7583, Universit\'{e} Paris-Est-Cr\'eteil, Universit\'e de Paris, Institut Pierre Simon Laplace, Cr\'{e}teil, France
             }

   \date{Received TBD; accepted TBD}

 
  \abstract
   {}
   {We investigate the atmospheric composition of the long period ($P_{\rm orb}=$~10 days), eccentric exo-Saturn WASP-117b. WASP-117b could be in atmospheric temperature and chemistry similar to WASP-107b. In mass and radius WASP-117b is similar to WASP-39b, which allows a comparative study of these planets.}
   { We analyze a near-infrared transmission spectrum of WASP-117b taken with Hubble Space Telescope/WFC3 G141, which was reduced with two independent pipelines. High resolution measurements were taken with VLT/ESPRESSO in the optical.}
   {We  report the robust ($3\sigma$) detection of a water spectral feature. Using a 1D atmosphere model with isothermal temperature, uniform cloud deck and equilibrium chemistry, the Bayesian evidence of a retrieval analysis of the transmission spectrum indicates a preference for a high atmospheric metallicity  ${\rm [Fe/H]}=2.58^{+0.26}_{-0.37}$ and clear skies. The data are also consistent with a lower-metallicity composition ${\rm [Fe/H]}<1.75$ and a cloud deck between $10^{-2.2} - 10^{-5.1}$~bar, but with weaker Bayesian preference. We retrieve a low CH$_4$ abundance of $<10^{-4}$ volume fraction within $1 \sigma$ and $<2\cdot 10^{-1}$ volume fraction within $3 \sigma$. We cannot constrain the equilibrium temperature between theoretically imposed limits of 700 and 1000~K. Further observations are needed to confirm quenching of CH$_4$ with $K_{zz}\geq 10^8$~cm$^2$/s. We report indications of Na and K in the VLT/ESPRESSO high resolution spectrum with substantial Bayesian evidence in combination with HST data.}
   {}

\titlerunning{Very high metallicity and absence of methane in WASP-117b}
\authorrunning{L. Carone et al.} 
   \maketitle
%

\section{Introduction}

The past years have revealed a large diversity in the atmospheres of transiting extrasolar gas planets colder than $1000$~K and smaller than Jupiter as well as a lack in methane in their atmospheres \citep[e.g.][]{Kreidberg2015,Kreidberg2018,Wakeford2017,Wakeford2018,Benneke2019,Chachan2019}. Whereas the atmospheric chemistry of hot Jupiters can be explained mainly with equilibrium chemistry, disequilibrium chemistry is expected to become more important for these cooler planets via vertical quenching, which results in an under-abundance of \ce{CH4} compared to predictions from equilibrium chemistry \citep{Crossfield2015}. In principle, \ce{CH4} is readily detectable along-side \ce{H2O} in the near to mid-infrared.
 
 While the quenching of \ce{CH4} has been confirmed first in brown dwarfs and later in directly imaged exoplanets \citep{Barman2011,Barman2011b,Moses2016,Miles2018,Janson2013}, observing disequilibrium chemistry in transiting, tidally locked extrasolar gas planets has been more challenging. Quantifying \ce{CH4} quenching reliably in transiting exoplanets to compare these disequilibrium chemistry processes to those occurring in brown dwarfs and directly imaged planets will be very illuminating to explore dynamical differences in different substellar atmospheres.
 
 Disequilibrium chemistry depends on vertical atmospheric mixing ($K_{zz}$), which provides a link between the observable atmosphere and deeper layers \citep[see e.g.][]{Agundez2014b}. Dynamical processes are further expected to be very different in tidally locked exoplanets compared to brown dwarfs just due to the different rotation and irradiation regime \citep{Showman2014,Showman2019}. For example, tidal locking slows down the rotation periods of transiting exoplanets to a few days, which is very slow in comparison to e.g. brown dwarfs with rotation periods less than 1~day \citep[see e.g.][]{Apai2017}. Disequilibrium chemistry processes across a whole range of substellar atmospheres could thus shed light on deep atmospheric processes and on to why the detection of methane has been proven to be very difficult so far in transiting exoplanets. 
 
 To this date, both methane and water have only been reliably detected for one mature extrasolar gas planet: at the day side of the warm (600--850~K) Jupiter HD 102195b \citep{Guilluy2019}. For HAT-P-11b, the presence of methane is inferred with 1D atmosphere models for retrieval by \citet{Chachan2019}, where the authors find that the observed steep rise in transit depth for wavelengths $\geq 1.5 \mu$m in their HST/WFC3 G141 data is not present when they simulate the spectrum after removing \ce{CH4} opacities from the model. The authors could, however, not constrain \ce{CH4} abundances further \citep{Chachan2019}. For the warm Super-Neptune WASP-107b \citep{Kreidberg2018}, methane quenching is suggested due to the absence of methane (no discernible opacity source that would translate to an increase in transit depth for $\lambda \geq 1.5 \mu$m in their HST/WFC3 G141 data), while water could be clearly detected. \citet{Benneke2019} also report for the mini-Neptune GJ~3470b methane depletion compared to disequilibrium chemistry models.

Many of the transiting exoplanets, for which disequilibrium chemistry may play a role, were also found to be less massive than Jupiter, ranging from mini-Neptune to Saturn-mass. These objects show a spread in metallicity that ranges from very low, $\leq 4 \times$~solar,  for the Neptune HAT-P-11b \citep{Chachan2019} to very high values, $>100 \times$ metallicity, for the exo-Saturn WASP-39b \citep{Wakeford2018}. 

In this paper, the transiting exo-Saturn WASP-117b joins the rank of super-Neptune-mass exoplanets with atmospheric composition constraints. WASP-117b is in mass ($0.277 M_{Jup}$) and radius ($1.021 R_{Jup}$) close to WASP-39b, which was found to be metal-rich \citep{Wakeford2018}. In temperature it could be 900~K or colder and thus similar in atmospheric chemistry to WASP-107b \citep{Kreidberg2018}. Therefore, this planet will aid a comparative analysis of methane content and metallicity from the Neptune to Saturn-mass range. 

\begin{figure}
   \centering
   \includegraphics[width=0.5\textwidth]{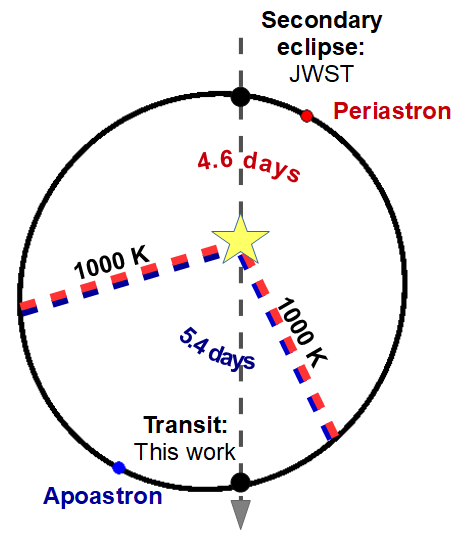}
      \caption{The orbit orientation of WASP-117b and the expected equilibrium temperature, assuming albedo $\alpha=0$ and to first order instant adjustment of atmospheric temperature to changes of incoming stellar flux as used in \citet{Lendl2014}. Temperatures around 1000~K for solar metallicity denote the transition in equilibrium chemistry between \ce{CO} (for the hot regime $T_{\rm eq}\gtrsim 1000$~K) and \ce{CH4} (for the warm $T_{\rm eq}\lesssim 1000$~K) \citep{Crossfield2015}. In the latter case, however, formation of \ce{CH4} is expected to be quenched due to disequilibrium chemistry. The planet is expected to remain 4.6~days in the hot regime and 5.4 days in the colder regime, respectively.
              }
         \label{fig: WASP-117b_orbit}
   \end{figure}

Furthermore, WASP-117b orbits its quiet F-type main sequence star on an eccentric orbit ($e=0.3$) with a relatively large orbital period of $\approx 10$~days \citep{Lendl2014,Mallonn2019}. Using the same formalism as \citet{Lendl2014}\footnote{\citet{Lendl2014} report equilibrium temperatures of 900~K and 1200 K during apoastron and periastron, respectively, assuming an albedo of $\alpha=0$ and efficient, uniform redistribution of absorbed stellar energy over the whole planet.} to calculate atmospheric temperatures during one orbit, the planet would then reside on its eccentric orbit for several days in the hot ($T > 1000$~K) temperature regime, characterized by CO as the main carbon-bearing species at $p=1$~bar for solar metallicity. The planet would also reside several days in the warm ($T < 1000$~K) regime, for which CH$_4$ becomes dominant in the observable atmosphere, and for which, depending on the strength of vertical mixing ($K_{zz}$) and atmospheric metallicity, disequilibrium chemistry of methane potentially becomes observable (Figure~\ref{fig: WASP-117b_orbit}). This is a different situation than compared to WASP-107b and WASP-39b that reside on tighter circular orbits all the time in the same temperature and thus chemistry regime.

The atmospheric properties of WASP-117b will thus shed further light on the diversity in basic atmospheric properties of transiting Super-Neptunes like atmospheric metallicity and \ce{CH4} quenching at pressures deeper than 0.1 bar that can lead to depletion by several orders of magnitude compared to equilibrium chemistry for atmospheric temperatures colder than 1000~K. In addition, due to the relatively long eccentric orbit, this exo-Saturn will also allow us to understand how exoplanets on wider non-circular orbits that are subjected to varying irradiation and atmospheric erosion differ from exoplanets on tighter, circular orbits.

Last but not least, we will show that it is possible and worthwhile to characterize exo-Saturns on orbital periods of 10~days with single-epoch observations from space. Thus, our WASP-117b observations are prototype observations for other transiting exoplanets with orbital periods of 10~days and longer like K2-287~b \citep{Jordan2019a} and similar objects that were discovered after WASP-117b \citep{Brahm2018, Jordan2020, Rodriguez2019}.

We report here the first observations to characterize the atmosphere of the exo-Saturn WASP-117b (Section \ref{sec: Obs}) in the near-infrared with HST/WFC3 (Section~\ref{sec: HST_obs}) and in the optical with VLT/ESPRESSO (Section~\ref{sec: VLT_obs}), respectively.
We investigate the significance of the water detection with HST/WFC3 and Na and K with VLT/ESPRESSO in Section~\ref{sec: Significance}.
We perform atmospheric retrieval with an atmospheric model (Section~\ref{sec: retrieval}) mainly on the HST/WFC3 data to constrain basic properties of the atmosphere of the exo-Saturn WASP-117b (Sections~\ref{sec: atmosphere models} and \ref{sec: Significance_model}). We then present improved stellar rotation and spin-orbit alignment parameters for WASP-117b via the Rossiter-McLaughlin effect measured with VLT/ESPRESSO (Section~\ref{sec: RM}).

We also compared our data with broadband transit observations obtained by the Transiting Exoplanet Survey Satellite (TESS) in the optical (Section~\ref{sec: TESS}) and discussed influence of stellar activity to explain the discrepancy between TESS transit depth and our WFC3/NIR transmission spectrum, which were obtained at different times (Section~\ref{sec: stellar contamination}). To provide better context for the low \ce{CH4} abundances that we find, we calculated a small set of representative chemical models for different temperatures and metallicities (Section~\ref{sec: Disequi}). To guide future observations to further characterize the system we also generated synthetic spectra based on our retrieved atmospheric models for wavelength ranges covered by HST/WFC3/UVIS and JWST (Section~\ref{sec: W117_prediction}). We investigated the benefit of additional measurements to constrain the presence of a haze layer and to distinguish between high ($\rm [Fe/H] > 2.2$) and low ($\rm [Fe/H] < 1.75$) metallicity models. We discuss the atmospheric properties of WASP-117b and how they compare to other Super-Neptunes like WASP-39b and WASP-107b in Section~\ref{sec: dicuss}. We provide a summary of our findings in Section~\ref{sec: Conclusion} and outline steps for future observations of the eccentric exo-Saturn WASP-117b in Section~\ref{sec: Outlook}.


\section{Observations}
\label{sec: Obs}

We report here a near-infrared transmission
spectrum measured with HST (Program GO 15301, PI: L.\ Carone) and a high resolution optical spectrum measured with VLT/ESPRESSO (PI: F.\ Yan). In the following, both set of observations and data reduction are described. For the HST/WFC3 raw data, we employed two completely independent data reduction pipelines for a robust retrieval of the atmospheric signal.

The relatively long transit duration of 6 hours required a long and stable observation by both HST and VLT/ESPRESSO, which were successfully carried out with both instruments. 

\subsection{HST observation and data reduction}
\label{sec: HST_obs}

We  observed  one  transit  of  WASP-117 b  with HST’s Wide Field Camera 3 (WFC3) instrument on UT 20-21 September 2019. The transit observation consists of eleven consecutive HST orbits. This HST observation benefited from re-observations of the WASP-117b transit in July 2017 with broad band photometry using two small telescopes. One in Chile, the Chilean-Hungarian Automated Telescope 0.7m (CHAT, PI: Jord\'an) and one in South Africa (1m, Los Cumbres Observatory). The combined observations allowed us to improve the uncertainties in mid-transit time from 2 hours (based on \citealt{Lendl2014}) to 2 minutes \citep{Mallonn2019}.

For the HST measurement, at the start of each orbit, an image of the target with the F126N filter, using NSAMP=3, rapid readout mode was taken. This image was used to anchor the wavelength calibration for each orbit. Otherwise, we obtained time series spectra with the G141 grism, which covers the  wavelength range  1.1--1.7~$\mu$m. We used  the NSAMP=15, SPARS 10 readout mode for these, in  spatial  scanning  mode with a scan rate of  0.12 arcseconds/sec and scan direction ``round trip''.

For raw data reduction, we employed two separate and completely independent pipelines to ensure that the derived atmospheric spectra are robust. We call the two pipelines henceforth nominal pipeline and CASCADE pipeline; their respective data reduction procedures are described in Appendix~\ref{sec: HST_pipelines}.

\subsubsection{Final transmission spectra}

\begin{figure}
   \centering
   \includegraphics[width=0.49\textwidth]{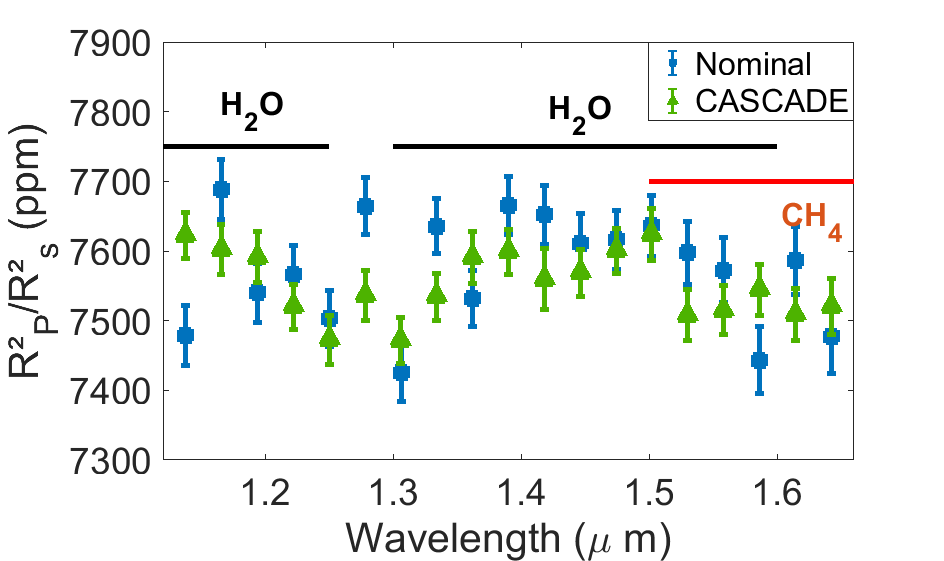}
      \caption{Final HST/WFC3 transmission spectra from the nominal (blue) and CASCADE (green) pipeline, covering a wavelength range of 1.125 -1.65 $\mu$m. Average transit depth is  $7573 \pm 43$ ppm and $7553 \pm 39$ ppm, respectively.
      Black and red lines indicate, respectively, the positions of the \ce{H2O} and \ce{CH4} absorption bands in the observed spectral range. 
              }
         \label{fig: Final_spectrum}
   \end{figure}

After data reduction, as described in Sections \ref{sec: Zhou} and \ref{sec: CASCADE}, we derived two transmission spectra (Figure~\ref{fig: Final_spectrum}). We excluded points outside of the wavelength range 1.125 - 1.65 $\mu$m because the instrument transmission has steeper gradients in those regions compared to the centre of the spectrum. Jittering in the wavelength direction together with the strong variation in transmission profile change can introduce large systematic errors. Since a strong methane absorption feature covered by the WFC3/G141 grism is centered at 1.62~$\mu$m, where \ce{H2O} has no absorption, we can still make judgements about the presence or absence of methane based on a spectrum covering 1.125 - 1.65~$\mu$m. This wavelength range was used for our further analysis.

The WASP-117b HST/WFC3 data indicate the presence of a muted water absorption spectrum with a slight drop in transit depth between 1.4 and 1.55~$\mu$m, which indicates \ce{H2O} absorption centered at 1.45~$\mu$m. There is potentially a second peak towards the shorter wavelengths at the edge of the spectral window. There is no strong \ce{CH4} absorption feature towards the longest wavelengths in the observed spectral window. The feature is present in both spectra (nominal and CASCADE), using two independent reduction pipelines for the same data set. 

However, it also appears that the nominal spectrum is slightly shifted towards deeper transit depths than the CASCADE spectrum. Further, some data points appear to deviate more than 3$\sigma$ from each other. We attribute these deviations to differences at the raw light curve extraction as described in the next subsection.


\subsubsection{Comparison of employed data pipelines}
\label{sec: Pipeline_comparison}

We find that the transmission spectra from the two pipelines demonstrate discrepancies (Figure~\ref{fig: Final_spectrum}). First, there is disagreement in the average transit depth. Calculating the standard error of the mean $\sigma/\sqrt{n}$ (SEM) gives values of $\mathrm{SEM_{nominal}}=18.2$~ppm for the nominal and $\mathrm{SEM_{CASCADE}}=11.0$~ppm for the CASCADE pipeline, respectively. The deviation between the mean transit depths is thus $1.1 \cdot \mathrm{SEM_{nominal}}$ or $1.9 \cdot \mathrm{SEM_{CASCADE}}$, which is a significant but not substantial ($>3\mathrm{SEM}$) deviation. Second, there are more significant inconsistencies at several wavelength channels. At 1.28 $\mu$m, the difference in transit depths is nearly $3\sigma$.

We discuss further the possible sources of these inconsistencies and the implications for our retrieval results. The uncertainties in our transmission spectra include observational noise (photon noise, readout noise, dark current, etc.) and noise correlated with the detector systematics. The two data reduction pipelines, nominal, and CASCADE, are fully independent in the treatment of these uncertainties. In general, each pipeline consists of two modules: first, the light curve extraction module, which reduces the time series of spectral images into spectral light curves; second, the light curve modeling module, which fits instrument systematic and transit profile models to the light curve and measures transit depths. Both modules can cause discrepancies in the transmission spectrum. 

To locate the source of the inconsistencies, we perform the following analysis: We use the light curve modeling module of the nominal pipeline (RECTE) to fit the uncalibrated light curves extracted in the CASCADE pipeline. We label the results as ``CASCADExRECTE''.

\begin{figure}
  \center
  \includegraphics[width=0.48\textwidth]{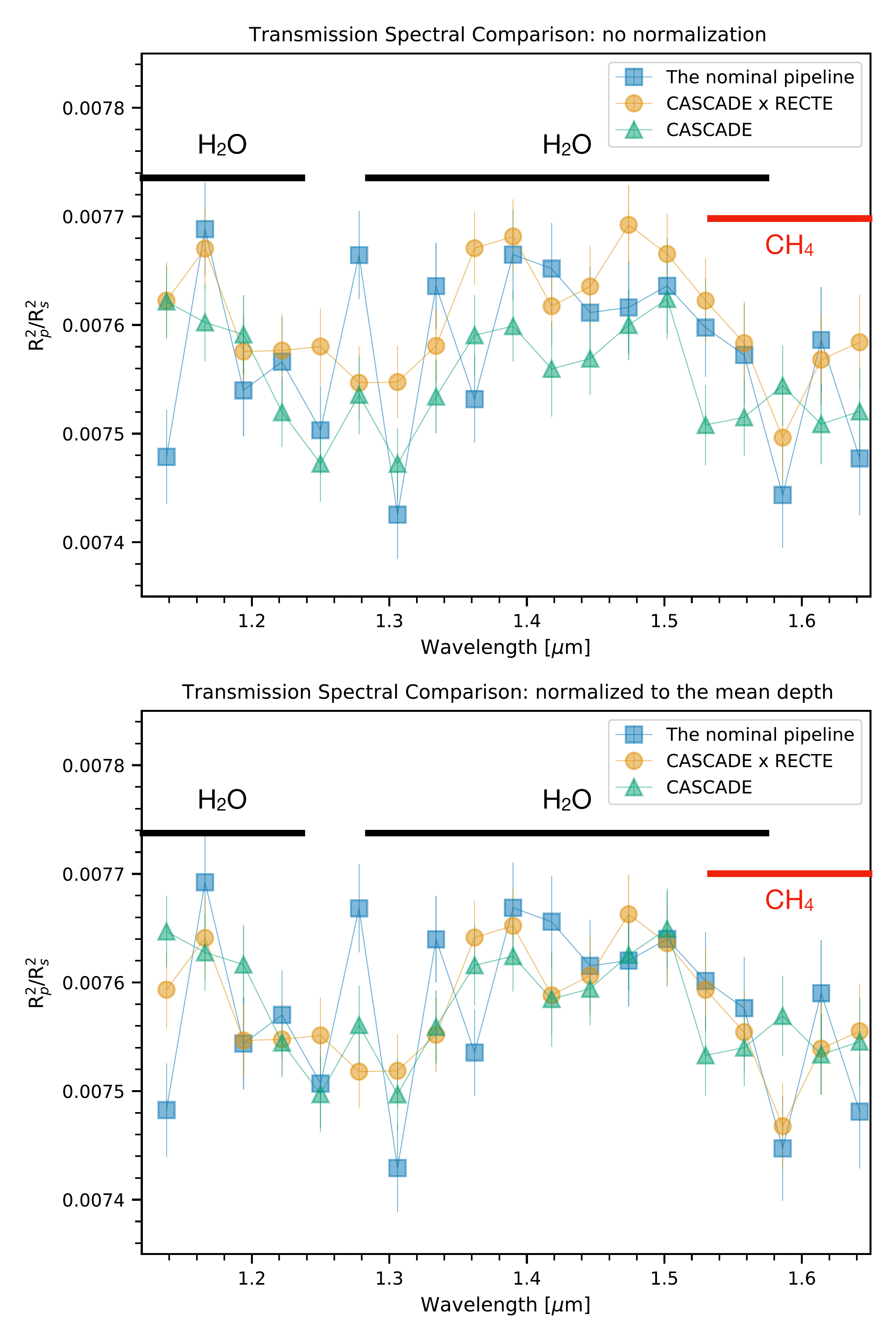}
  \caption{Comparison of transmission spectra obtained with different pipelines. The upper panel shows the original spectra, and the lower panel shows the spectra scaled to the same average depth. After scaling, the spectra using the same raw data extraction than CASCADE but different systematic treatment (CASCADE and CASCADExRECTE) are consistent with each other.}
  \label{fig:pipeline comparison}
\end{figure}

In Figure~\ref{fig:pipeline comparison} (top panel), we compare the transmission spectra of nominal, CASCADE, and RECTExCASCADE pipelines without accounting for the different transit depths. In Figure~\ref{fig:pipeline comparison} (bottom panel), we scale the three spectra to the same average transit depth. The second comparison, thus, shows more clearly deviations on the shape of the spectra. Furthermore, weighted residual sums of squares (RSS)
\begin{equation}
  \label{eq:rss}
  \mathrm{RSS} = \sum_i^n \frac{(\mathrm{spec(i)_{1}} - \mathrm{spec(i)_{2}})^{2}}{(\mathrm{uncertainty^2(i)_{1}} + \mathrm{uncertainty^2(i)_{2}})},
\end{equation}
where we sum over the total number $n$ of wavelength bins $i$, are derived to evaluate the difference statistically. For non-normalized spectra, the RSS between nominal and CASCADE is 35.2 (degrees of freedom, DOF=19), and the RSS between CASCADExRECTE and CASCADE is 33.4 (DOF=19). After normalization, the RSS between nominal and CASCADE is 32.4 (degrees of freedom, DOF=18), and the RSS between CASCADExRECTE and CASCADE is 11.6 (DOF=19). 

Thus, with the same light curve extraction module and scaling the spectra to the same average depth, the light curve fitting modules in the two pipelines provide fully consistent results. This can also be seen in Figure~\ref{fig:pipeline comparison} (bottom panel) by comparing the spectra CASCADExRECTE and CASCADE. All data points, except at 1.58~$\mu$m agree within 1~$\sigma$ with each other.

From these comparison results, we conclude: 1. difference in light curve extraction introduces random errors and causes the transmission spectra to differ at several wavelength channels; 2. difference in systematic correction causes a slight offset ($<1\sigma$) in the average transit depths; 3. after scaled to the same average depth, the spectra from the two pipelines agree with each other, i.e., the shape of the spectra are consistent. The random errors and the offset yield apparent 3$\sigma$ differences in single points in the final spectra (Figure~\ref{fig: Final_spectrum}).

\begin{figure}
  \center
  \includegraphics[width=0.48\textwidth]{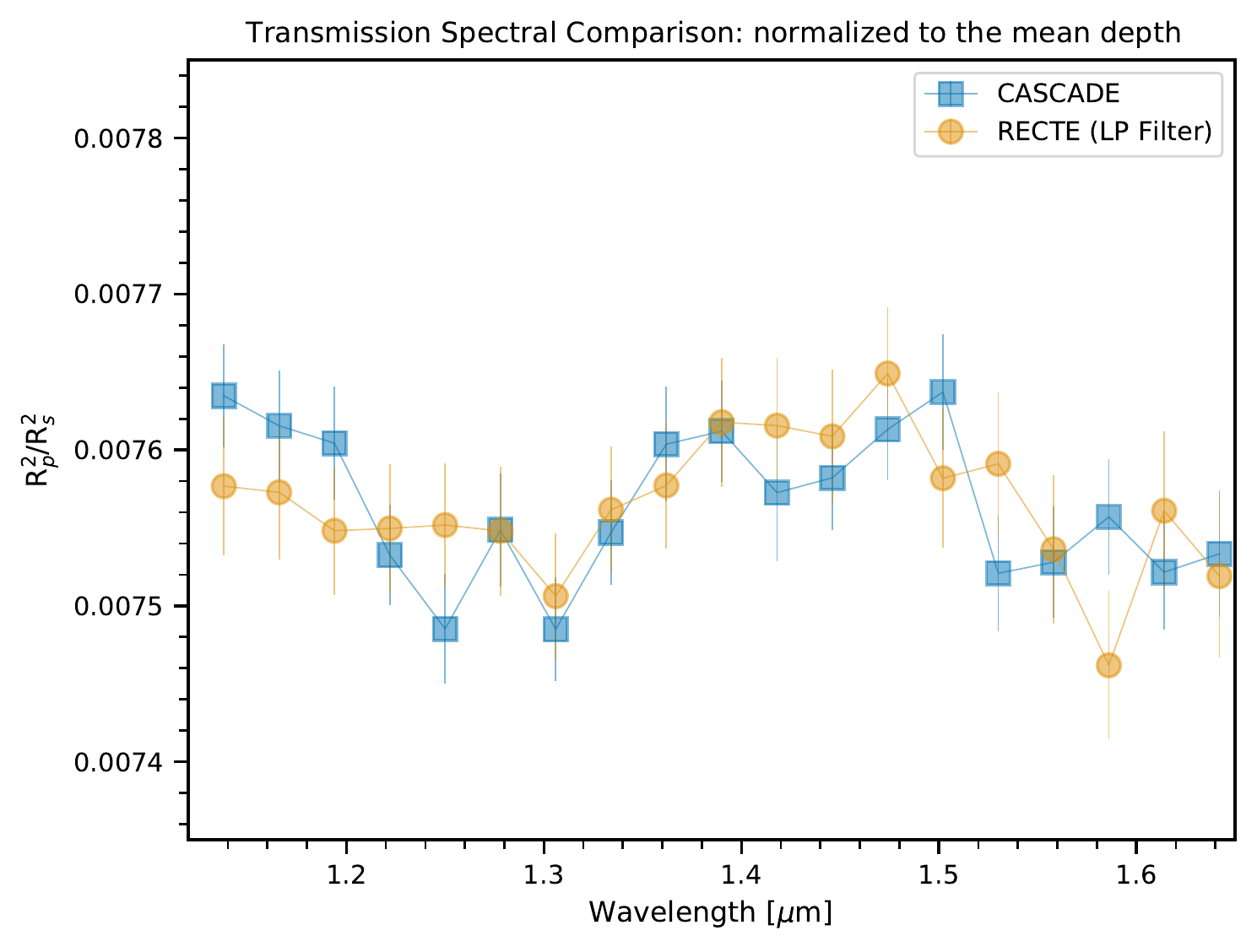}
  \caption{Comparison of transmission spectra of a slightly modified nominal pipeline and the CASCADE pipeline. In the nominal pipeline, a Gaussian filter was applied to the images before wavelength registration to address the issue of telescope pointing drifts in the wavelength direction. The two spectra are fully consistent, supporting that discrepancies between two pipeline are due to their variations in wavelength registration in the raw light curve extraction stage.}
  \label{fig:pipeline_comp_last}
\end{figure}

The largest inconsistency stems from the extracted light curves and is likely a consequence of the relatively strong telescope pointing drift during the WASP-117b observations. In the wavelength dispersion direction, the peak-to-peak drift distance is  ${\sim}0.4$ pixels, greater than the sub-0.1-pixel level normally seen in HST scanning mode observations \citep[e.g.,][]{Kreidberg2014}. Different treatments in aligning the spectra between the two pipelines can lead to the inconsistency of the final transmission spectra. To demonstrate this point, we performed another reduction with a slightly modified nominal pipeline, in which we applied a Gaussian filter to the images
  before aligning them in the dispersion direction. The kernel of the Gaussian filter was the same as the spectral resolution element of WFC3/G141. This treatment reduced the column-to-column pixel value variations and thus decreased the influence of pointing drift. We derived the transmission spectrum using this pipeline and compared it with the
  CASCADE one (Figure~\ref{fig:pipeline_comp_last}). The two spectra are fully consistent with an RSS value of 15.6
  (DOF=19).
  
 Regardless of the treatment of the pointing drift, the shape of the spectra deriving from the nominal and CASCADE pipelines, respectively, are consistent with each other. Thus, we do not expect substantial discrepancies in the outcomes of the atmospheric retrieval for the two spectra.
 
  Our goal is to robustly determine the planet properties, which are not dependent upon the choice of systematics models. We will thus use both data pipelines for a most robust interpretation of the single epoch transit observation with HST/WFC3. In the following, we will use mainly results derived from the nominal pipeline that is well established for further analysis, e.g., with VLT/ESPRESSO (Section~\ref{sec: VLT_obs}) and TESS data (Section~\ref{sec: TESS}).

\subsection{VLT/ESPRESSO observations}
\label{sec: VLT_obs}
To also probe gas phase absorption of Na and K in the atmosphere of WASP-117b, we observed one transit of the planet on 24/25 October 2018 under the ESO program 0102.C-0347 with the ultra-stable fibre-fed echelle high-resolution spectrograph ESPRESSO \citep{Pepe2010} mounted on the VLT. The observation was performed with the 1-UT high-resolution and fast read-out mode. We set the exposure time as 300\,s and pointed fiber B on sky. We observed the target continuously from 23:56 UT to 09:28 UT and obtained 98 spectra in total. These spectra have a resolution of R $\sim$ 140\,000 and a wavelength coverage of 380-788\,nm.

\subsubsection{Data reduction}
We reduced the raw spectral images with the ESPRESSO data reduction pipeline (version 2.0.0). The pipeline produced spectra with sky background corrected by subtracting the sky spectra measured from fiber B from the target spectra measured from fiber A. 
The pipeline also calculated the radial velocity (RV) of the stellar spectrum with the cross correlation technique.
We discarded six spectra that have relatively low signal-to-noise ratios. Among the remaining 92 spectra, 59 spectra were observed during transit and 33 spectra were observed during the out-of-transit phase. 

\subsubsection{Transmission spectra of sodium and potassium}

To obtain the planetary transmission spectra of sodium (Na) and potassium (K), we implemented the following procedures.

\textit{(1) Removal of telluric lines}\\
There are telluric Na emission lines around the sodium doublet, and these emission lines were corrected using fiber B spectra. We further corrected the telluric $\mathrm{H_2O}$ and $\mathrm{O_2}$ absorption lines by employing the theoretical $\mathrm{H_2O}$ and $\mathrm{O_2}$ transmission model described in \cite{Yan2015b}. The corrections were performed in the Earth's rest frame.

\textit{(2) Removal of stellar lines}\\
In order to remove the stellar lines, we firstly aligned all the spectra into the stellar rest frame by correcting the barycentric Earth radial velocity (BERV) and the stellar systemic velocity. 
We then generated a master spectrum by averaging all the out-of-transit spectra and divided each observed spectrum with this master spectrum.
The residual spectra were then filtered with a Gaussian function ($\sigma$ $\sim$ 3\,\AA) to remove large scale features.

\begin{figure*}[t!]
   \centering
   \includegraphics[width=0.95\textwidth, height=0.6\textwidth]{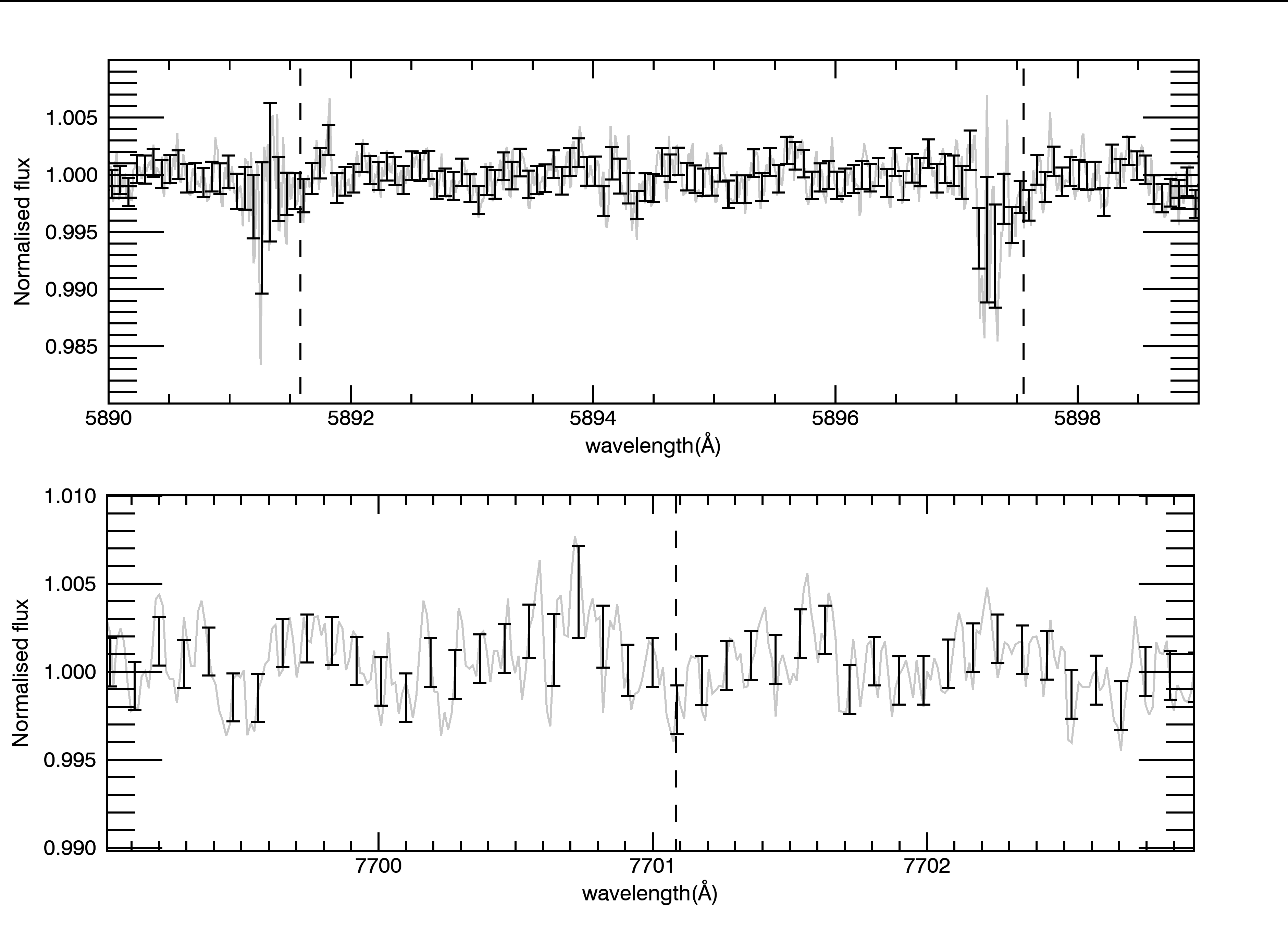}
      \caption{Top: normalised transmission spectrum around the Na D doublet lines. Bottom: normalised transmission spectrum around the K $\mathrm{D_1}$ line. The center wavelengths of these absorption lines are indicated as dashed lines. The black data points are spectra binned every seven wavelength points and the grey lines are original spectra. The stellar line profile change due to the CLV and RM effects is corrected. It should be noted that here vacuum wavelengths are displayed. In normalized flux, planetary Na and K absorption appear as weak depressions around the dashed line.
              }
         \label{fig: High_res_data}
   \end{figure*}

\textit{(3) Correction of the CLV and RM effects}\\
During the planet transit, the observed stellar line profile has variations originating from several effects. The Rossiter–McLaughlin effect \citep{Queloz2000} and the center-to-limb variation (CLV) effect \citep{Yan2015a, Czesla2015, Yan2017} are two main effects. We followed the method described in \cite{Yan2018} and \cite{Yan2019} to model the RM and CLV effects simultaneously. The stellar spectrum was modelled with the \texttt{Spectroscopy Made Easy} tool \citep{Piskunov2016} and the Kurucz \texttt{ATLAS12} model \citep{Kurucz1993}. We used the stellar parameters from \cite{Lendl2014} except the $v\,\mathrm{sin}i_\star$ and $\lambda$ values, which are taken from the RM fit of the ESPRESSO RVs.
The simulated line profile change due to the CLV and RM effects is weak and below the errors of the observed data. We subsequently corrected the CLV and RM effects for the obtained residual spectra.

\textit{(4) Obtaining the transmission spectrum}\\
We shifted all the in-transit residual spectra to the planetary rest frame and added up all these shifted spectra. Because of the high orbital eccentricity, the planetary orbital velocity changes from +10 km\,s$^{-1}$ to +20 km\,s$^{-1}$ during transit. Therefore, in the planetary rest frame, the position of the stellar Na/K line is $\sim$ +15 km\,s$^{-1}$ away from the expected planetary signal.

We investigated the Na D doublet lines (5891.584 $\AA$ and 5897.555 $\AA$) and K $\mathrm{D_1}$ line (7701.084 $\AA$). The K $\mathrm{D_2}$ line (7667.009 $\AA$) is heavily affected by the dense $\mathrm{O_2}$ lines and therefore is not used in the analysis. The listed wavelengths are in vacuum.
The final transmission spectra are presented in Fig.~\ref{fig: High_res_data}. There are no strong absorption features (dip) at the expected wavelengths of the planetary Na/K lines within $3 \sigma$ (indicated as dashed lines in the figure).

At the wavelength regions where the stellar Na lines are located (i.e. $\sim$ 0.3\,$\AA$ away from the expected planetary signal), there are some spectral features. But we attribute these features to the large errors of these points, because the flux inside the deep stellar Na line is significantly lower than the adjacent continuum. 

The planetary Na and K lines in the VLT/ESPRESSO data are below $3 \sigma$ significance. Thus, we decided not to perform atmospheric retrieval on the VLT/ESPRESSO data alone. Instead, we used the statistically significant atmosphere signal in the HST/WFC3 data in conjunction with VLT/ESPRESSO to strengthen the case that planetary Na and K lines are indeed present in the VLT/ESPRESSO data. This analysis will be presented in Section~\ref{sec: Significance}.

\section{WASP-117b atmospheric properties and improved planetary parameters}

In this section, we first explore the significance of the water detection in the HST/WFC3 and the possible existence of weak Na and K in the VLT/ESPRESSO data. We then perform atmospheric retrieval with different models to interpret the atmospheric properties of the exo-Saturn WASP-117b.

\begin{table*}[t]
\caption{Results of Bayesian model comparison for \ce{H2O} in HST/WFC3 and Na/K in VLT/ESPRESSO}             
\label{table: HST_H2O_Na_K_significance}     
\centering
{\renewcommand{\arraystretch}{1.2}%
\begin{tabular}{c |c | c | c }       
\hline
Retrieval model & Parameters & Evidence ${\rm ln} Z$ & Bayes factor $B$\\
\hline
\multicolumn{4}{c}{HST/WFC3 for \ce{H2O}}\\
\hline
Full hypothesis & $P_0,P_{\rm clouds}$,\ce{H2O},\ce{CO2},\ce{CH4},\ce{CO},\ce{N2} & \begin{tabular}[l]{@{}l@{}} nominal:$-32.973$\\ CASCADE:$-19.394$ \end{tabular}&  \begin{tabular}[l]{@{}l@{}} nominal: \textbf{baseline}\\ CASCADE:  \textbf{baseline} \end{tabular}\\
\hline
\ce{H2O} removed & $P_0,P_{\rm clouds}$,\ce{CO2},\ce{CH4},\ce{CO},\ce{N2} & \begin{tabular}[l]{@{}l@{}} nominal: $-36.042$\\ CASCADE: $ -22.971$\end{tabular}&  \begin{tabular}[l]{@{}l@{}} nominal: $21.5$ ($3 \sigma_{\ce{H2O}}$)\\ CASCADE: \textbf{$35.8$ ($3 \sigma_{\ce{H2O}}$)} \end{tabular}\\
\hline
\ce{CH4} removed & $P_0,P_{\rm clouds}$,\ce{H2O},\ce{CO2},\ce{CO},\ce{N2} & \begin{tabular}[l]{@{}l@{}} nominal:$-33.093$\\ CASCADE:$-19.300$\end{tabular}&  \begin{tabular}[l]{@{}l@{}} nominal: 1.1\\ CASCADE:  0.9 \end{tabular}\\
\hline
no clouds & $P_0$,\ce{H2O},\ce{CO2}, \ce{CH4},\ce{CO},\ce{N2} & \begin{tabular}[l]{@{}l@{}} nominal:$-33.828$\\ CASCADE: $-19.271$\end{tabular}&  \begin{tabular}[l]{@{}l@{}} nominal: 2.35\\ CASCADE:  0.88 \end{tabular}\\ \hline
\multicolumn{4}{c}{HST/WFC3 \& VLT/ESPRESSO for \ce{Na} and \ce{K}}\\
\hline
Full hypothesis & $P_0,P_{\rm clouds}$,\ce{H2O},\ce{CO2},\ce{CH4},\ce{CO},\ce{N2}, Na/K & \begin{tabular}[l]{@{}l@{}} nominal:$-174.929$\\ CASCADE:$-162.158$ \end{tabular}&  \begin{tabular}[l]{@{}l@{}} nominal: \textbf{baseline}\\ CASCADE:  \textbf{baseline} \end{tabular}\\
\hline
Na/K removed & $P_0,P_{\rm clouds}$,\ce{H2O},\ce{CO2},\ce{CH4},\ce{CO},\ce{N2} & \begin{tabular}[l]{@{}l@{}} nominal:$-176.918$\\ CASCADE: $-163.849$ \end{tabular}&  \begin{tabular}[l]{@{}l@{}} nominal: 7.3 ($2.5 \sigma_{Na,K}$)\\ CASCADE: 5.4 ($2.4 \sigma_{Na,K}$) \end{tabular}\\
\hline                      
\end{tabular}}
\end{table*}

\subsection{Significance of water and Na/K detection}
\label{sec: Significance}

To identify the significance of the \ce{H2O} detection in the HST/WFC3 spectrum, we follow the approach of \citep{Benneke2013} and run a number of forward models of similar complexity against each other, using \texttt{petitRADTRANS} \citep{Molliere2019}. In this model, we assume isothermal temperatures and a gray uniform cloud, which is modeled by setting the atmospheric opacity to infinity for $P > P_{\rm cloud}$. Thus, $P_{\rm cloud}$ can be treated as the pressure at the top of a fully opaque cloud. Furthermore, we included opacities of the following absorbers: \ce{H2O}, \ce{CH4}, \ce{N2}, \ce{CO} and \ce{CO2}. The mass fractions of absorbers are free parameters with priors in log space ranging from -10 to 0. For the remaining atmospheric mass, a mixture of \ce{H2} and \ce{He} is assumed with a ratio of 3:1. We retrieve the atmospheric reference pressure $P_0$ to reproduce the apparent size of the planet in the WFC3 wavelength range.

We quantify the significance of the observed molecular absorption features by using  the \texttt{MultiNest} sampling technique that enables to quantify and compare model parameters and their significance. \texttt{MultiNest} is implemented within the python wrapper \texttt{PyMultiNest} \citep{Buchner2014}. 

\begin{figure*}
   \centering
   \textbf{Full model}\par \medskip
    \textbf{Nominal}  \hspace{7.5 cm}  \textbf{CASCADE}\par
   \includegraphics[width=0.48\textwidth]{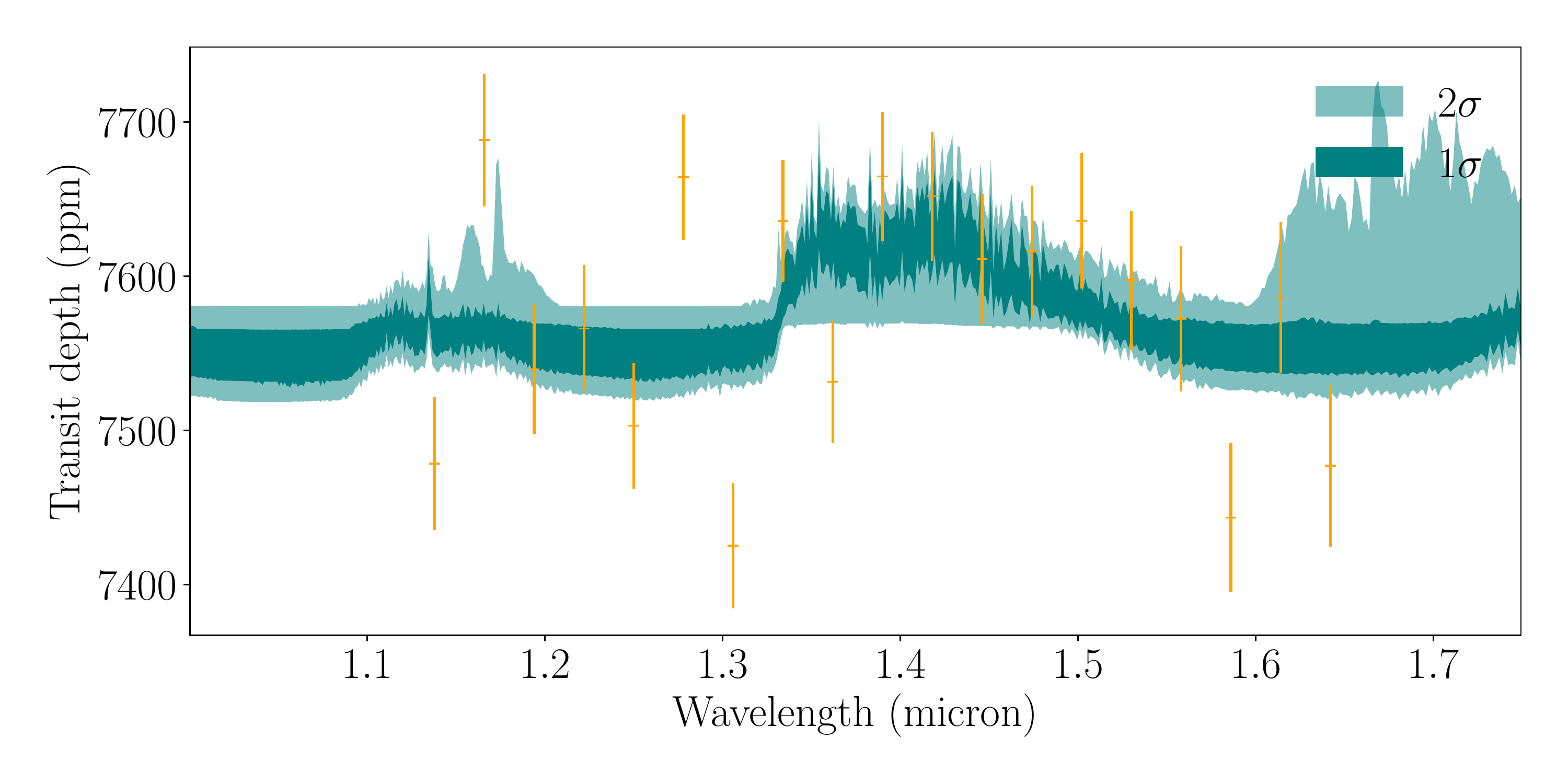}
   \includegraphics[width=0.48\textwidth]{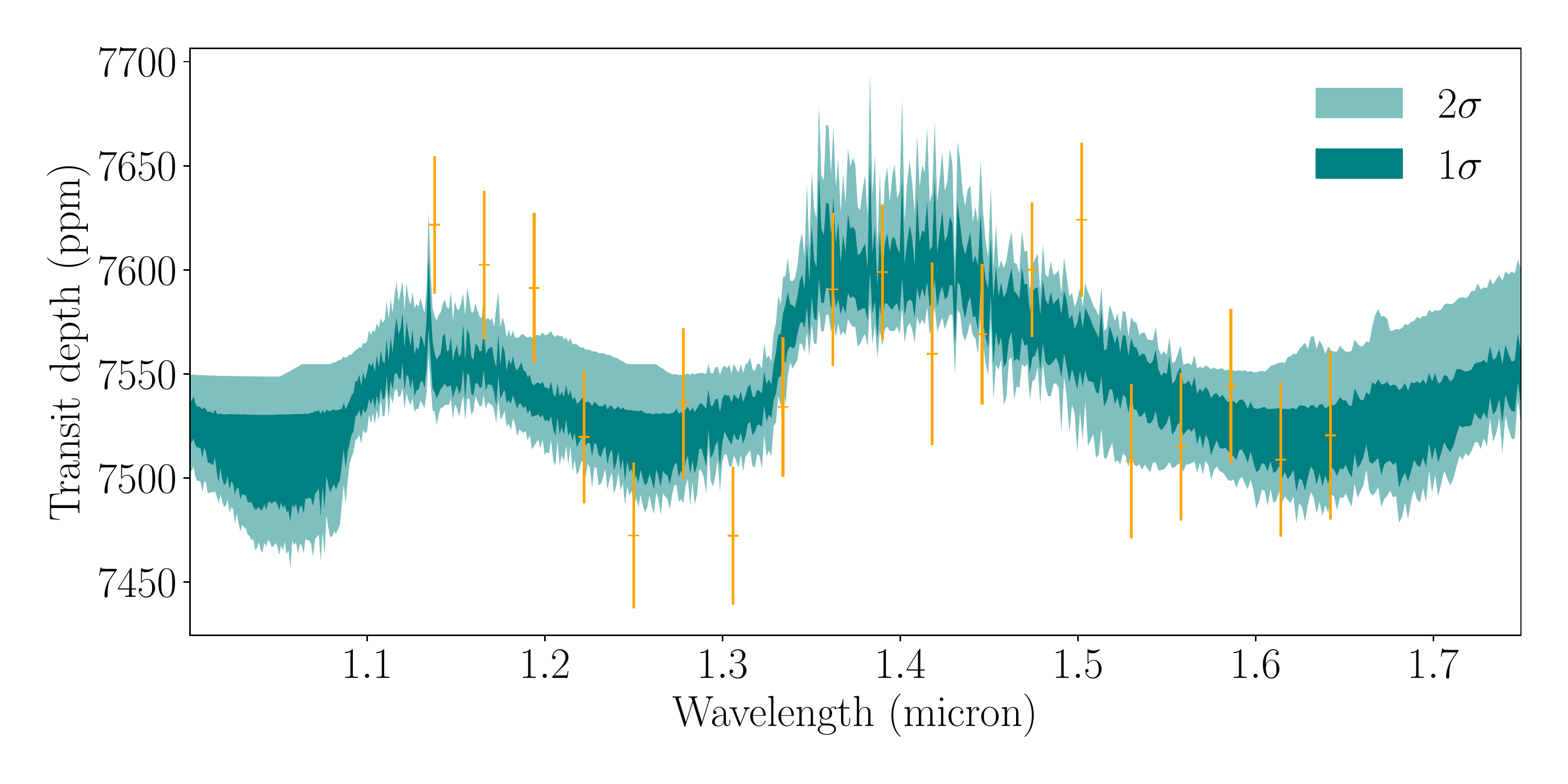}
    \textbf{\ce{H2O} removed}\par \medskip
   \includegraphics[width=0.48\textwidth]{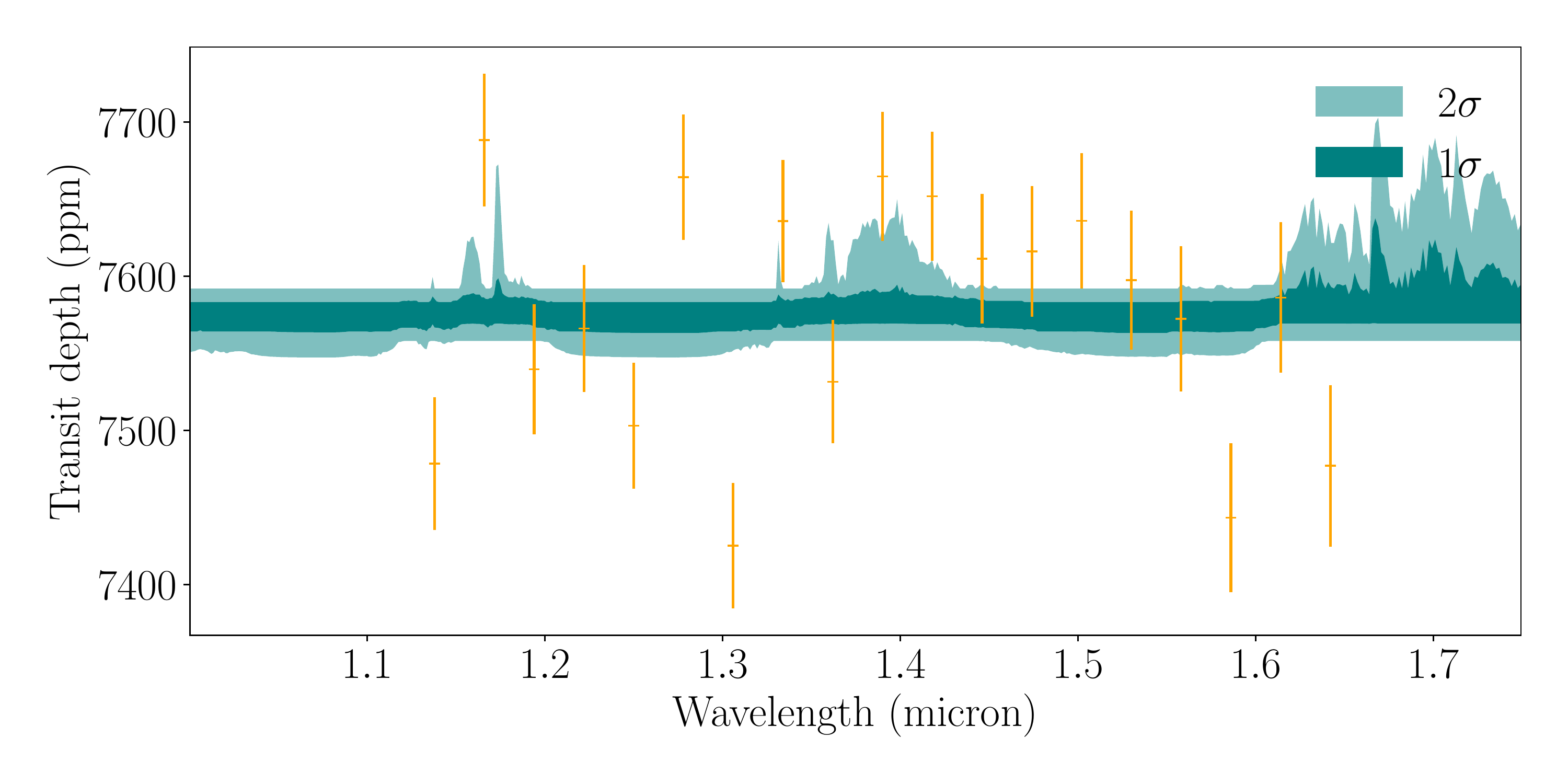}
   \includegraphics[width=0.48\textwidth]{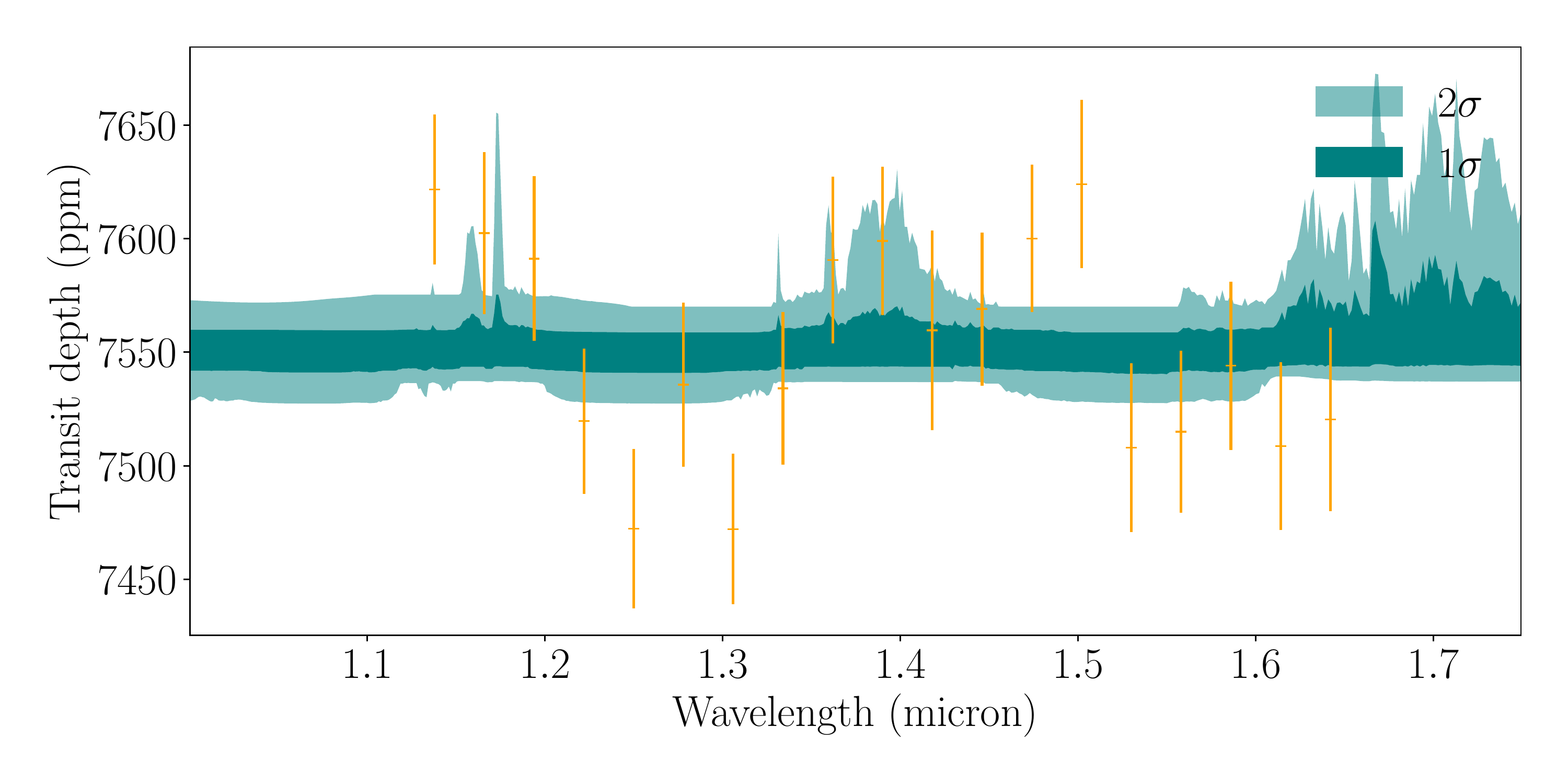}
      \caption{\textbf{Top:} Full retrieval model is fit to WASP-117b transmission spectrum reduced with the nominal pipeline (left) and CASCADE (right). \textbf{Bottom:} Retrieval model with \ce{H2O} removed fit to WASP-117b transmission spectrum reduced with the nominal (left) and CASCADE pipeline (right). \newline 
      For all spectra, retrieved atmosphere models with confidence within $1\sigma$ (dark green) and  $2\sigma$ (light green) are shown.}
         \label{fig: H2O_significance}
   \end{figure*}

Table~\ref{table: HST_H2O_Na_K_significance} lists the (natural) log evidences (${\rm ln} Z$) and the Bayes factor $B$ for different hypotheses. $B$ is calculated via $B = \exp(\ln Z_{\rm base}-\ln Z)$ from the \texttt{MultiNest} output for each model. Figure~\ref{fig: H2O_significance} shows the full model, including \ce{H2O}, versus the  model without \ce{H2O} for data derived with both pipelines.

Following \citet{Kass1995}, we regard $B$ values of 1--3, 3--20, 20--150, and $>$150 as 'weak'\footnote{Or 'not worth more than a bare mention'}, 'substantial', 'strong', and 'very strong' preference for a given hypothesis, respectively. \citet[][Table 2]{Benneke2013}, \textbf{adopted} from \citet{Trotta2008} allows us further to translate $B$ to lower limits on $\sigma$ confidence levels\footnote{We note that \citet{Benneke2013} adopt a different scale, where they label $B=$ 3--12 as weak, $B=$ 12--150 as ‘moderate’ and and $B>150$ as ‘strong’.}.

Our statistical analysis yields ``strong'' Bayesian preference ($B>$20) for the detection of water in the WASP-117b observational data for both pipeline. This corresponds to a $3 \sigma$ detection. 

The same approach is also used for the combined HST/WFC3 (nominal) and VLT/ESPRESSO spectrum and the possible detection of Na and K. Here, the full model is extended to also include K and Na opacities again assuming their combined mass fraction to be a free parameter, but fixing the Na/K abundance ratio to the solar value \citep{Asplund2009}.

\begin{figure}
   \centering
   
   \includegraphics[width=0.48\textwidth]{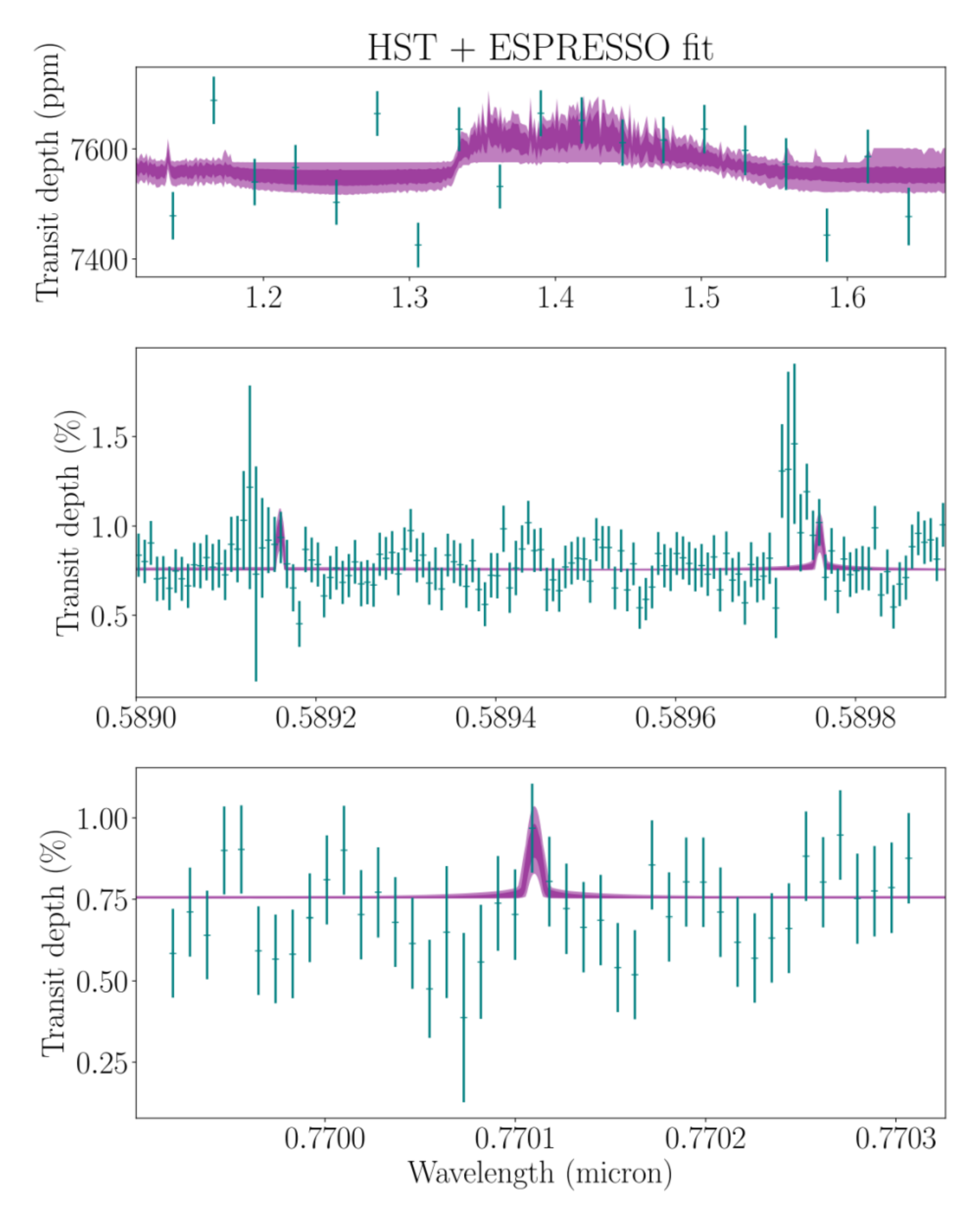}

      \caption{ Full retrieval model fit to the WASP-117b transmission spectrum in WFC3 and VLT/ESPRESSO with the nominal pipeline with \ce{Na} and \ce{K}. The WFC3 data are shown in the top panel, the middle planek shows the Na doublet and the bottom the K D${}_1$ line. The retrieval model without Na and K (not shown) is a flat line in the optical and indistinguishable from the full model without Na and K in the WFC3 NIR range.  For all spectra, retrieved atmosphere models with confidence within $1\sigma$ (dark green) and  $2\sigma$ (light green) are shown.}
         \label{fig: Na_K_significance}
   \end{figure}

Figure~\ref{fig: Na_K_significance} shows the full model, including Na and K, versus the  model without Na and K for data derived with the nominal pipeline. The  combined analysis of the HST/WFC3 and VLT/ESPRESSO yields still substantial ($B>3$) evidence for the presence of Na and K in the data, which would correspond to $\approx 2.4 \sigma$, using \citet[][Table 3]{Benneke2013}.

   Corner plots of the full set of retrieved parameters for the full retrieval models and those without \ce{H2O} and without Na/K ($B>3$) can be found in the appendix (Figures~\ref{fig: Corner_plot_Full_H2O_nom}, \ref{fig: Corner_plot_no_H2O_nom}, \ref{fig: Corner_plot_H2O_CASCADE}, \ref{fig: Corner_plot_no_H2O_CASCADE}, \ref{fig: Corner_plot_VLT_nom}, \ref{fig: Corner_plot_VLT_no_Na_K}, \ref{fig: Corner_plot_VLT_CASCADE}, and \ref{fig: Corner_plot_VLT_no_Na_K_CASCADE}).

\subsection{Atmospheric retrieval}
\label{sec: retrieval}

To constrain the atmospheric parameters of WASP-117b further, we again ran forward atmospheric models using the open-source code \texttt{petitRADTRANS} \citep{Molliere2019}, this time retrieving the planetary metallicity and C/O value directly. Our retrieval setup was applied on only the HST/WFC3 data first, and then on the combination of the HST/WFC3 and ESPRESSO data.

We constructed the retrieval forward model with the following rationale: in principle, retrievals of transmission spectra allow for many properties of the atmosphere to be constrained, such as the average terminator temperature and temperature gradient, abundances of absorbers, as well as the cloud properties such as cloud base position, scale height, average particle sizes and cloudiness fraction of the terminator \citep[e.g.,][]{barstowaigrain2013,rocchettowaldmann2016,lineparmentier2016,macdonaldmadhusudhan2017,Molliere2019,barstow2020}. Also differences between morning and evening terminators can likely be constrained or lead to erroneous conclusions, if ignored \citep{macdonaldgoyal2020}. The same holds for variations between the day and nightside, probed across the terminator \citep[][]{caldasleconte2019,plurielzingales2020}. The number of free parameters, that is, the complexity of the retrieval model, needs to be justified by the quality of the data, however. Too complex models should be avoided for observations of low S/N, and in general the number of free parameters should be less than the number of data points. This prevents overfitting of the data. A useful criterion to judge whether one model is better than another, given the data, is the Bayes factor \citep{Kass1995}. The Bayes factor analysis will penalize those models which are too complex, given the data. However, a Bayes factor analysis should not be applied blindly. A simple (few parameters) model, that gives a good fit to the data, will always be favored, even if the model assumptions are highly unphysical. The caveat in using Bayesian factors analysis is that Bayes factors do not know physics.

Given the limited quality of our data, we decided to keep the complexity of the retrieval 1D~forward model low, with a limited number of free parameters (see Table~\ref{table: models}). More specifically, we assumed an isothermal temperature structure, whereas the absorber abundances are here not kept as free parameters but are modelled with a chemical equilibrium model, using the chemistry code described in the appendix of \citet{Molliere2017}. Our abundance treatment for WASP-117b is thus analogous to \citet{Kreidberg2018} for WASP-107b. WASP-107b could be similar in temperature and atmosphere chemistry to WASP-117b during transit. 

While \texttt{petitRADTRANS} offers a wide range of different cloud parameterizations, we decided to only retrieve a gray cloud deck pressure and, optionally, a scattering-slope opacity to model small-particle hazes. The haze opacity is parameterized as
\begin{equation}
\kappa_{\rm haze}= \kappa_0 \left( \frac{\lambda}{\lambda_0}\right)^{\gamma}   
\end{equation}
in the retrieval model, where $\gamma$ determines the steepness of the scattering slope, which we constrain to $\gamma >-6$ to avoid unphysically steep slopes \citep{pinhasmadhu2017,barstow2020}. The parameters $\kappa_0$ and $\lambda_0$ are reference opacities and wavelengths, where the first is freely retrieved and the latter is set to $\lambda_0=0.35$~$\mu$m. The unit of $\kappa_0$ is cm$^2$~g$^{-1}$ and the haze opacity is assumed to be vertically constant.

Assuming a 1D temperature structure for transmission spectroscopy during transit could lead to underestimating the atmospheric temperature at the limbs as indicated by \citet{macdonaldgoyal2020}. In addition, patchy clouds could mimic situations of high enrichment \citep{lineparmentier2016}, which is another possible limitation to keep in mind.

\subsection{HST/WFC3 Atmosphere retrieval}
\label{sec: atmosphere models}

In the following, we will discuss constraints on the atmospheric composition of  WASP-117b,  based  on atmospheric retrieval of the HST/WFC3 transmission spectrum using the nominal (Section~\ref{sec: Zhou}) and the CASCADE pipeline (Section~\ref{sec: CASCADE}), respectively.

We employed several models with physically motivated priors to constrain the atmospheric properties of WASP-117b. The models and their priors are listed in Table~\ref{table: models}. \textit{Model 1} and \textit{Model 2} represent atmospheric models with isothermal temperature, equilibrium chemistry and a gray cloud deck. \textit{Model 1} sets weak constraints on metallicity within $[-1.5, +3]$, \textit{Model 2} sets constraints to lower metallicity $\rm{[Fe/H]}<1.75$. \textit{Model 3} assumes the same parameters and priors as \textit{Model 2} and adopts additionally a haze layer. 

\begin{table}[t!]
\caption{List of models used for WASP-117b atmosphere retrieval}             
\label{table: models}     
\centering 
{\renewcommand{\arraystretch}{1.2}%
\begin{tabular}{c | p{3.5 cm} | p{2.5 cm}}       
\hline
\hline
Model name  & prior in parameters & short description \\   
\hline 
\hline
  \textit{Model 1} &  \begin{tabular}[c]{@{}c@{}}T $\in [500, 1500]$~K\\ $\log(P_0/\mathrm{1 bar}) \in [-12, +2]$ \\ $\mathrm{C/O} \in [0.1, 1.6]$  \\${\rm [Fe/H]} \in [-1.5, +3]$ \end{tabular} & high $\rm [Fe/H]$  \& gray cloud\\
  & $\log(P_{\rm cloud}) \in [-6, +2]$ & \\
  \hline
  \textit{Model 2} & \begin{tabular}[c]{@{}c@{}} as in Model 1 except \\$  {\rm [Fe/H]} \in [-1.5, +1.75]$ \end{tabular}  & low $\rm [Fe/H]$  \& gray cloud\\
  \hline
  \textit{Model 3} &  \begin{tabular}[c]{@{}c@{}} as in Model 1 except \\${\rm [Fe/H]} \in [-1.5, +1.75]$ \\  $\gamma \in [-6,0]$ \\ $\log(\kappa_0) \in [-10,+10]$ \end{tabular}  & low $\rm [Fe/H]$   \& gray cloud \& haze layer\\
\hline                      
\end{tabular}}
\end{table}

\begin{table*}
\caption{Median Parameters and 68\% Confidence Intervals for \textit{Models 1, 2} and \textit{3} retrieved with HST/WFC3 spectra based on two different data reduction pipelines.}             
\label{table: HST_retrieval}     
\centering
{\renewcommand{\arraystretch}{1.2}%
\begin{tabular}{p{2.35 cm} ||c c || c c || c c}       
\hline

& \multicolumn{2}{c}{\textit{Model 1}} & \multicolumn{2}{c}{\textit{Model 2}} & \multicolumn{2}{c}{\textit{Model 3}}\\
\hline
Parameter & Nominal & CASCADE & Nominal & CASCADE & Nominal & CASCADE\\   
\hline 
\hline
   Temperature [K] & $1129^{+228}_{-289}$ &  $821^{+264}_{-226}$ & $1002^{+348}_{-350}$ & $674^{+392}_{-135}$ & $860^{+371}_{-257}$ & $738^{+360}_{-157}$\\ 
   Pressure $\log(P_0/\mathrm{1 bar})$ & $-6.69^{+0.99}_{-1.21}$ & $-7.20^{+1.11}_{-1.51}$ & $-4.44^{+1.52}_{-0.97}$ & $-4.45^{+1.47}_{-1.01}$ & $-4.49^{+1.33}_{-1.13}$ & $-4.89^{+1.23}_{-1.09}$\\%
    $\rm{[Fe/H]}$ & $2.58^{+0.26}_{-0.37}$ & $2.55^{+0.28}_{-0.35}$ & $0.66^{+0.85}_{-1.64}$ &  $0.54^{+0.97}_{-1.62}$& $0.41^{+0.97}_{-1.26}$ & $0.67^{+0.81}_{-1.33}$\\
   C/O & $0.25^{+0.17}_{-0.10}$ & $0.27^{+0.24}_{-0.11}$ & $0.37^{+0.30}_{-0.19}$ & $0.36^{+0.41}_{-0.19}$ & $0.40^{+0.48}_{-0.20}$ & $0.38^{+0.40}_{-0.19}$ \\      %
   Cloud top \par $\log(P_{\rm cloud}/\mathrm{1 bar})$ & $-1.32^{+2.24}_{-2.44}$ & $-0.78^{+1.77}_{-1.77}$ & $-3.69^{+1.42}_{-0.97}$& $-3.30^{+1.29}_{-1.03}$& $-3.69^{+1.27}_{-1.15}$ & $-2.97^{+2.32}_{-1.49}$ \\
    $\log(\kappa_0)$ & - & - & - &- & $-4.91^{+3.61}_{-3.10}$ &  $-1.99^{+2.84}_{-5.22}$ \\
   Scattering slope $\gamma$ & - & - & - & - & $-3.55^{+1.89}_{-1.57}$ &  $-2.61^{+1.59}_{-2.09}$ \\
   \hline
   & \multicolumn{6}{c}{volume fractions within $1 \sigma$ }\\
   \hline
   log(H$_2$O) at \par $10^{-4}$~bar& $-0.62_{-0.51}^{+0.26}$ & $-0.75^{+0.34}_{-0.66}$ & $-2.70_{-1.63}^{+1.15}$ & $-2.68^{+1.17}_{-1.77}$ & $-2.96_{-1.50}^{+1.21}$ & $2.66^{+1.07}_{-1.66}$ \\
   log(CH$_4$) at \par $10^{-4}$~bar  & $-13.19_{-6.66}^{+3.87}$ & $-9.34^{+5.38}_{-5.92}$ &  $-9.89^{+5.39}_{-2.88}$ & $-4.75^{+1.58}_{-5.43}$ & $-7.55^{+3.63}_{-4.41}$ & $-5.28^{+2.07}_{-5.05}$\\
\hline 
  & \multicolumn{6}{c}{ $3 \sigma$ upper limits of CH$_4$ volume fractions }\\
  \hline
   CH$_4$ at \par $10^{-4}$~bar  & $ 8.6 \cdot 10^{-2}$ & $2.0\cdot 10^{-1}$  &  $1.6 \cdot 10^{-2}$  & $1.9 \cdot 10^{-2}$ & $1.6 \cdot 10^{-2}$  & $1.7 \cdot 10^{-2}$ \\
\end{tabular}}
\end{table*}

\subsubsection{\textit{Model 1} - condensate clouds and weakly constrained $\rm{[Fe/H]}$}
\label{sec: Model 1}

 \begin{figure*}
   \centering
   \textbf{\textit{Model 1}}\par \medskip
    \textbf{Nominal}  \hspace{7.5 cm}  \textbf{CASCADE}\par
   \includegraphics[width=0.48\textwidth]{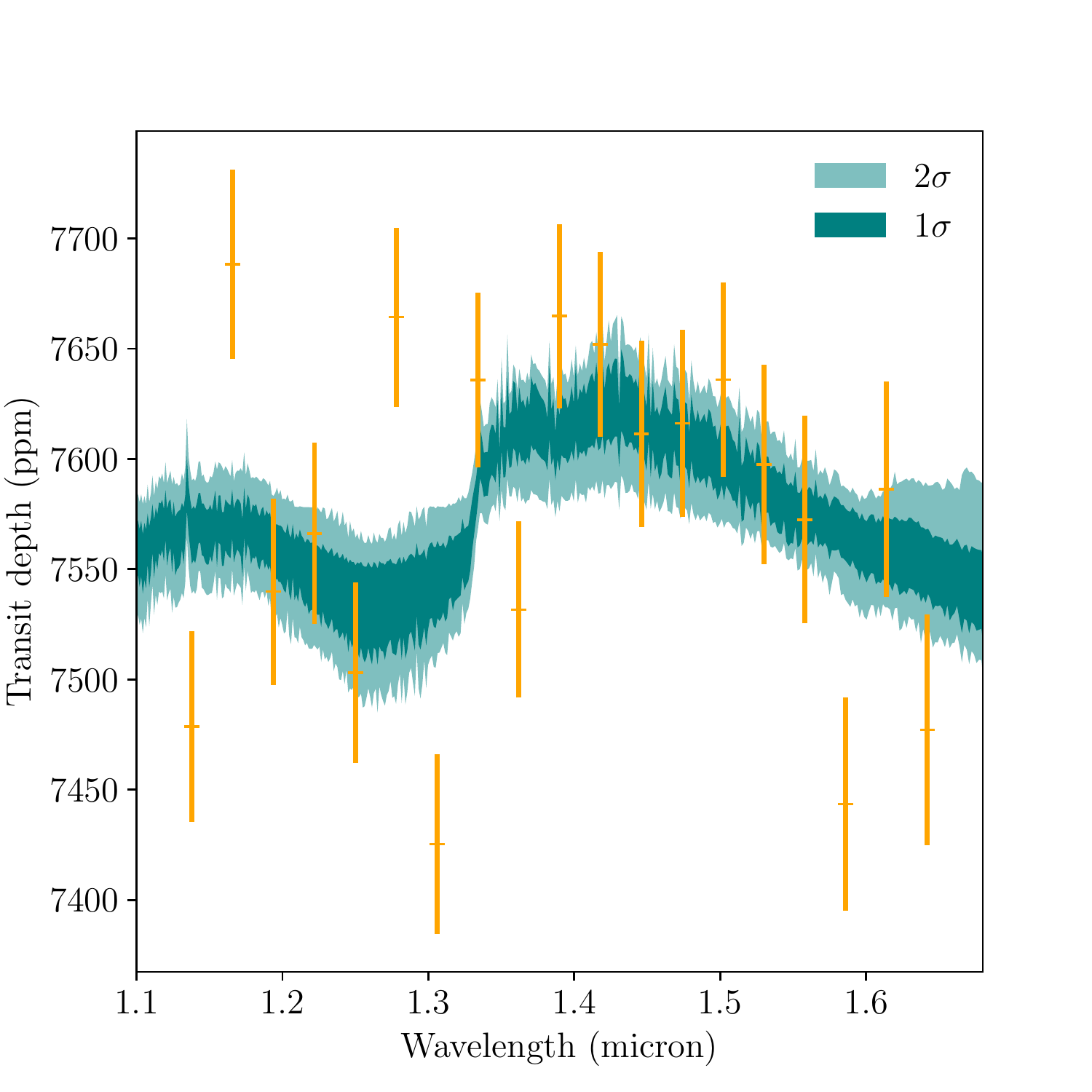}
   \includegraphics[width=0.48\textwidth]{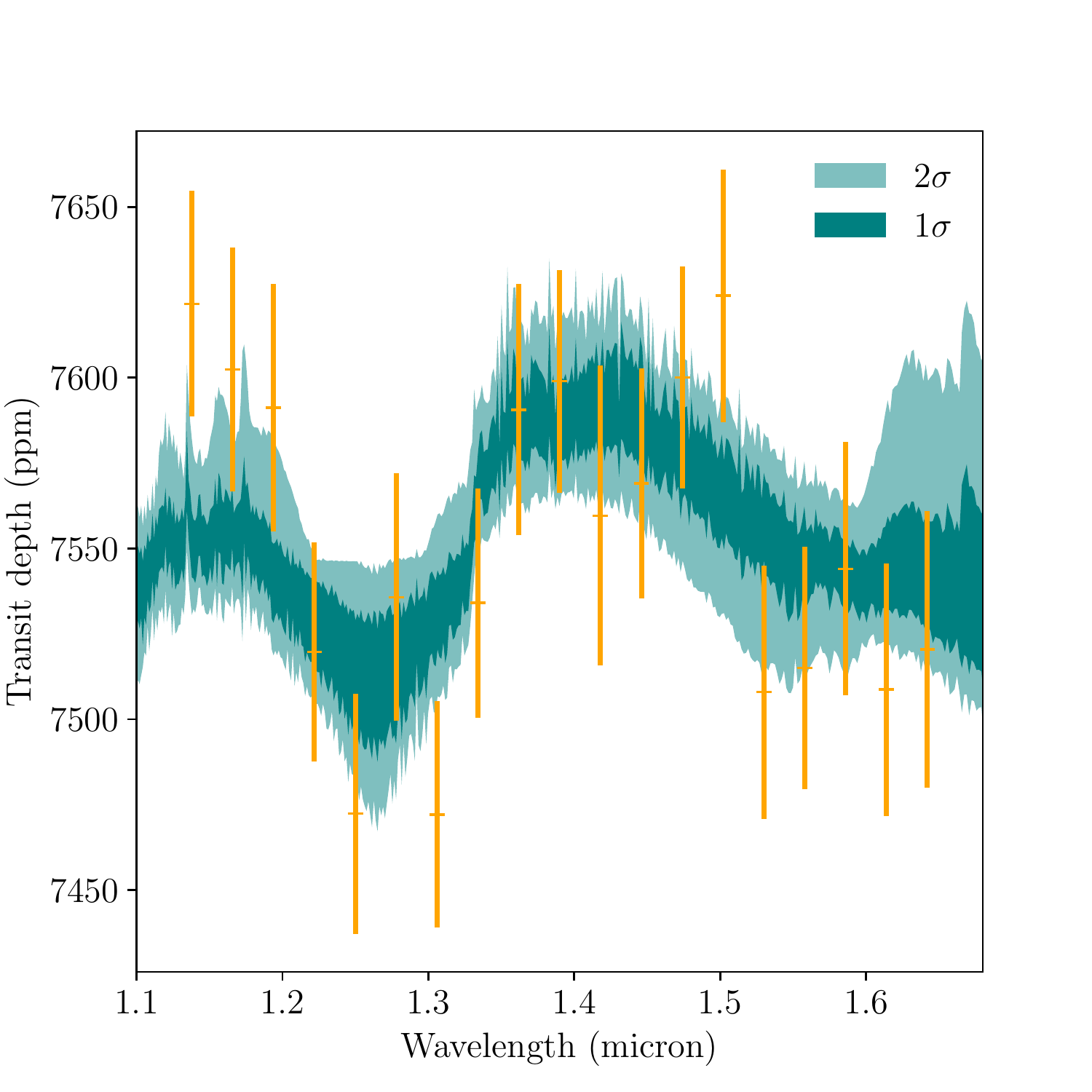}
      \caption{\textbf{Left:} \textit{Model 1} (unconstrained $\rm{[Fe/H]}$ and grey cloud) fit to WASP-117b transmission spectrum reduced with the nominal data pipeline by \citet{Zhou2017}. \textbf{Right:} \textit{Model 1} (unconstrained $\rm{[Fe/H]}$ and grey cloud) fit to WASP-117b transmission spectrum reduced with the data pipeline CASCADE. \newline 
      For both spectra, retrieved atmosphere models with confidence within $1\sigma$ (dark green) and  $2\sigma$ (light green) are shown. These models were derived with \texttt{petitRADTRANS} \citep{Molliere2019} and indicate a clear atmosphere of high metallicity, where the large mean molecular weight leads to a muted water spectrum.}
         \label{fig: Best_fit_Model1}
   \end{figure*}

Figure~\ref{fig: Best_fit_Model1} displays the 2-98 and 16-84~percentile envelopes of the retrieved transmission spectra of \textit{Model 1}, together with the WASP-117b HST data, reduced with the nominal and CASCADE pipeline, respectively. The corner plots with the retrieved properties based on \textit{Model 1} are displayed in Figures~\ref{fig: Corner_plot_Zhou_m1}, \ref{fig: Corner_plot_Bouwman_m1} and~\ref{fig: Corner_plot_H2O_CH4_m1}, and the results are listed in Table~\ref{table: HST_retrieval}. The retrieval based on the nominal spectrum yields a relatively high temperature of $1129^{+228}_{-289}$~K, whereas retrieval based on the CASCADE spectrum would indicate a rather low temperature of $821^{+264}_{-226}$. The discrepancy in temperature can be explained by data reduction differences, that lead in the nominal spectrum towards an overall deeper transit depth, that is, a more inflated atmosphere, which can be achieved by attaining higher temperatures compared to the CASCADE spectrum. However, both temperatures are within one sigma of each other, which again indicates that this shift is not substantial. We thus conclude that the atmospheric temperature is not well constrained with our data, at least when using \textit{Model 1}.

Retrieval with \textit{Model~1} further suggests very large metallicities ${\rm [Fe/H]}>2.2$ based on both spectra. At the same time, a cloud top located relatively deep in the atmosphere ($p>10^{-4}$~bar) is inferred and the reference pressure $P_0$ is set high in the atmosphere ($P_0<10^{-5.5}$) to reconcile the relatively large transit depth of this inflated exo-Saturn with the high metallicities. The high metallicity solutions thus indicate relatively clear skies with very small atmospheric scale heights. These results indicate that the water feature is strongly muted and that condensate cloud modelling alone can not properly account for the shape of the water signal. Instead, the model assumes high mean molecular weight with more than $100\times$ solar metallicity, irrespective of which data reduction pipeline is used.

Furthermore, retrieval based on both the nominal and CASCADE spectra tend to yield subsolar C/O values with an upper limit of 0.42 and 0.51 respectively. Due to this, the \ce{CH4} content is retrieved within $1 \sigma$ to be below $1.1\times 10^{-4}$ volume fraction at $p=10^{-4}$~bar, when using data retrieved with the CASCADE spectrum and even below $5.7\times 10^{-10}$ volume fraction when using the nominal pipeline. However, we note that the $3\sigma$ range is very large (see Figure~\ref{fig: Corner_plot_H2O_CH4_m1}) such that we cannot claim strong significance for these low \ce{CH4} abundances. 

Similar low C/O values were also found for WASP-107b by \citet{Kreidberg2018}. These authors likewise attribute the low C/O values to the absence of \ce{CH4} within the WFC3/G141 wavelength range, which is consistent with our retrieved low abundances of \ce{CH4}. If WASP-117b is colder than 800~K during transit and thus in a similar temperature range to WASP-107b then the very low abundance of \ce{CH4} could indicate disequilibrium chemistry. We explore this possibility in Section~\ref{sec: Disequi}.

We further note that we find a tail of solutions with lower metallicity and high cloudiness in the retrieved parameters based on \textit{Model~1} fitted to both spectra, the nominal and the CASCADE spectrum (See Figures~\ref{fig: Corner_plot_Zhou_m1} and \ref{fig: Corner_plot_Bouwman_m1} in the ${\rm [Fe/H]} - P_{cloud}$~corners, indicated with a black ellipse). We will thus explore lower metallicity solutions more carefully in the next sections.

\subsubsection{\textit{Model 2} - condensate clouds and ${\rm [Fe/H]}<1.75$}
\label{sec: Model 2}

\begin{figure*}
   \centering
   \textbf{\textit{Model 2}}\par \medskip
    \textbf{Nominal}  \hspace{7.5 cm}  \textbf{CASCADE}\par
   \includegraphics[width=0.48\textwidth]{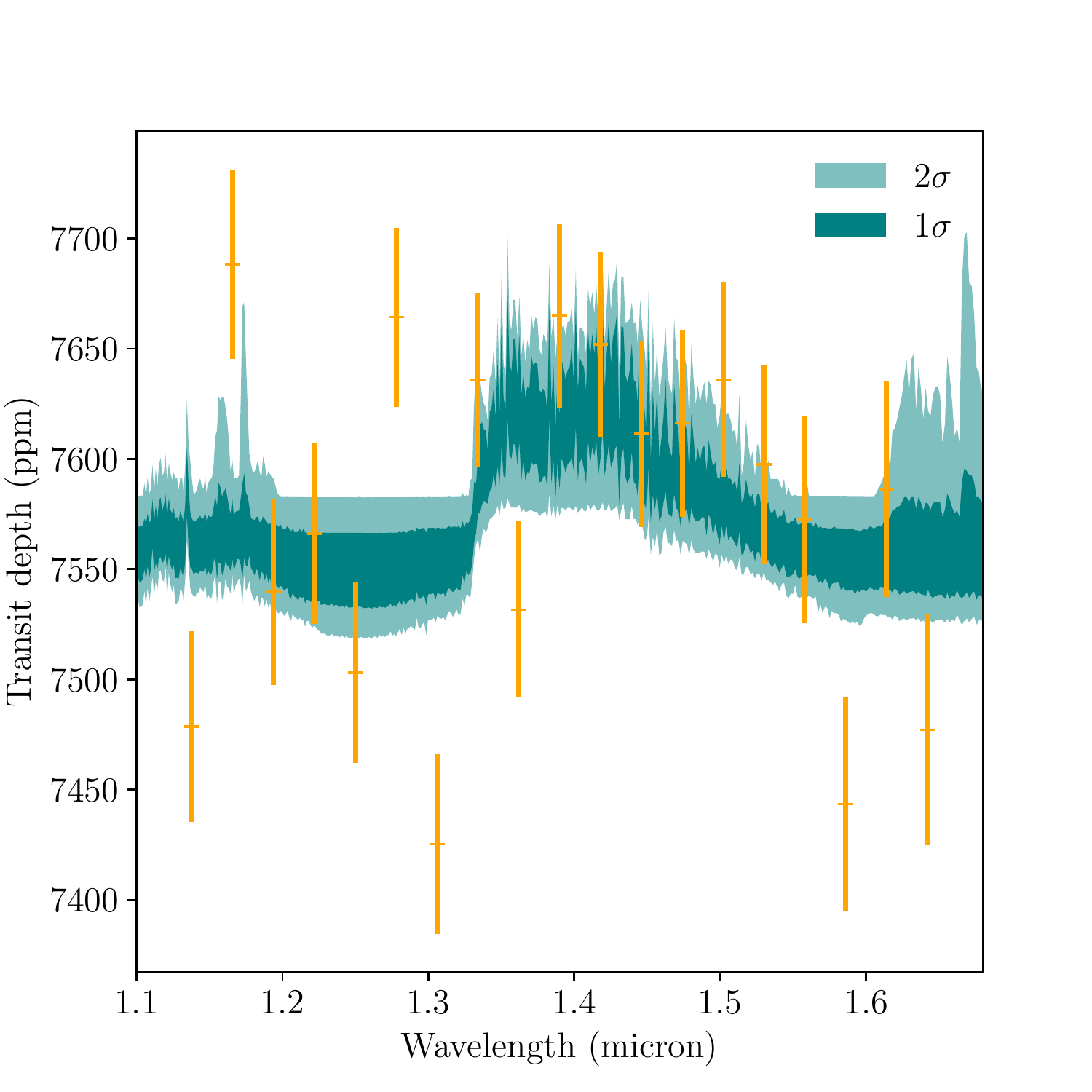}
   \includegraphics[width=0.48\textwidth]{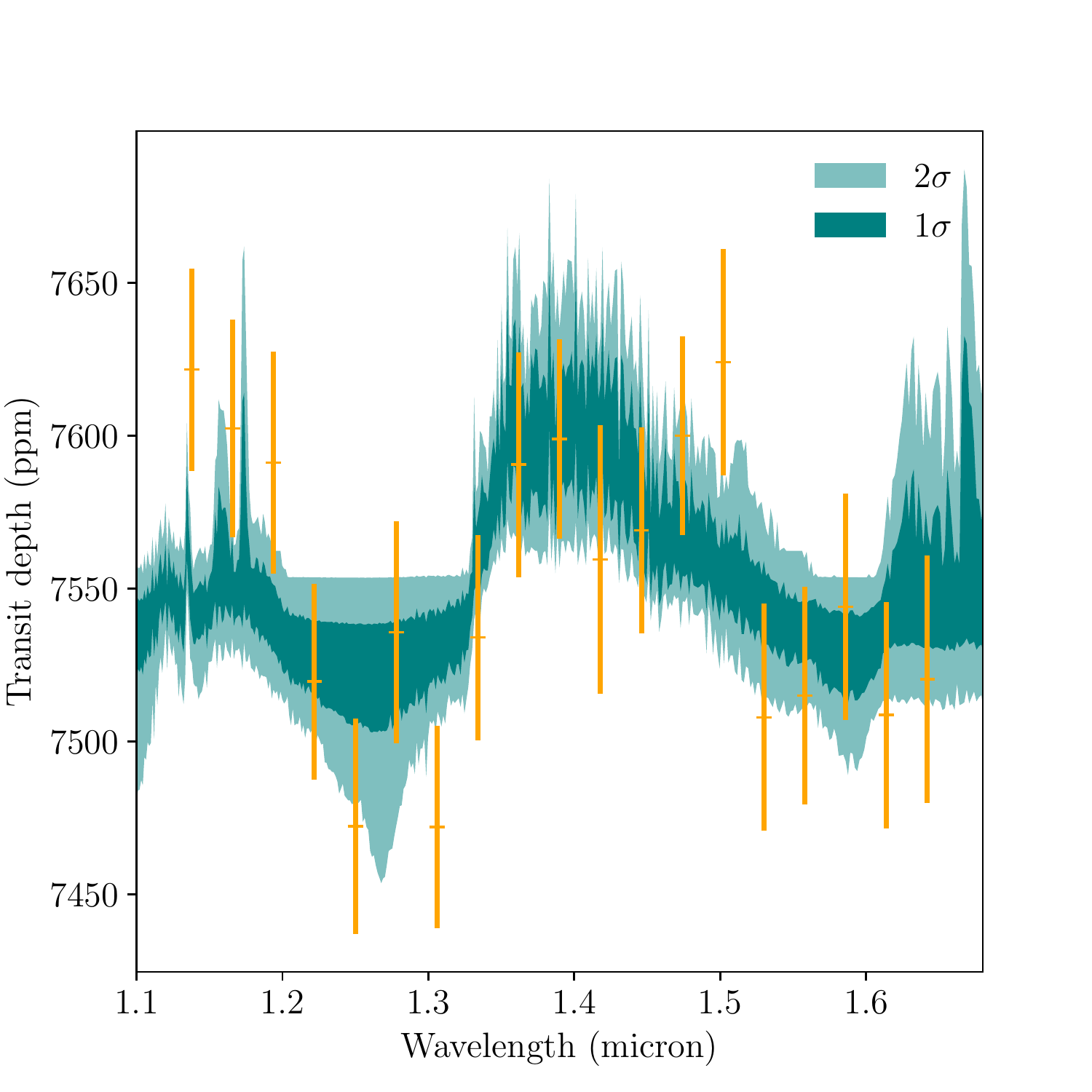}
      \caption{\textbf{Left:} \textit{Model 2} (prior ${\rm [Fe/H]}<1.75$ and grey cloud) fit to WASP-117b transmission spectrum reduced with the data pipeline by \citet{Zhou2017}. \textbf{Right:} \textit{Model 2} (prior ${\rm [Fe/H]}<1.75$and grey cloud) fit to WASP-117b transmission spectrum reduced with the data pipeline CASCADE. \newline 
      For both spectra, retrieved atmosphere models with confidence within $1\sigma$ (dark green) and  $2\sigma$ (light green) are shown. These models were derived with \texttt{petitRADTRANS} \citep{Molliere2019} and indicate a cloudy atmosphere with a muted water feature.
              }
         \label{fig: Best_fit_Model2}
   \end{figure*}

We designed \textit{Model 2} with a constraint on allowed metallicities as agnostically as possible to select for the tail of cloudy low metallicity solutions of \textit{Model 1}. We adapted again the simplest (uniform) cloud model for retrieval, because the information content in the muted observed water spectrum observed with HST/WFC3 is too limited to warrant a more complex cloud model with e.g. patchy clouds as proposed by \citet{lineparmentier2016}.

We set an upper limit of $56\times$ solar~metallicity, that is, we imposed a prior for $\rm{[Fe/H]}<1.75$. Figure~\ref{fig: Best_fit_Model2} displays the best fit of \textit{Model 2} to the WASP-117b HST data, reduced with the nominal and CASCADE pipelines, respectively. The detailed corner plots with the retrieved properties based on \textit{Model 2} are displayed in Figures~\ref{fig: Corner_plot_Zhou_m2}, \ref{fig: Corner_plot_Bouwman_m2} and \ref{fig: Corner_plot_H2O_CH4_m2}. Table~\ref{table: HST_retrieval} lists concisely the results of the retrieval.

We note that the retrieved atmospheric metallicity for WASP-117b is not well constrained for \textit{Model 2} with a tendency to still favor solutions at the higher end of the imposed prior. On the other hand, no matter which data reduction pipeline is used, nominal or CASCADE, in both cases we retrieve solutions with better constrained cloud deck located higher in the atmospheric compared to \textit{Model 1}. With \textit{Model 2}, the cloud deck is retrieved to lie between $p=10^{-5.1}$ and $p=10^{-2.2}$~bar. Since we excluded higher metallicities that reduce the scale height to very low values, the reference pressure $P_0$ is deeper in the atmosphere compared to \textit{Model 1}. In other words, in \textit{Model 2} cloudy solutions are favored to explain the muted water feature.

Sub-solar C/O values that indicate low abundances of CH$_4$ are also favored by \textit{Model 2}. However, we report a division between possible solutions within our constrained metallicity range. There is a tail of high $\rm{C/O} > 1 $ solutions (Figures~\ref{fig: Corner_plot_Zhou_m2} and \ref{fig: Corner_plot_Bouwman_m2}), which are correlated in the posterior of our retrieval with low metallicities ($\rm[Fe/H]<-0.5$). 

For the $\rm{C/O} > 1$ solutions, we further find that the $\ce{CH4}$ volume fraction at $p=10^{-4}$~bar is greater than that of \ce{H2O}. For $\rm{C/O} < 1$ solutions, the \ce{CH4} abundances are always lower than the \ce{H2O} abundances. A dominance of \ce{CH4} over \ce{H2O} is expected for high C/O ratios \citep{Maddu2012,Molliere2015}. However, we stress that these high C/O solutions only represent a small subset of solutions, associated with low metallicities [Fe/H]~$< -0.5$, and only occur when we impose a metallicity constraint of ${\rm [Fe/H]}<1.75$ on retrieval. The majority of the solutions are high metallicity solutions ($\rm[Fe/H]>1$)) that still strongly favor solar to sub-solar C/O ratios also in \textit{Model 2}. 

The low C/O ratios appear again to correspond to low \ce{CH4} abundances (in any case lower compared to \ce{H2O} abundances), where we retrieved within $1 \sigma$ \ce{CH4} abundances below $6.7\times 10^{-4}$ volume fraction at $10^{-4}$~bar. Also for \textit{Model 2}, the $3\sigma$ envelope of \ce{CH4} abundances is very large (Figure~\ref{fig: Corner_plot_H2O_CH4_m2}). 

\subsubsection{\textit{Model 3} - condensate clouds, prior ${\rm [Fe/H]}<1.75$ and haze layer}
\label{sec: Model 3}
\begin{figure*}
   \centering
   \textbf{\textit{Model 3}}\par \medskip
    \textbf{Nominal}  \hspace{7.5 cm}  \textbf{CASCADE}\par
   \includegraphics[width=0.48\textwidth]{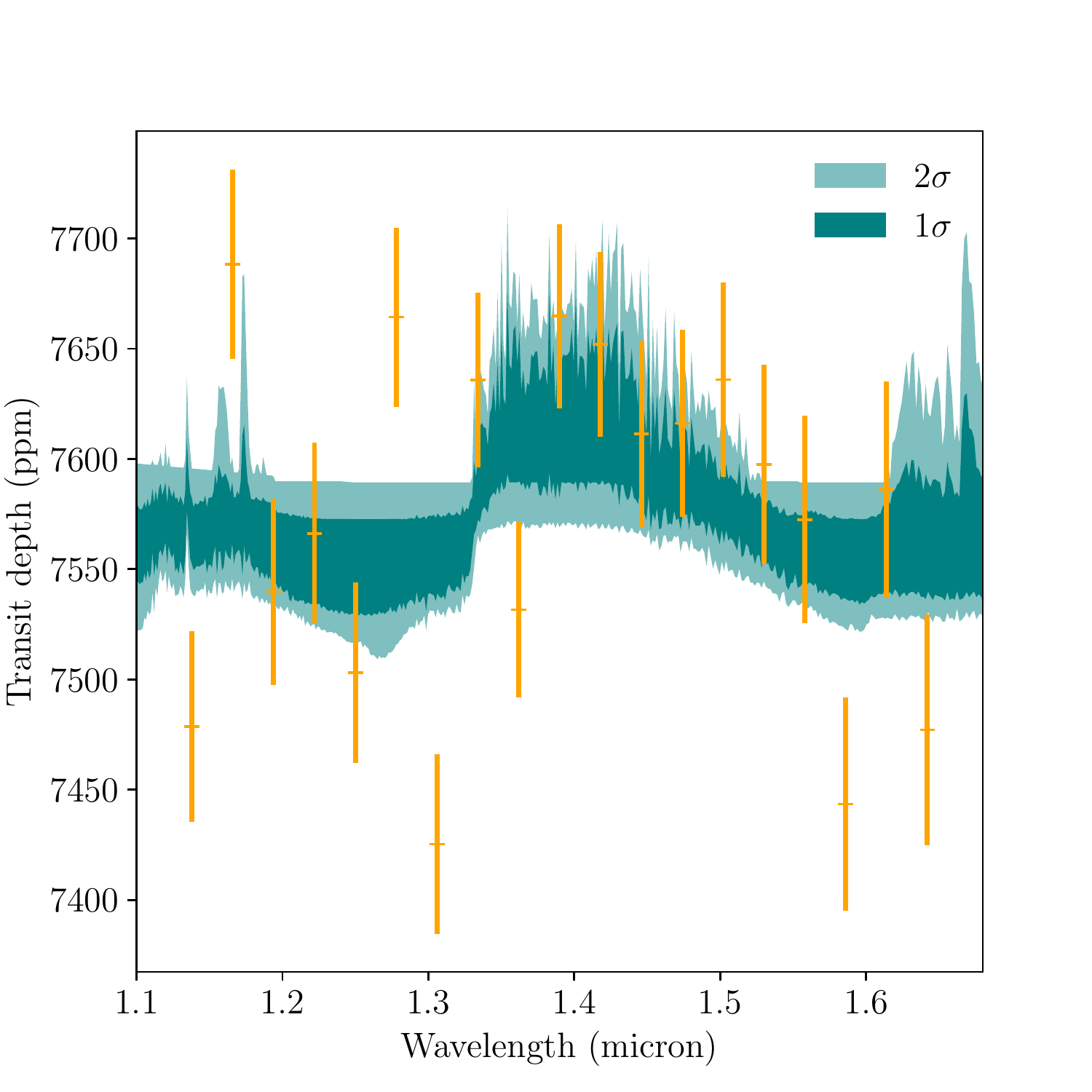}
   \includegraphics[width=0.48\textwidth]{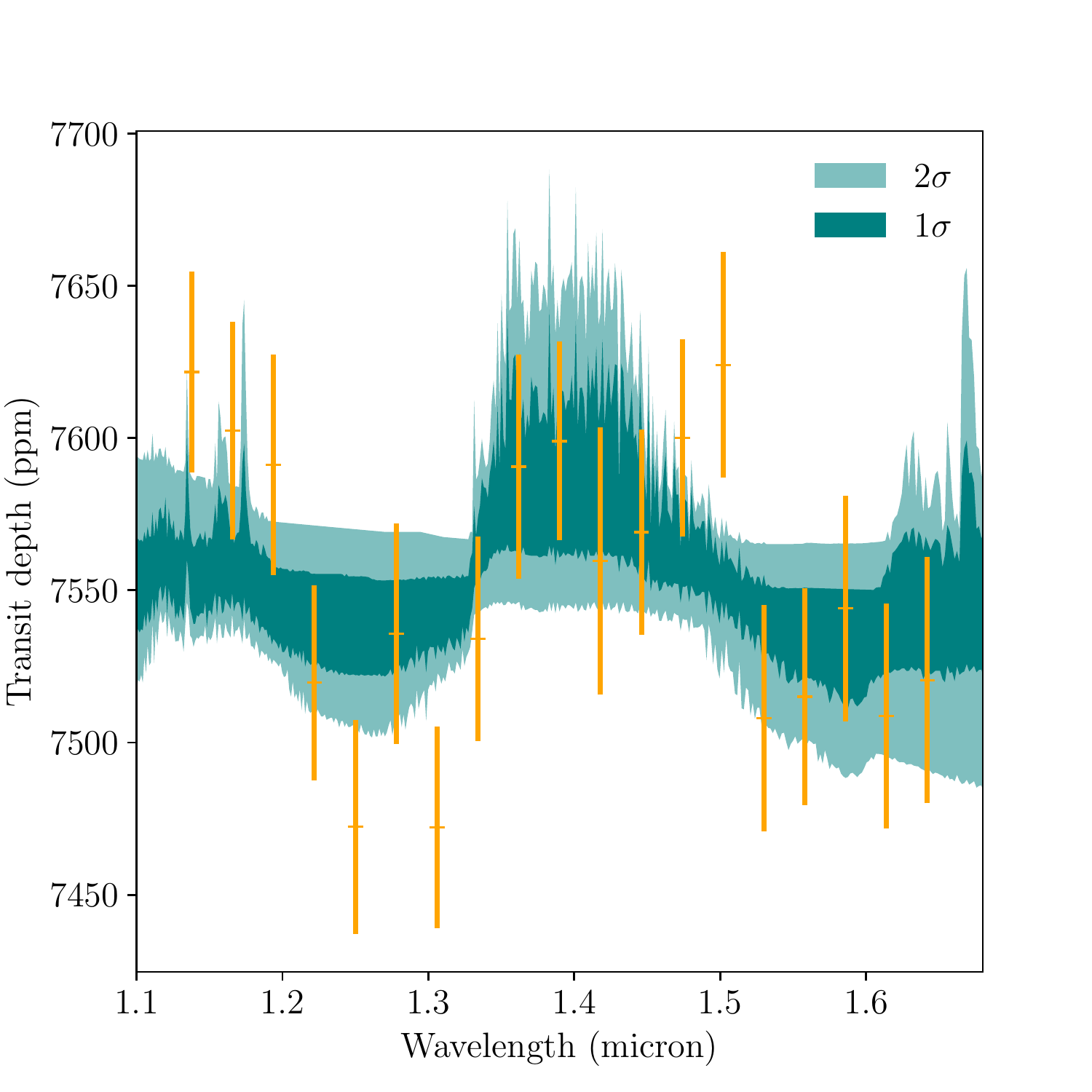}
      \caption{\textbf{Left:} \textit{Model 3} (grey cloud and haze layer with prior ${\rm [Fe/H]}<1.75$ and $\gamma >-6$) fit to WASP-117b transmission spectrum reduced with the nominal data pipeline by \citet{Zhou2017}. \textbf{Right:} \textit{Model 3} (grey cloud and haze layer with prior ${\rm [Fe/H]}<1.75$ and $\gamma >-6$) fit to WASP-117b transmission spectrum reduced with the data pipeline CASCADE. \newline 
      For both spectra, retrieved atmosphere models with confidence within $1\sigma$ (dark green) and  $2\sigma$ (light green) are shown. These models were derived with \texttt{petitRADTRANS} \citep{Molliere2019} and indicate a cloudy atmosphere with a muted water feature.
              }
         \label{fig: Best_fit_Model3}
   \end{figure*}

Another possibility, which may explain the muted observed water spectrum, could be the existence of a haze layer on top of the cloudy atmosphere. This possibility was explored with \textit{Model 3}. Figure~\ref{fig: Best_fit_Model3} displays the best fit of \textit{Model 3} to the WASP-117b HST data, reduced with the nominal and CASCADE pipelines, respectively. The detailed corner plots with the retrieved properties based on \textit{Model 3} are displayed in Figures~\ref{fig: Corner_plot_Zhou_m3}, \ref{fig: Corner_plot_Bouwman_m3}, and \ref{fig: Corner_plot_H2O_CH4_m3}. A summary of the parameters retrieved with this model is given in Table~\ref{table: HST_retrieval}.

For the CASCADE spectrum, \textit{Model 3} yields a bi-modal posterior distribution for the haze opacity. The median value is ${\rm log}(\kappa_0)=-1.99^{+2.84}_{-5.22}$, with the high-$\kappa_0$ peak ocurring at ${\rm log}(\kappa_0)=0.15$ and the low peak ocurring at ${\rm log}(\kappa_0)=-5.8$. The low peak is consistent with the posterior of Model 3 for the nominal spectrum while the high peak indicates some of the models are hazy with a moderate scattering slope of $\gamma=-2.61^{+1.59}_{-2.09}$ (Figure~\ref{fig: Best_fit_Model3}, right). High values of $\kappa_0$ also correspond to a gray cloud deck that can lie deeper in the atmosphere compared to \textit{Model 2}. The water feature is for some models inside the \textit{Model 3} framework muted by the combined effect of condensed clouds and the haze layer. 

 For the nominal spectrum, however, \textit{Model 3} is clearly a less good fit (Figure~\ref{fig: Best_fit_Model3}, left). Comparison between the corner plots Figure~\ref{fig: Corner_plot_Zhou_m3} and Figure~\ref{fig: Corner_plot_Bouwman_m3} show that the retrieved haze layer properties $\kappa_0$ and $\gamma$ in the nominal spectrum are more uniformly spread across the allowed parameter range compared to the CASCADE spectrum. Generally, the retrieved parameters based on the nominal spectrum are, however, for \textit{Model 3} well within $1 \sigma$ of the parameters retrieved with the CASCADE spectrum as a basis. The sub-set of hazy solutions identified in the posterior of retrieval with \textit{Model 3} based on the CASCADE spectrum are thus not substantial enough to indicate a strong disagreement between data reduction pipelines that are used in this work.
 
 The metallicity is like in \textit{Model 2} unconstrained in \textit{Model 3} within imposed low metallicity limits of ${\rm [Fe/H]}=[-1.5, 1.75]$, irrespective of which pipeline is used. The temperature is again not well constrained in \textit{Model 3}, as can be seen when comparing Figure~\ref{fig: Corner_plot_Bouwman_m3} and Figure~\ref{fig: Corner_plot_Bouwman_m2}. 

In \textit{Model 3}, we identify again a tail of high C/O ratio solutions as in \textit{Model 2}. Most solutions, especially those with $\rm[Fe/H]>1$, favor, however, subsolar C/O ratios. As noted in previous subsections, low C/O values are indicative of low abundances of methane. This is reflected again by retrieved \ce{CH4} abundances smaller than $10^{-4}$ volume fraction at $p=10^{-4}$~bar within $1 \sigma$ for the nominal reduction pipeline. We note that fitting \textit{Model 3} to the CASCADE spectrum, allows even up to  $10^{-3.2}$ volume fraction at $p=10^{-4}$~bar within 1$\sigma$.

The significance of the molecular absorption feature as well as that of a haze layer is investigated in the following.

\subsection{Significance of differences in metallicity and haze layer}
\label{sec: Significance_model}

To quantify the different models that we used for retrieval, we compare now \textit{Model 1, 2} and \textit{3} with each other, using the same method as in Section~\ref{sec: Significance}. In Table~\ref{table: HST_evidence}, we list again the (natural) log evidences (${\rm ln} Z$) and the Bayes factor $B$. 

\begin{table}
\caption{Statistical analysis of WASP-117b HST data for different models.}             
\label{table: HST_evidence}     
\centering                        
\begin{tabular}{c c c |c}       
\hline\hline                
Model & ln Z & Bayes factor B & $\chi^2$\\   
\hline 
\multicolumn{4}{c}{Nominal spectrum}\\
\hline
   \textit{Model 1} & $-31.59$  &  \textbf{baseline} & \textbf{47.9}\\
   \textit{Model 2} & $-33.05$ & 4.3 & \\
   \textit{Model 3} & $-34.5$ & 18.4 & \\
   \hline 
\multicolumn{4}{c}{CASCADE spectrum}\\
\hline
   \textit{Model 1} & $-17.83$  & \textbf{baseline}& \textbf{19.1}\\
   \textit{Model 2} & $-20.46$ & 13.9 &\\
   \textit{Model 3} & $-21.28$ & 31.5 &\\
\hline                      
\end{tabular}
\end{table}

According to this statistical analysis,  \textit{Model 1}, a model with high atmospheric metallicity  ($\rm{[Fe/H]} \geq 2.2$) and clear skies fits the data best compared to the lower metallicity, cloudy models \textit{Model 2}  and \textit{Model 3}, where the latter also assumes a haze layer on top. This is true for both pipelines.

The Bayesian preference is, however, larger for the CASCADE spectrum compared to the nominal spectrum. Based on data reduced with the CASCADE pipeline, the preference is ``substantial'' to ``strong''. Based on data reduced with the nominal spectrum, the preference is still ``substantial'', that is, $1 - 2 \sigma$.

Theoretical predictions of \citet{Thorngren2019} indicate that up to $200 \times$ solar metallicity are in principle possible for Saturn-mass exoplanets like WASP-117b. We again point out the possible degeneracy of high metallicity with patchy cloud solutions \citep{lineparmentier2016}.

 We also added a $\chi^2$ analysis for \textit{Model 1} to quantify yet again agreement of the data derived from different pipelines in comparison with the best fit physical model (Table~\ref{table: HST_evidence}). The $\chi^2$ deviation as well as the scatter in data points below 1.35~$\mu$m in Figure~\ref{fig: WASP-117b_orbit} indicates some problem with the nominal pipeline. We demonstrated in Section~\ref{sec: Pipeline_comparison} that the discrepancy below 1.35~$\mu$m is likely caused by the difference in calibrating the telescope’s pointing movement in the dispersion direction. Therefore, we performed retrieval analyses on spectra obtained using both pipelines. Since we found fully consistent results, this confirms the high fidelity of our conclusions.
 
 Our work illustrates that even with the well established nominal pipeline, one has to be aware of instrumental effects that may yield a worse performance as expected. It further highlights the importance of using independent data reduction pipelines to confirm the analysis results. Still, better and additional data e.g. in the optical range are needed for further constraints of cloud and haze coverage of WASP-117b and thus atmospheric metallicity. We already obtained additional observations of WASP-117b with VLT/ESPRESSO (Section~\ref{sec: VLT_obs}) to support the HST/WFC3 observation. In addition, we can analyze WASP-117b transit data obtained by the TESS satellite (Section~\ref{sec: TESS}).

\subsection{Rossiter–McLaughlin effect measured with VLT/ESPRESSO}
\label{sec: RM}

\begin{figure}
   \centering  
   \includegraphics[width=0.48\textwidth]{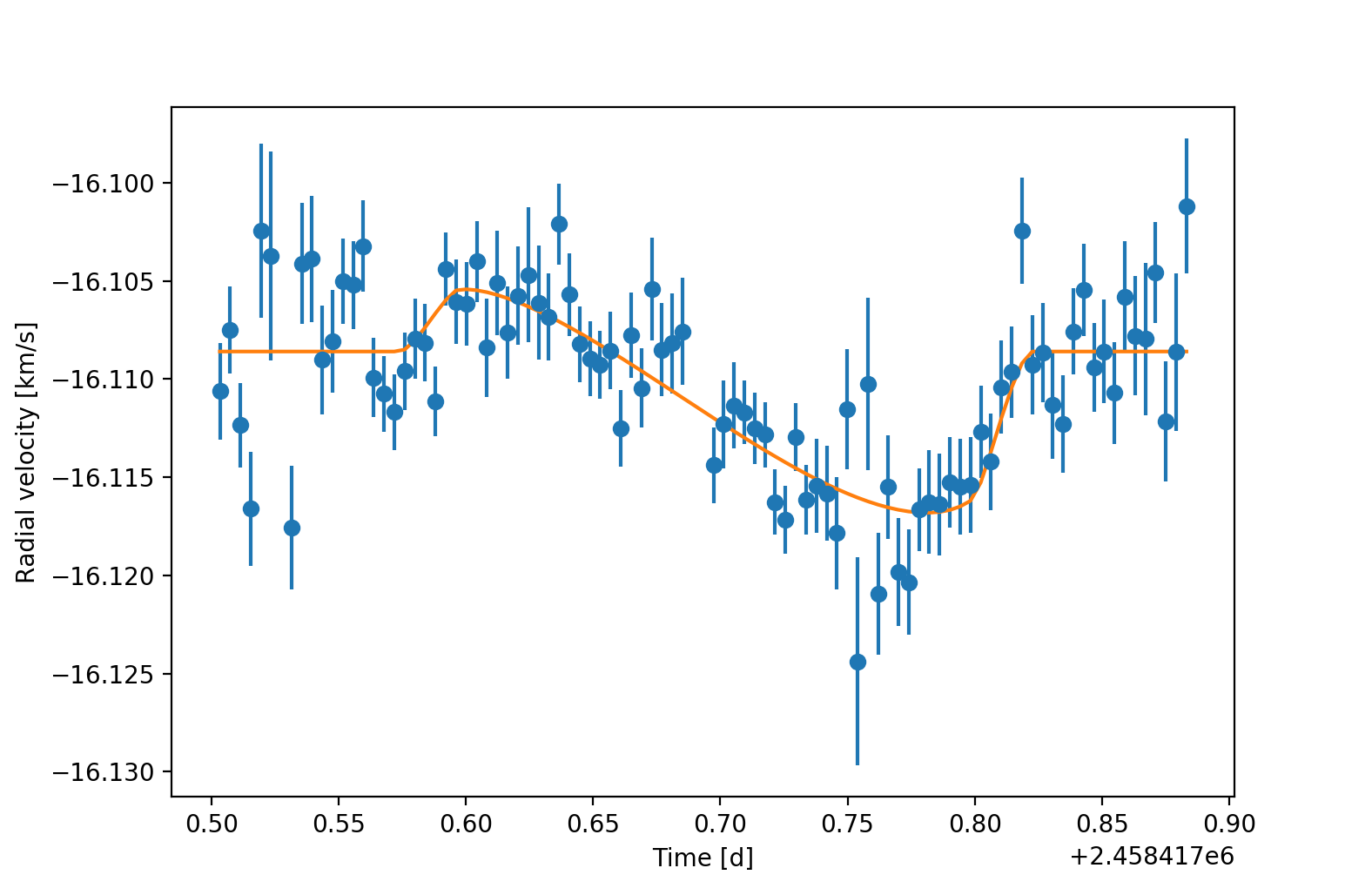}
      \caption{ESPRESSO  RVs  obtained  on  25  October  2018 during a transit of WASP-117b. The barycentric Earth radial velocity and the stellar orbital velocity are corrected. The orange line shows the best-fit model of the Rossiter-McLaughlin effect.
              }
         \label{fig: RM_effect}
\end{figure}

From our high resolution ESPRESSO data we could also improve the constraints on the stellar rotation and spin-orbit alignment of WASP-117b, via the Rossiter–McLaughlin (RM) effect. The RM of the system was initially measured by \cite{Lendl2014} using one transit dataset from the HARPS observation. We fitted the RM effect using our ESPRESSO data to update the obliquity parameters. We applied a Markov chain Monte Carlo (MCMC) approach using the \texttt{emcee} tool \citep{Mackey2013}. 
The RM effect was modelled with the python code \texttt{RmcLell} from the PyAstronomy library\footnote{https://github.com/sczesla/PyAstronomy} \citep{Czesla2019}.
The projected stellar rotation velocity ($v\,\mathrm{sin}i_\star$), projected spin-orbit angle ($\lambda$), and systemic velocity ($V_\mathrm{sys}$) were set as free parameters and we fixed other parameters to the values in \cite{Lendl2014}.
The RV curve together with the best-fit model are presented in Fig.~\ref{fig: RM_effect}. The retrieved parameters are listed in Table \ref{table:RM}.

We did not use these updated parameters in the analyses for the rest of this work, for which the original data by \citet{Lendl2014} was already sufficient to reduce the HST/WFC3 and TESS data before the updated VLT/EPRESSO data was available to us. Still, we present the updated RM values here for completeness sake and for inclusion in future work.

\begin{table}
\caption{Retrieved parameters from the Rossiter-McLaughlin effect.}             
\label{table:RM}      
\begin{tabular}{c | c }        
\hline\hline                 
\multicolumn{2}{c}{This work} \\
\hline
$v\,\mathrm{sin}i_\star$ & $1.46\pm0.14$ km\,s$^{-1}$\\
projected spin-orbit angle $\lambda$ & $-46.9 _{-4.8}^{+5.5}$ deg\\
systemic velocity  $V_\mathrm{sys}$ & $-16.1086\pm0.0004$ km\,s$^{-1}$\\    
\hline 
\multicolumn{2}{c}{\citet{Lendl2014}} \\
\hline
$v\,\mathrm{sin}i_\star$ & $1.67 _{-0.24}^{+0.31}$ km\,s$^{-1}$ \\
projected spin-orbit angle $\lambda$ & $-44\pm11$~deg\\
\hline      
\hline                                   
\end{tabular}
\end{table}

\subsection{TESS transit data}
\label{sec: TESS}

We further found that the Transiting Exoplanet Survey Satellite (TESS) has acquired four observations of WASP-117 between August 26 2018 and October 13 2018 during Sectors 2 and 3 of TESS' primary mission, covering four consecutive transits of WASP-117b\footnote{https://exofop.ipac.caltech.edu/tess/target.php?id=166739520}. These transit data are more accurate than the transit depth measured from the ground as reported in the discovery paper \citep{Lendl2014} from the WASP South survey and could potentially put further constraints on the cloud properties of WASP-117b.

 Interestingly, the reported TESS transit depth of $7200\pm 150$~ppm was found to be very shallow (albeit within $3 \sigma$) compared to the value reported by \cite{Lendl2014} ($8030^{+550}_{-480}$~ppm) and by our HST/WFC3 nominal transit depth ($7573 \pm 43$~ppm). We thus decided to perform an independent analysis on both the published TESS photometry and the target-pixel files (TPFs) of this target, to verify this slight discrepancy. For the former, we used the PDC lightcurve published at MAST\footnote{\url{https://archive.stsci.edu/}}, which we fitted using \texttt{juliet} \citep{Espinoza2019}. As priors for this fit, we used the eccentricity and argument of periastron as reported by \citet{Lendl2014} as priors ($ e=0.302 \pm 0.023$, argument of periastron $\omega = 242\pm2.7$~deg). To account for the systematic trends in the data, we used a Gaussian Process (GP) with a Mat\`ern 3/2 kernel, whose parameters had wide priors --- the time-scale of the process had a prior between $10^{-5}$ and 1000 days, whereas the square-root of the variance of the process had a prior between $10^{-3}$ to 100 ppm. Independent GPs were used for each sector, which also had an added white-noise component modelled as a added gaussian-noise to the GP with a large prior on the square-root of its variance between 0.01 and 1000 ppm, and independent out-of-transit flux offsets with gaussian priors centered around 0, but with a standard deviation of $10^{5}$ ppm. For the transit depth and impact parameter we used the uninformative sampling scheme proposed in \cite{Espinoza2018}, which samples the entire range of physically plausible values for those parameters. A wide log-uniform prior was defined for the stellar density between 100 and 10,000 kg/m$^3$, and the efficient sampling scheme of \cite{kipping} was used to model the limb-darkening effect through a quadratic law, where the coefficients were left as free parameters of the fit. Finally, relatively wide priors were defined for the ephemerides of the orbit: a normal distribution centered around 10 with a standard deviation of 0.1 days for the period, and a normal distribution centered around 2458357.6 with a standard deviation of 0.1 days for the time-of-transit center. No dilution contamination was applied in this fit, as this is corrected in the PDC photometry.

 Our \texttt{juliet} analysis on the PDC lightcurve reveals a transit depth within 1-sigma with the ExoFOP reported transit depth. 
 Next, we went on to test if the PDC algorithm might be diluting the transit signal due to, e.g., too strong detrending. To this end, we used the TPFs reported in MAST to extract our own photometry for this target. We used apertures consisting of 1, 2 and 3 pixels around the brightest pixel, which accounts for apertures of about 30'', 40'' and 50'' around the target. To detrend the data we used Pixel-Level Decorrelation \citep[PLD][]{pld}, initially used for Spitzer but which has been successfully applied to photometric data from missions like K2 \citep{pldk2}. The detrending was performed simultaneously with the transit model defined above, in order to account for all the uncertainties simultaneously on each of the apertures. The results from this fit were in excellent agreement with the transit depths obtained from the PDC algorithm; for example, for our smallest aperture, we obtained a transit depth of $7144 \pm  177$ ppm. Given no strong dilution is expected in this smaller aperture (judging from Gaia sources around WASP-117), this analysis gives us confidence in the fact that the relatively shallower TESS transit depths are in fact real and not related to dilution or detrending methods. This result implies that the TESS transit depth is shallow due to a physical mechanism: either spectral contamination due to stellar activity of the host star WASP-117 or due to clouds and hazes in the atmosphere of WASP-117b.

\begin{table}
\caption{WASP-117b transit depths}             
\label{table: transits}     
\centering                        
\begin{tabular}{c c c}       
\hline
\hline
Survey  & transit depth &  wavelength range \\   
\hline 
\hline
 WASP South & $8030^{+550}_{-480}$ ppm & 0.4 - 0.7 \\
 TESS (ExoFOP) & $7200 \pm 150$ ppm & 0.6 -1 $\mu$m\\
 TESS (JULIET) &  $7267 \pm 135$ ppm & 0.6 -1 $\mu$m  \\ 
 WFC3 (nominal) &  $7573 \pm 43$ ppm & 1.125 -1.65 $\mu$m \\
 
\hline                      
\end{tabular}
\end{table}

\subsubsection{Stellar activity and clouds as a possible explanation for TESS data discrepancy}
\label{sec: stellar contamination}

Stellar activity can add systematic offsets to the measured transit depth of an exoplanet like WASP-117b, which can make it difficult to compare observations at different wavelength ranges and taken at different epochs \citep{Rackham2019}. 

In the WASP-117b discovery paper, \citet{Lendl2014} did not detect any variability greater than 1.3 mmag with 95\% confidence. We independently investigated the variability of WASP-117 using ASAS-SN Sky Patrol\footnote{\url{https://asas-sn.osu.edu/}} \citep{Shappee2014, Kochanek2017}, which includes $V$-band photometric monitoring of WASP-117 for the 2014–2017 observing seasons taken with two cameras, `bf' and `bh'. We considered only the data for the 2014 and 2015 observing seasons using the `bf' camera.  Data collected using the `bh' camera, including some of the 2015 season and all later seasons, appear to suffer from a strong instrumental systematic. The RMS scatter of the data included in this analysis is 11.5\,mmag, while the mean measurement uncertainty is 4.5\,mmag, suggesting that some variability is present. A Lomb-Scargle periodigram analysis \citep{Lomb1976,Scargle1982}, implemented in \texttt{astropy}\footnote{http://www.astropy.org. Astropy is a community-developed core Python package for Astronomy \citep{astropy:2013, astropy:2018}}, consistently reveals a periodicity with a full amplitude of 9.4\,mmag, though the period with the largest power depends on the observing season included. Considering the ASAS-SN data, we conservatively adopt 9.4\,mmag or 1\% as the reference variability level for WASP-117 in this analysis.

We modeled the variability of WASP-117 using the method detailed by \citet{Rackham2018}, including considerations for FGK dwarfs \citep{Rackham2019}. 
In brief, the approach involves adding spots and faculae to a large set of model photospheres to establish the probabilistic relationship between active region coverage and rotational variability for a set of stellar parameters.
The observed variability of a star is then used to estimate its active region coverage and the related stellar contamination of transmission spectra from the system.
We set the photosphere temperature to the stellar effective temperature determined by \citet{Lendl2014}. We further adopted the surface gravity, stellar surface gravity, and metallicity from that work. We determined spot and facula temperatures from the scaling relations detailed by \citet{Rackham2019}. We used the spot size and initial facula-to-spot areal ratio\footnote{The method we use allows the facula-to-spot areal ratio to drift to smaller values as a star becomes more spotted, as is observed on the Sun \citep{Shapiro2014}. See \citet{Rackham2018} for further details.} from the `solar-like case' outlined by \citet{Rackham2018}. Table~\ref{tab:stellar_contamination} summarizes the adopted parameters.

\begin{figure}
   \includegraphics[width=0.5\textwidth]{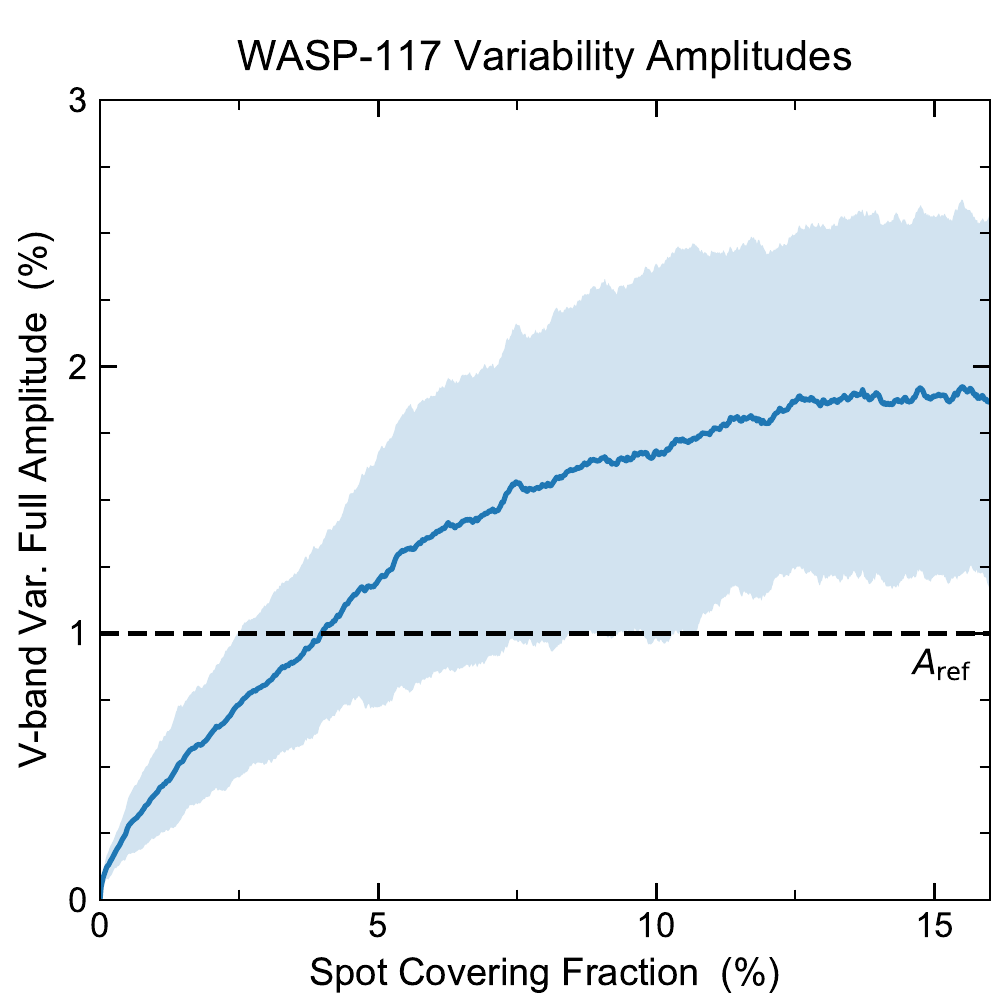}
\caption{Modeled rotational variability amplitudes of WASP-117 in V-band as a function of spot covering fraction. The median estimate (blue line) and $1\sigma$ confidence interval (blue shaded region) are shown.}
    \label{fig:variability}
\end{figure}

\begin{figure}
   \includegraphics[width=0.5\textwidth]{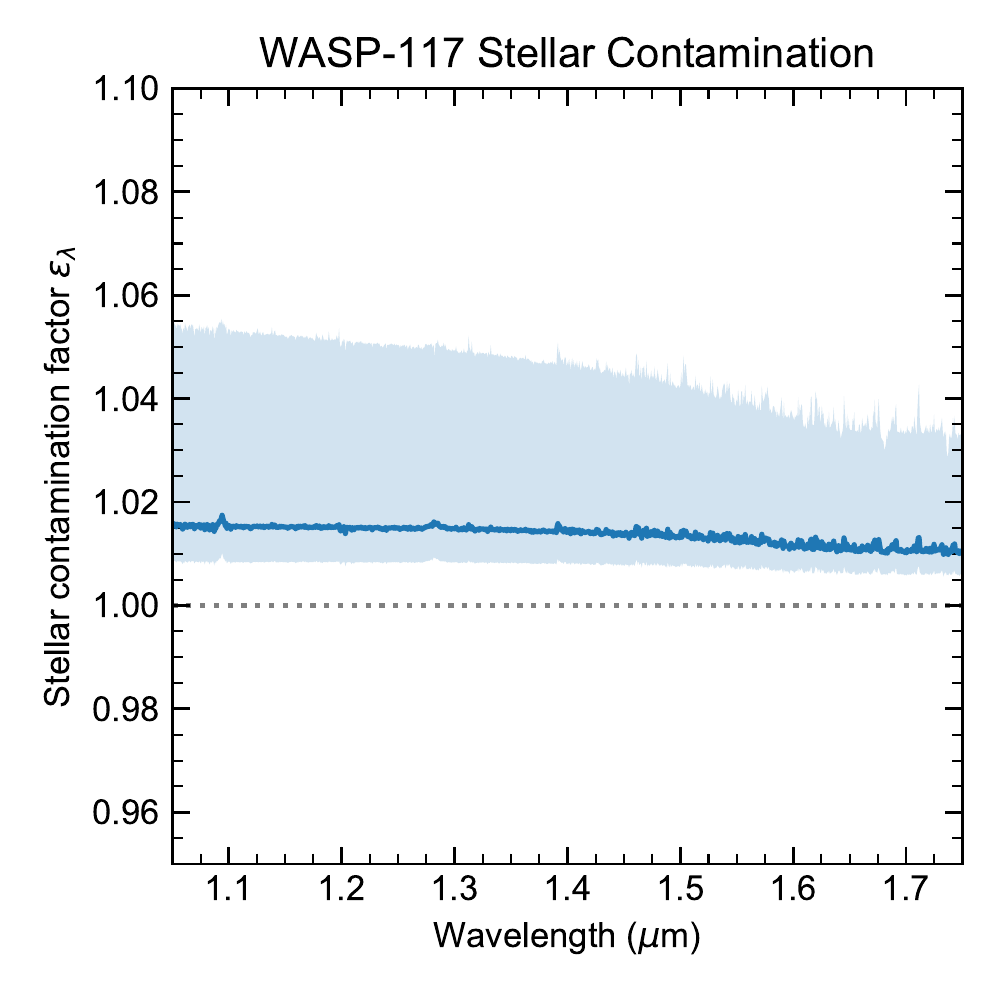}
\caption{Spectroscopic stellar contamination signal in the HST/WFC3 G141 bandpass produced by unocculted active regions on WASP-117. The median estimate (blue line) and $1\sigma$ confidence interval (blue shaded region) are shown.}
    \label{fig:epsilon}
\end{figure}

Figure~\ref{fig:variability} illustrates the modeled variability of WASP-117 as a function of spot covering fraction. We find that the adopted variability amplitude of WASP-117 corresponds to full-disk spot and facula covering fractions of $F_\mathrm{spot} = {4}^{+6}_{-2}\%$ and $F_\mathrm{fac} = {29}^{+19}_{-9}\%$, respectively, at $1\sigma$ confidence. If present outside of the transit chord, these active regions could alter the observed transit depths. Figure~\ref{fig:epsilon} illustrates the wavelength-dependent stellar contamination factor $\epsilon_{\lambda}$, the ratio of the observed transit depth to the true transit depth \citep[see Eq.~2 of][]{Rackham2018}, produced by these active region coverages over the HST/WFC3 G141 bandpass. We find the effect of unocculted spots dominates over that of unocculted faculae, producing a net increase in transit depths ($\epsilon_{\lambda} > 1$). Over the complete G141 bandpass, we estimate that transit depths can be inflated by $1.4^{+3.1}_{-0.6}\%$ at $1\sigma$ confidence. This inflation is relative to a measurement in the same wavelength range that is completely unaffected by stellar activity. No strong spectral features are apparent in the contamination signal besides a slight increase toward shorter wavelengths: at 1.1 $\mu$m, transit depths may be increased by $1.5^{+3.8}_{-0.7}\%$, while at 1.7 $\mu$m the increase is limited to $1.1^{+2.3}_{-0.5}\%$. We note that while the scale of $\epsilon_{\lambda}$ is comparable to the precision of our HST/WFC3 G141 spectrum, the lack of strong spectral features in the contamination spectrum provides reassurance against spectral features we see in the near-infrared spectrum actually resulting from stellar contamination.

\begin{table}
    \centering
    {\renewcommand{\arraystretch}{1.2}
    \caption{Parameters used in stellar heterogeneity analysis.}
    \label{tab:stellar_contamination}
    \begin{tabular}{lcc} 
        \hline
        \hline
Parameter                  & Description                  & Value \\
        \hline
$T_\mathrm{phot}$          & Photosphere temperature      & 5460~K  \\
$[\mathrm{Fe}/\mathrm{H}]$ & Metallicity                  & 0.14 \\
$\log g$                   & Surface gravity              & 4.37 \\
$T_\mathrm{spot}$          & Spot temperature             & 4780~K \\
$T_\mathrm{fac}$           & Facula temperature           & 5600~K \\
$R_\mathrm{spot}$          & Spot radius                  & $2^\circ$ \\
$\alpha$                   & Initial facula-to-spot areal ratio   & 10 \\
$A_\mathrm{ref}$           & V-band variability amplitude & 2\% \\
        \hline
    \end{tabular}}
\end{table}

 \begin{figure}
   \centering
   \includegraphics[width=0.49\textwidth]{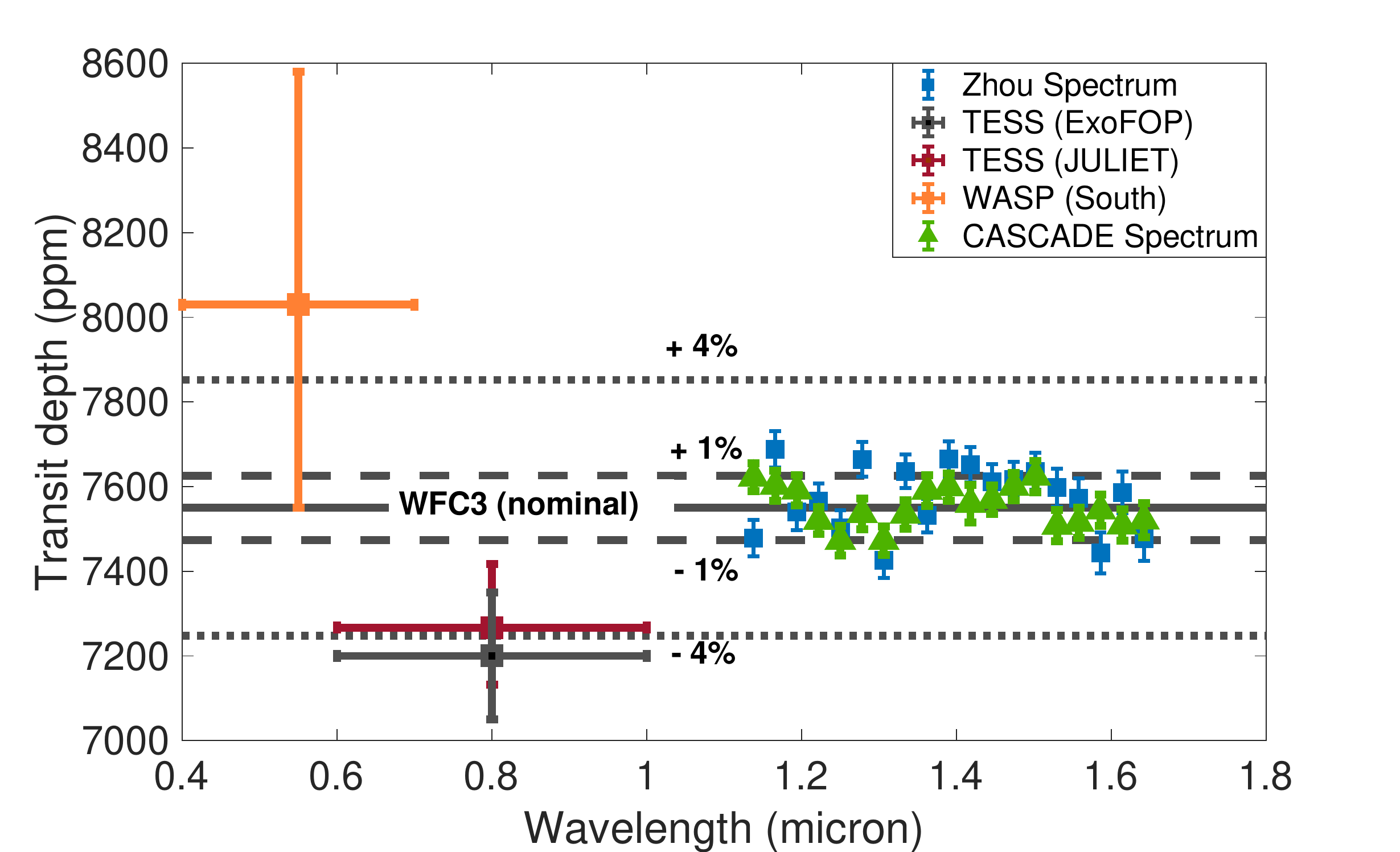}
      \caption{HST transmission spectra from the nominal and CASCADE pipeline compared with TESS transit depth, reduced by ExoFOP, our independent re-analysis with JULIET and the WASP South transit depth. Solid grey line denotes the nominal transit depth in the WFC3 wavelength range. Dashed line and dotted lines show radius inflation/deflation by 1\% and 4\% respectively.
              }
         \label{fig: TESS_compare}
   \end{figure}

\begin{figure}
   \includegraphics[width=0.49\textwidth]{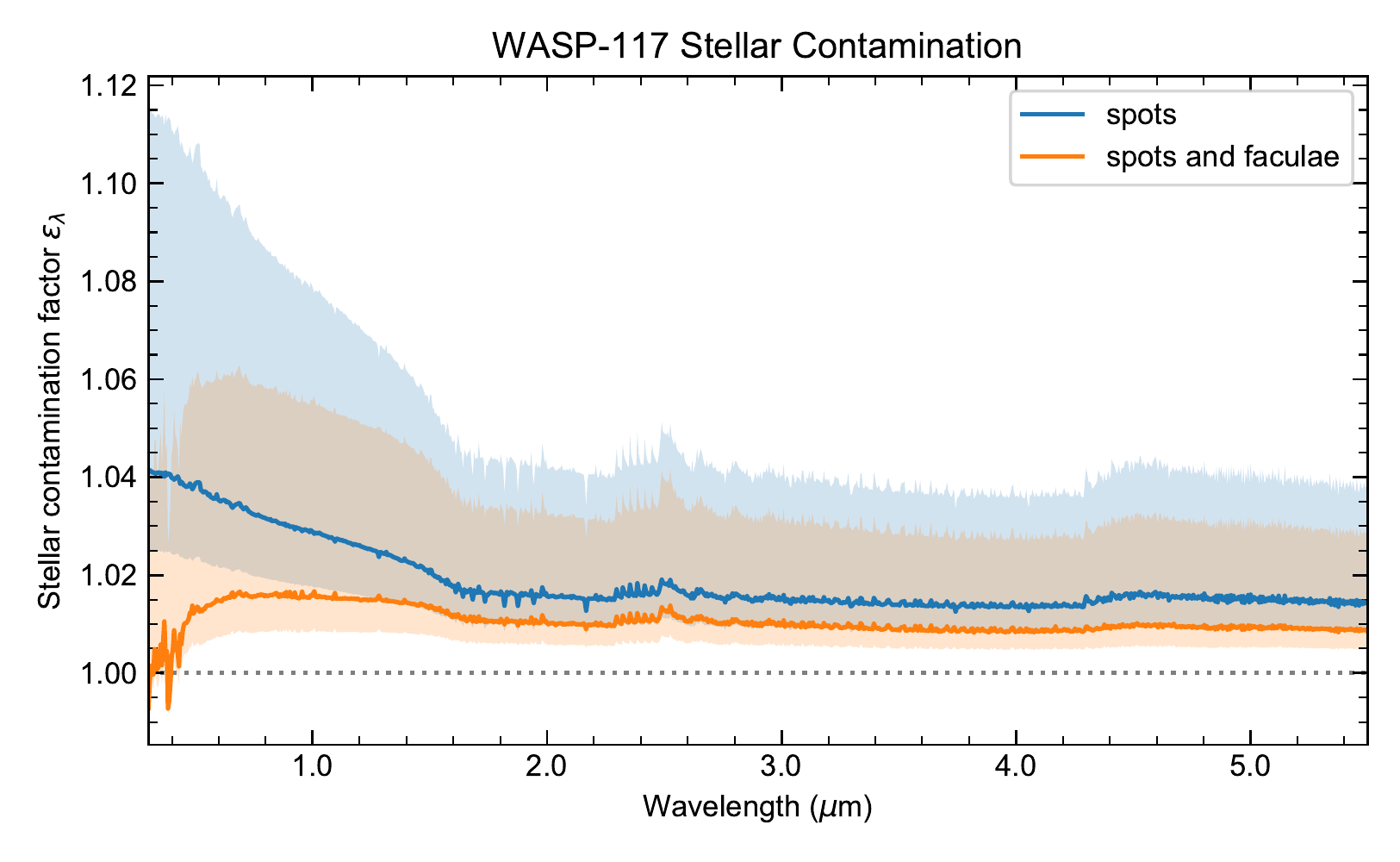}
\caption{Spectroscopic stellar contamination signal for 0.3--5.5\,$\mu$m produced by unocculted active regions on WASP-117. The median estimates (blue line for spots and orange for spots and faculae, respectively) as well as $1\sigma$ confidence intervals (shaded regions) are shown. }
    \label{fig:epsilon_larger}
\end{figure}

\begin{table}
    \centering
    \caption{Stellar contamination factors in the TESS and WFC3/G141 bandpasses for the spots-only models.}
    \label{tab:stellar_epsilon}
    \begin{tabular}{lcc} 
        \hline
        \hline
                  & Band pass                  & Value \\
        \hline

$\epsilon_{\rm WFC3}$  & $1.1 - 1.7 \mu$m & $1.014^{+0.031}_{-0.006}$ \\
$\epsilon_{\rm TESS}$  & $0.6 - 1.0 \mu$m & $1.03^{+0.06}_{-0.01}$ \\
        \hline
    \end{tabular}
\end{table}

Figure~\ref{fig: TESS_compare} shows the relative increase of the planetary radius in the near infrared ($1.1 \mu$m) between 1\% and 4\%, the TESS transit depths and HST/WFC3 spectrum. If the TESS data (observed in 9/2018) would have been taken at a time when the host star showed low activity, that is, very low coverage fraction of active regions compared to the time HST/WFC3 data was taken (observed in 9/2019), then this difference in stellar activity between the two epochs could in principle explain the discrepancy between the optical and the near-infrared spectrum. 

We note, however, that the raw HST/WFC3 data (Figure~\ref{fig:lc}) do not display any obvious signs of stellar activity, e.g., in form of spot-crossing events. Conversely, the VLT/ESPRESSO data (10/2019) taken at a few weeks after the TESS measurements do not show strong activity in the stellar Na and K lines. In other words, with the data at hand, there are no indications that the host star was more active in one epoch compared to the other.

Alternatively, if the stellar activity between the two epochs did not differ drastically, unocculted bright regions (faculae) could be invoked to explain decreases in transit depths at visual wavelengths (that is, $\epsilon <1$). We consider this unlikely , however, because the  TESS band pass is redder than the wavelength where we would expect faculae to start dramatically decreasing transit depths, which is typically around $0.5~\mu$m \citep[][Fig.~5]{Rackham2019}. Furthermore, the scale of the decrease is too large between the TESS and WFC3 band. The transit depth is shallower by about 4\% in the TESS band (Figure~\ref{fig: TESS_compare}). We expect, however, less than 1\% decrease in transit depth due to the chromaticity of bright faculae between the TESS and WFC3 G141 bands (Figure~\ref{fig:epsilon_larger}).

 As we will show in Section~\ref{sec: W117_prediction}, no atmospheric model fitted to the HST/WFC3 data yields transit depths shallower than 7400~ppm in the optical wavelength range - including the very cloudy (\textit{Model 2}) and hazy solutions (\textit{Model 3}). 
 Without additional measurements, it is thus for now not possible to determine why the TESS transit depth is shallow ($7200 \pm 150$) compared to the earlier WASP and the most recent HST/WFC3 measurements.

\section{WASP-117b temperatures and chemistry composition}
\label{sec: Disequi}

We could not use the TESS data to constrain atmospheric properties of WASP-117b. We further could not constrain the atmospheric temperature of WASP-117b during transit based on the muted water feature observed with HST/WFC3 (Table~\ref{table: HST_retrieval}). 

 We still think, however, that it is worthwhile to explore exemplary chemistry models for atmospheric compositions and temperature possible for WASP-117b based on the WFC3 data and theoretical reasoning. These physically based constraints a) act as sanity checks for the simplified 1D retrieval, b) still allow us to draw further conclusions and c) inform us about the benefit of future observations.

We calculated theoretical average planetary equilibrium temperatures $T_{Pl,inst}$, assuming that the planet instantly adjusts to the stellar irradiation, that the planet is in equilibrium with the incoming stellar irradiation and that the planet experiences global heat redistribution. This is the assumption that \citet{Lendl2014} used for their analysis, using albedo $\alpha=0$. Here, we use equation (2) in \citet{Mendez2017} with $\beta=1$ (efficient heat re-distribution over the whole planet) and broadband thermal emissivity $\epsilon=1$, using stellar and planetary parameters from \citet{Lendl2014} and setting albedos $\alpha=0,0.3,0.6$ within limits of cloud models for hot to warm Jupiters \citep{Parmentier2016}. $T_{Pl,inst}$ gives the possible range of global average  temperatures, assuming instant heat adjustment, global heat circulation and different albedos for WASP-117b during one orbit, including at transit (Table~\ref{table: W117_temperature}, temperatures during transit set in bold). These temperatures were compared to the assumption that the planet's temperature is constant and equal the time-averaged planetary temperature $T_{Pl,ave}$, where we used equation~(16) of \citet{Mendez2017} again with $\alpha=0,0.3,0.6$. We stress that these temperatures are first order assumptions to estimate the maximum possible temperature range during transit. A more comprehensive analysis of temperature variation during one orbit requires the comparison between radiative and dynamical time scales, where the latter has to be estimated from a 3D circulation model \citep{Lewis2014}.

\begin{table*}
\caption{Theoretical temperature constraints of WASP-117b during one orbit.}             
\label{table: W117_temperature}     
\centering                        
\begin{tabular}{c | c | c c c | c c c}       
\hline\hline                
Time & distance${}^a$ $a$ & \multicolumn{3}{c}{$ T_{Pl,inst}$} |& \multicolumn{3}{c}{$ T_{Pl,ave}$}\\ 
     & [AU] & $\alpha = 0$ &  $\alpha = 0.3$ &  $\alpha = 0.6$ &  $\alpha = 0$ &  $\alpha = 0.3$ &  $\alpha = 0.6$\\
\hline 
  apoastron  & 0.1230 & 899  & 822 & 715 & \multirow{4}{*}{\textbf{1019}} & \multirow{4}{*}{\textbf{932}} & \multirow{4}{*}{\textbf{810}} \\ 
  \textbf{transit}  & \textbf{0.1168} & \textbf{922} & \textbf{844} & \textbf{734} &  &  & \\ 
   periastron & 0.0663 & 1224  & 1120 & 974 & &  & \\ 
   secondary eclipse & 0.682 & 1207 & 1104 & 960 & &  & \\
\hline
${}^a$ based on \citet{Lendl2014}
\end{tabular}

\end{table*}

Table~\ref{table: W117_temperature} shows that depending on albedo and heat adjustment assumptions, WASP-117b could assume globally averaged equilibrium temperatures $T_{eq}$ between 700\,K to 1000\,K during transit. In the following, we investigated the implications of equilibrium and disequilibrium chemistry within this temperature range for \textit{Model 1} (high metallicity $\rm[Fe/H]>2.2$) and \textit{Model 2} (low metallicity $\rm[Fe/H]<1.75$). 

We computed a self-consistent pressure-temperature profile for the exo-Saturn WASP-117b using \texttt{petitCODE} \citep{Molliere2015,Molliere2017} and multiple one-dimensional chemical kinetics models \citep{Venot2012}, incorporating vertical mixing and photo-chemistry. In order to systematically represent the best-fit retrieval models (see Table~\ref{table: HST_retrieval}), as well as the parameter space of physically accepted solutions, we performed four base models with varying temperature and metallicity. We adopted values for atmospheric metallicities of 350 times and 5 times solar metallicity, consistent with \textit{Model 1} and \textit{Model 2} respectively. Further, we picked equilibrium temperatures $T_{eq}$ of 1000~K and 700~K to represent the theoretically possible range of temperatures during transit. A C/O ratio of 0.30 is adopted in accordance with the retrievals. 

We acknowledge that a C/O ratio of 0.3 may be too low, because we assumed equilibrium chemistry during retrieval. We argue, however, that higher C/O ratios produce even higher \ce{CH4} abundances for the same pressure and temperature range \citep[see e.g.][]{Maddu2012}. Thus, any constraints on \ce{CH4} quenching via vertical mixing ($K_{zz}$)  that we find for such a low C/O ratio of 0.3, will also hold for higher C/O ratios and provide e.g. a lower limit for possible $K_{zz}$. Therefore, we decided given the lack of other constraints that adopting  $\rm{C/O}=0.3$ is the best approach for the time being.

Another possible concern in basing disequilibrium chemistry on retrieved parameters is the difference in planetary atmospheric temperature structure used by both models. The former uses a self-consistent 1D atmospheric temperature profile (Figure~\ref{fig: Disequi}, bottom right  panel), based on \texttt{petitCODE}  \citep{Molliere2015,Molliere2017}, the retrieval code \texttt{petitRADTRANS} \citep{Molliere2019} uses isothermal temperatures. However, we note that for $p \leq 10^{-3}$~bar the temperature is approximately isothermal also in the self-consistent 1D exoplanet atmosphere model. Thus, we conclude that these differences in temperature do not play a strong role here.

\begin{figure}
    \centering
    \includegraphics[width=0.5\textwidth]{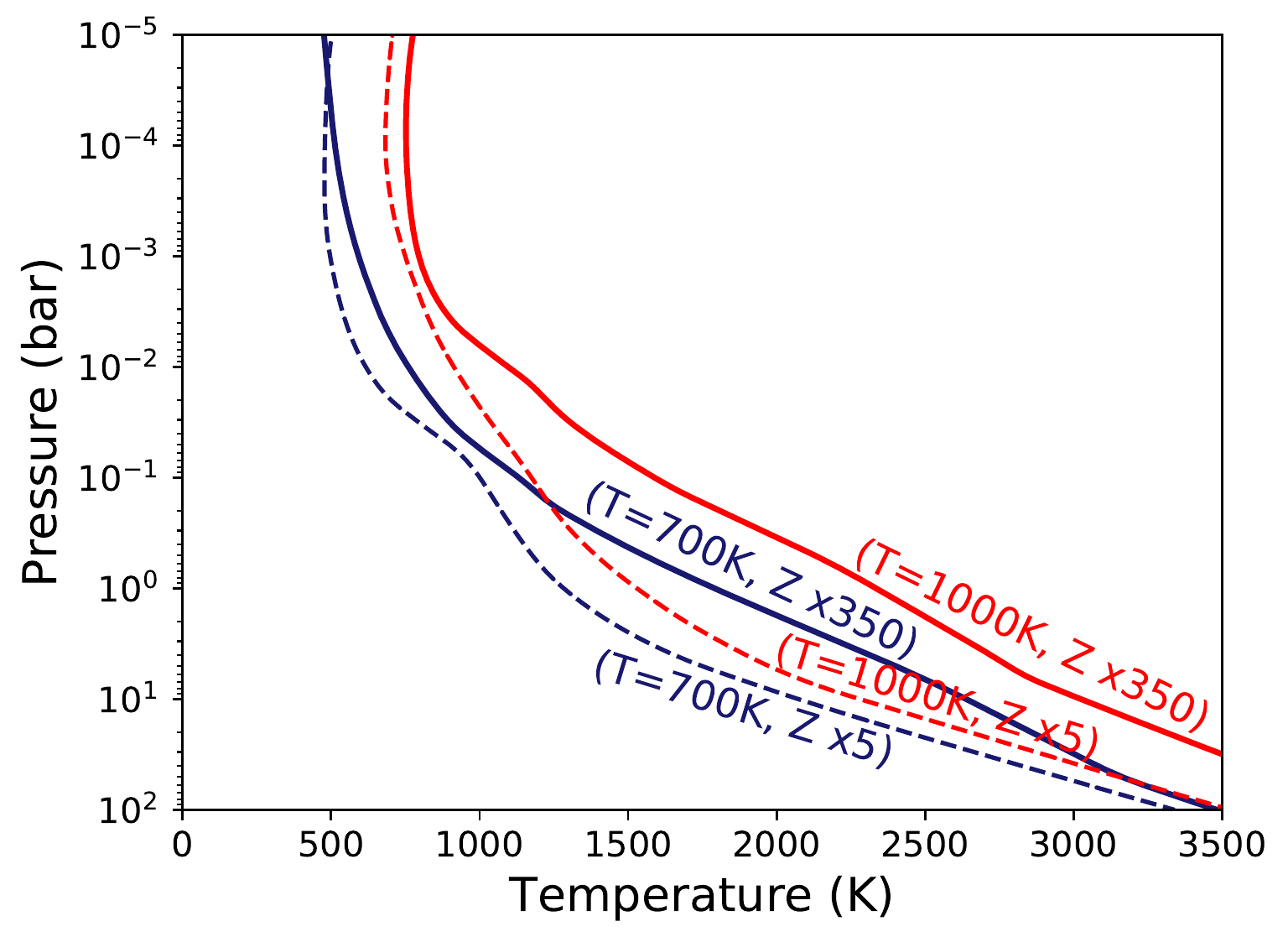}
    \caption{Pressure-temperature profiles computed with \texttt{petitCODE} for four different scenario's of WASP-117b: temperatures $T_{\rm eff}=1000$~K (red), $T_{\rm eff}=700$~K (blue), and metallicities $Z = Z_\odot \times 350$ (solid), $Z = Z_\odot \times 5$ (dashed). For all cases, a C/O ratio of 0.3 and intrinsic temperature of 400~K were adopted. The profiles serve as input for the chemical kinetics calculations of Figure~\ref{fig:diseq_bestfit_retrievals}.}
    \label{fig: pt_profiles}
   \end{figure}

Furthermore, for the pressure-temperature iteration we assumed a uniform day-side heat redistribution and a moderate intrinsic temperature of 400~K, in agreement with the planet's irradiation \citep{Thorngren2019_tint_and_RCB_depth}. Other stellar and planetary parameters were derived from \cite{Lendl2014}. The pressure-temperature profiles computed with \texttt{petitCODE}, which are used in the chemical kinetics calculation, are shown in Figure~\ref{fig: pt_profiles}. For the chemical kinetics model we applied a vertically constant eddy diffusion coefficient $K_{zz}= 10^{10}$~cm$^2$/s. We also applied photochemical reactions. To represent the spectral energy distribution of WASP-117, we have used an ATLAS stellar model ($T = 6000$~K, log$g = 4.5$) \citep{Castelli2003} between 168~nm and 80~$\mu$m, and a time-averaged solar UV spectrum \citep{Thuillier2004} between 1~nm and 168~nm.

\begin{figure*}%
    \centering%
    \includegraphics[width=0.99\textwidth]{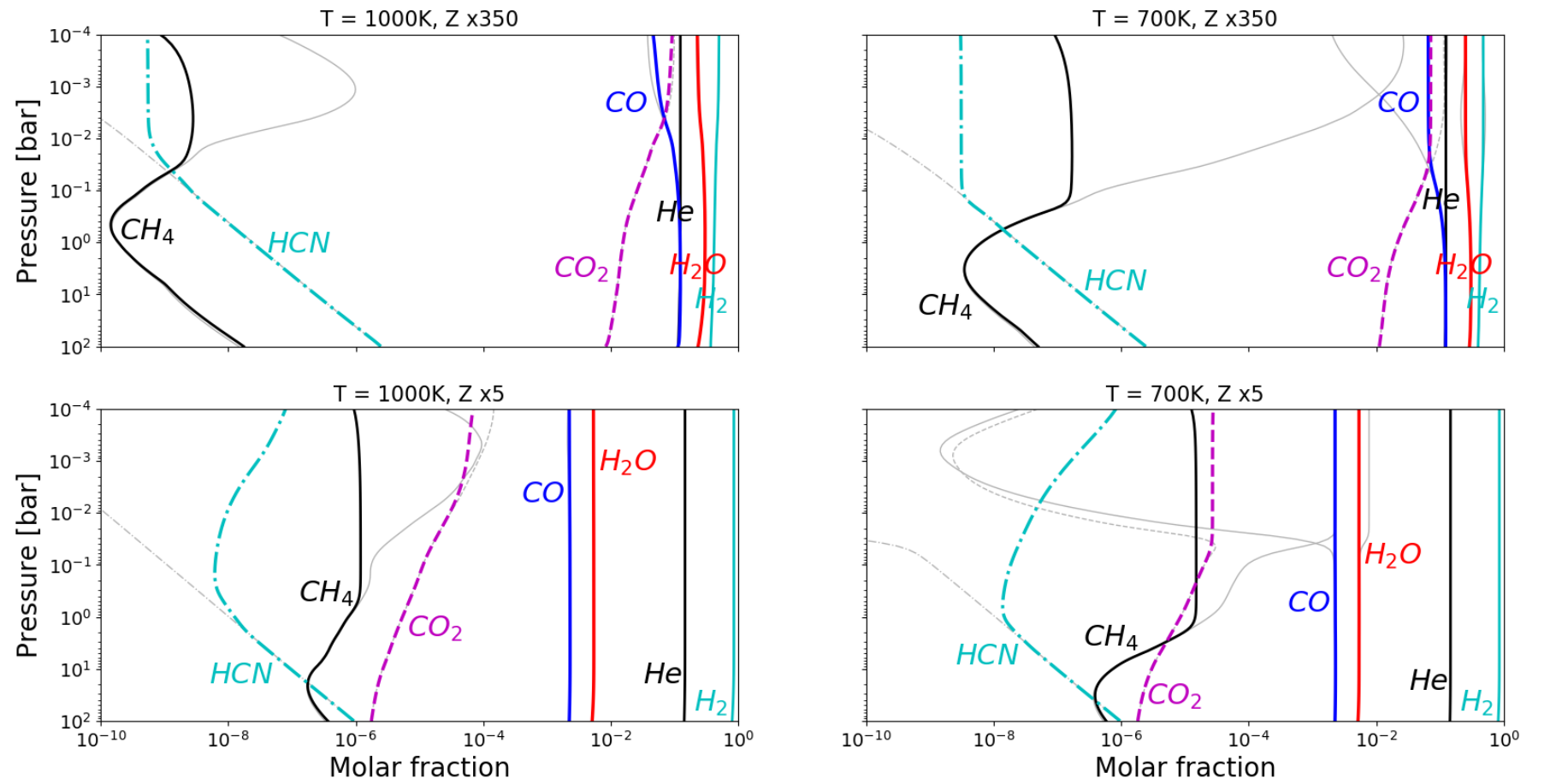}%
    \caption{Molecular abundances computed for hot ($T_{\rm eq}=1000$~K) or warm ($T_{\rm eq}=700$~K), and very ($Z_\odot \times 350$) or moderately ($Z_\odot \times 5$) enriched atmospheres, using the chemical kinetics code of \cite{Venot2012}. The colored lines denote the disequilibrium abundances for the main molecules in the planetary atmosphere. The gray lines indicate the chemical equilibrium state. A constant $K_{zz} = 10^{10}$~cm$^2$/s and C/O = 0.3 were adopted.}%
    \label{fig:diseq_bestfit_retrievals}%
\end{figure*}%

As can be seen in Figure~\ref{fig:diseq_bestfit_retrievals}, the hot models with $T_{eq}= 1000$~K naturally have a low methane abundance in chemical equilibrium. This is exacerbated when disequilibrium chemistry is taken into account, as methane gets quenched at even lower abundances. For the warm models with $T_{eq}=700~K$, methane starts to dominate over CO in chemical equilibrium in the higher atmospheric layers ($p < 1$~mbar for the high metallicity case and $p < 40$~mbar in the lower metallicity case). Also in these models, vertical mixing serves to reduce the methane abundance in the layers above the quenching level ($p < 100$~mbar for the high metallicity case and $p < 1$~mbar in the lower metallicity case). 

For low atmospheric temperatures ($T_{eq}=700$~K),  disequilibrium chemistry through vertical mixing is required to keep a low methane abundance ($<10^{-4}$ volume fraction) in the observable part of the atmosphere ($p<0.1$~bar in transmission) for both, low and high atmospheric metallicity. Vertical mixing would be even stronger for lower atmospheric metallicity composition compared to high metallicity, because in the former case, more \ce{CH4} is produced at a given pressure level. The reason for this is twofold. First, for a given equilibrium temperature the lower metallicity case results in lower atmospheric temperatures (Figure~\ref{fig: pt_profiles}), leading to a higher methane abundance. Second, the lower metallicity increases the temperature of transition between CO / \ce{CH4} \citep{Lodders2002}, resulting again in a higher methane abundance. Provided WASP-117b has indeed an equilibrium temperature of 700~K during transit, we could potentially further constrain $K_{zz}\geq 10^{8}$~cm$^2$/s (see Figure~\ref{fig: Disequi} and Section~\ref{sec: Vertical mixing}, where we present an abundance  study for this case, as a function of $K_{zz}$).
We note that this assessment is based on the median abundances that we retrieve for CH$_4$, with 1-$\sigma$ uncertainties on the upper limit of the CH$_4$ abundances as large as 2~orders of magnitude, see figures \ref{fig: Corner_plot_H2O_CH4_m1}, \ref{fig: Corner_plot_H2O_CH4_m2}, \ref{fig: Corner_plot_H2O_CH4_m3}.

 \begin{figure}
   \centering
   
   \includegraphics[width=0.48\textwidth, height=5 cm]{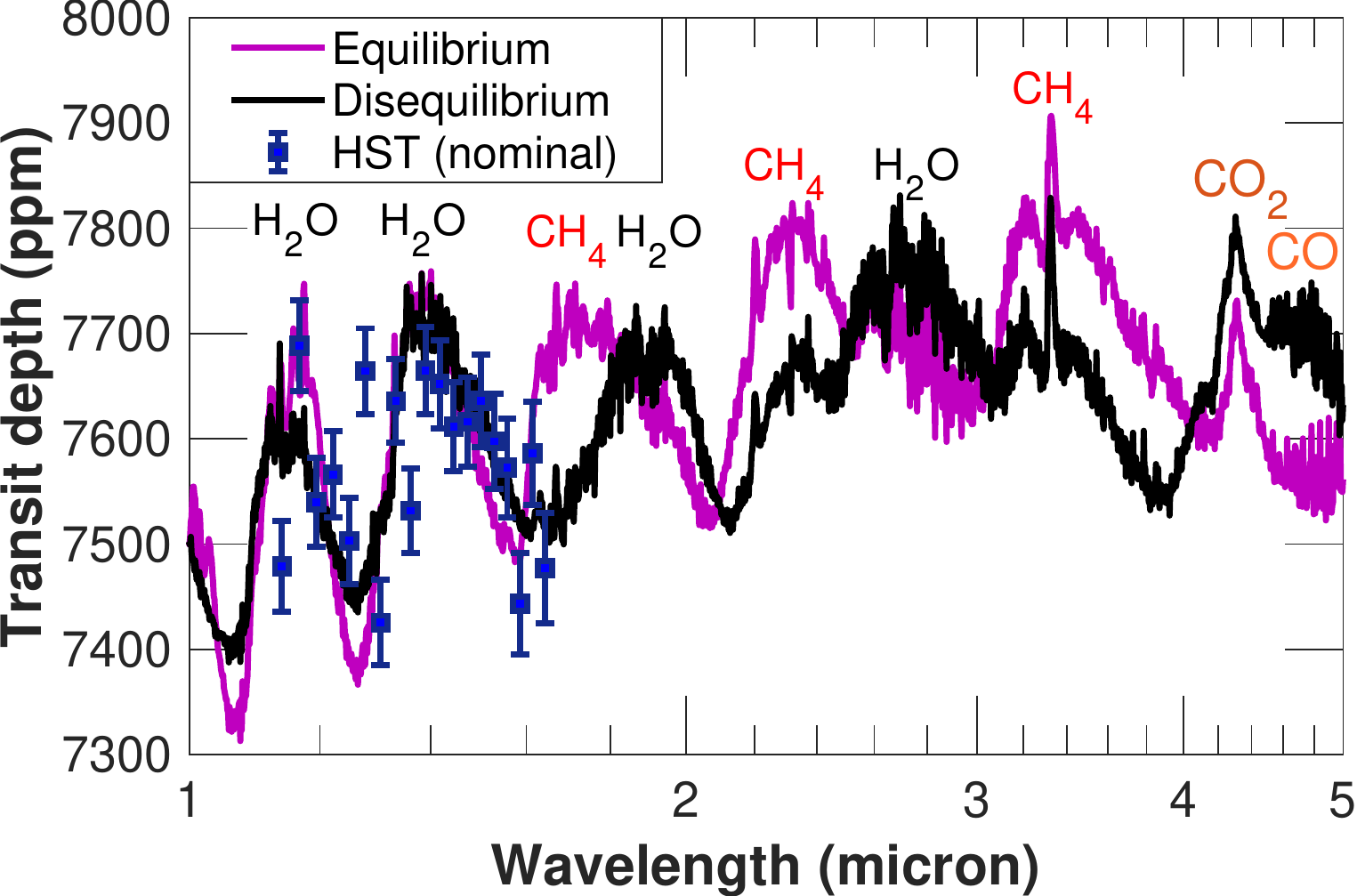}
      \caption{Prediction of molecular features in the 1 - 5~$\mu$m wavelength range based on equilibrium (purple) and disequilibrium chemistry (black) cloud-free self-consistent forward models with methane quenching (black) for WASP-117b, assuming $T_{eq}=700$~K and $K_{zz}=10^8$~cm${}^2$s and $Z=Z_0 \times 5$. }
         \label{fig: JWST_disequi}
   \end{figure}

A first disequilibrium versus equilibrium model comparison in the 1--5~$\mu$m range shows that JWST data can in principle distinguish between these two scenarios (Figure~\ref{fig: JWST_disequi}): The absence of strong \ce{CH_4} bands, but also the presence of \ce{CO} in the near infrared indicate disequilibrium chemistry.

We will investigate in the next section in more detail the prospect of JWST data to yield a tighter constraint on the \ce{CH4} abundances, atmospheric temperatures and disequilibrium chemistry in WASP-117b.

\section{Prospects for future investigations with HST and JWST}
\label{sec: W117_prediction}

We investigated to what extent future observations with HST in the UV and optical, and JWST in the near to mid-infrared could constrain the properties of WASP-117b.

\begin{figure*}
   \centering
   \textbf{Nominal}  \hspace{7.5 cm}  \textbf{CASCADE}\par
   \includegraphics[width=0.48\textwidth]{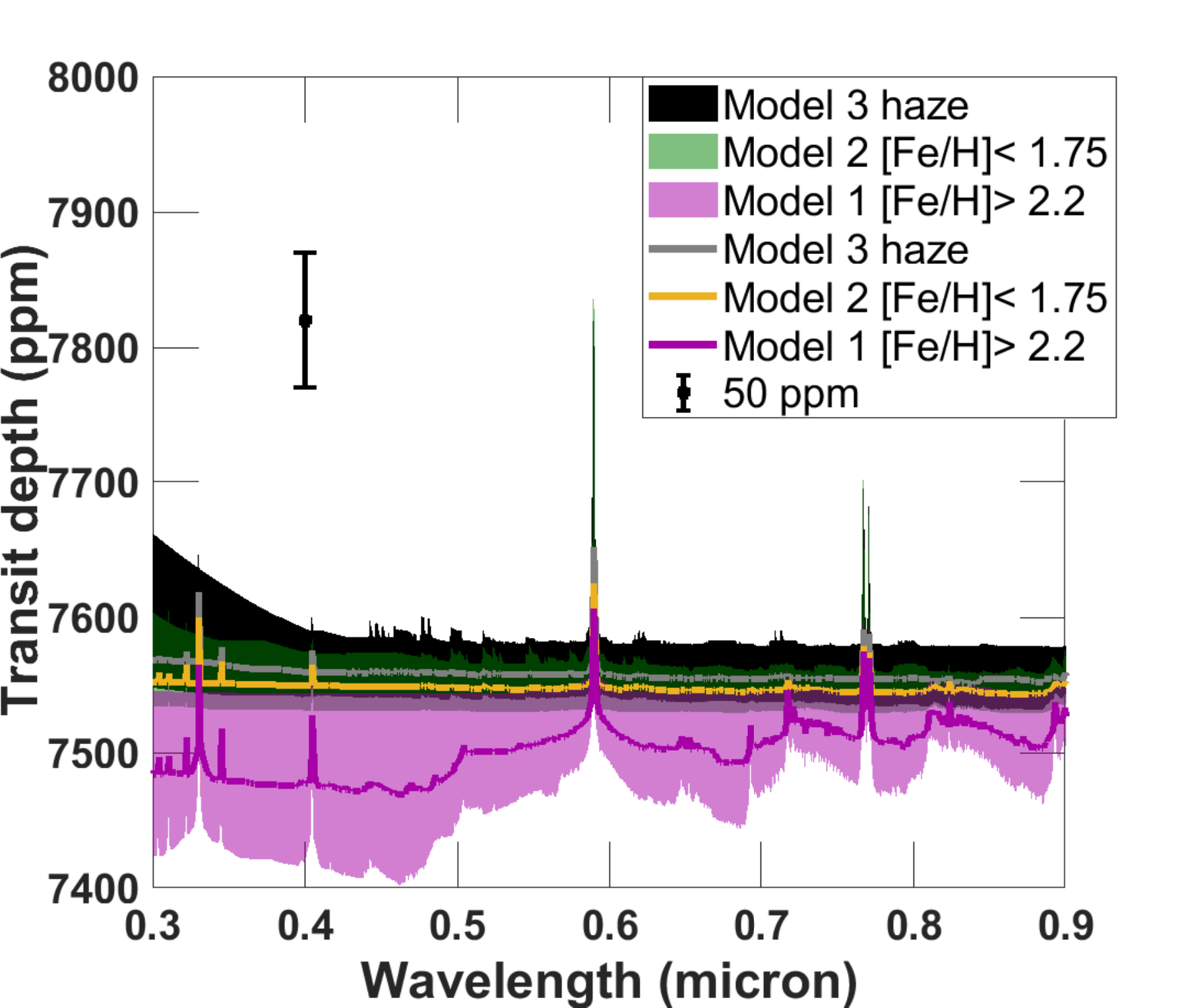}
   \includegraphics[width=0.48\textwidth]{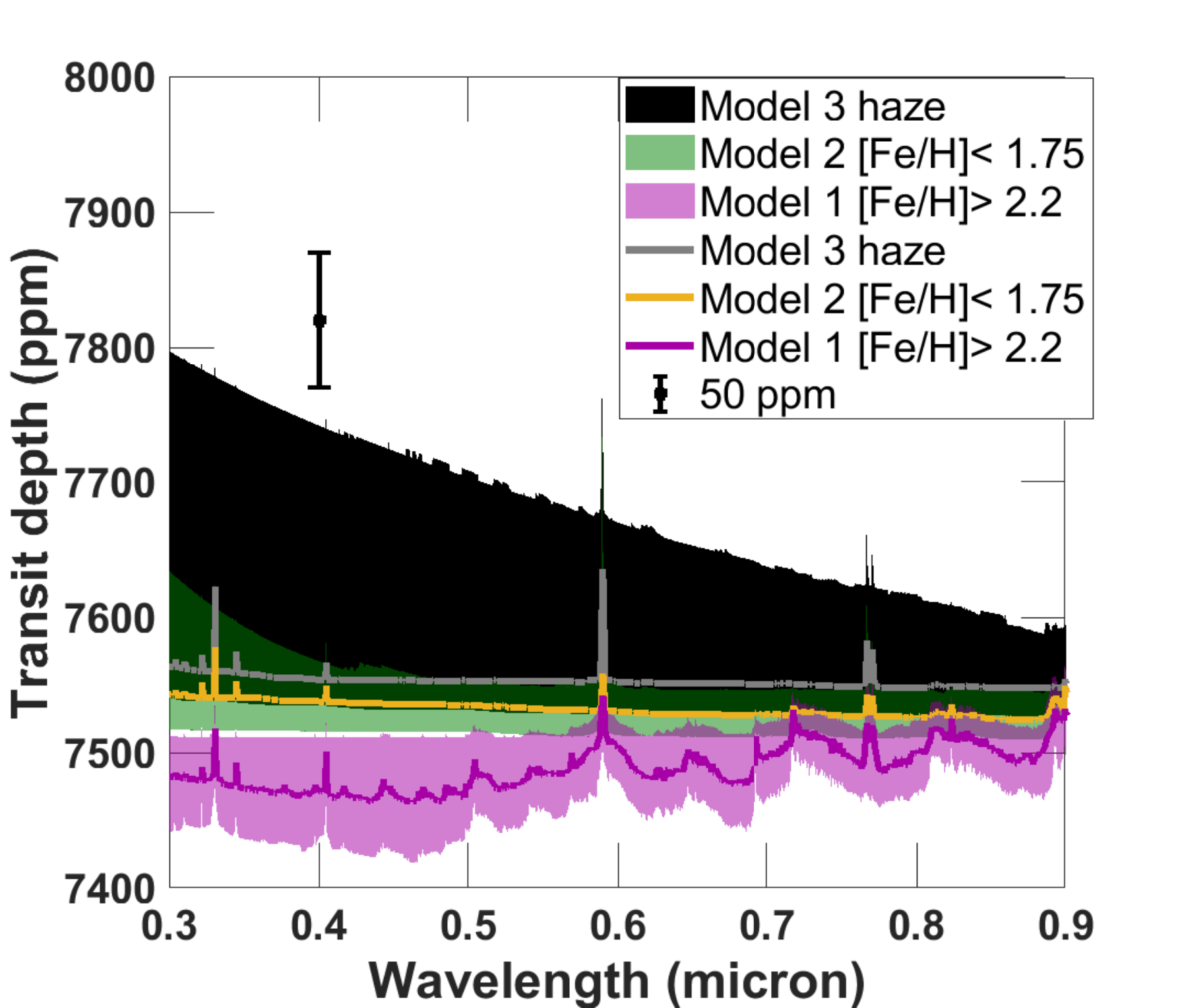}
      \includegraphics[width=0.48\textwidth]{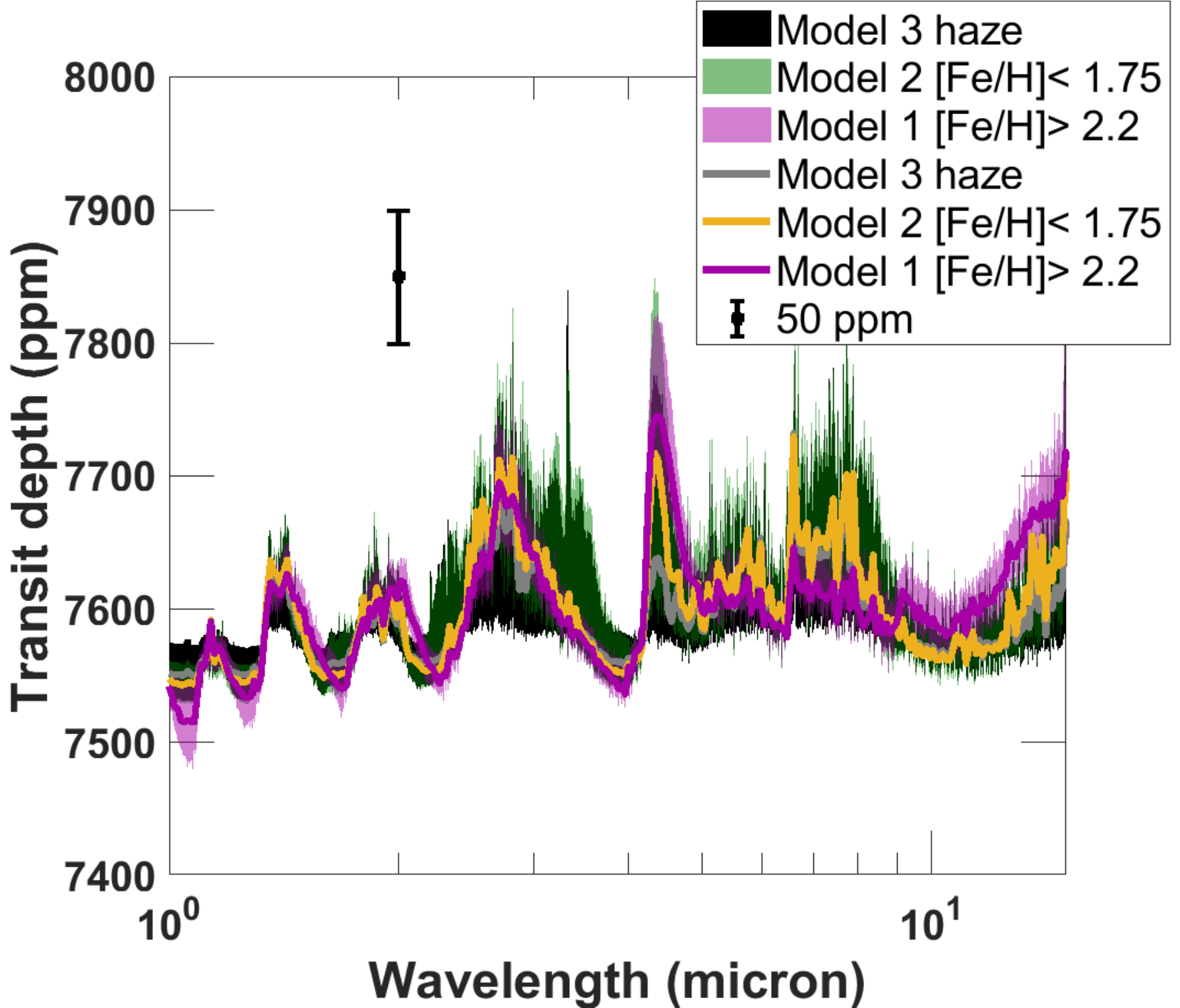}
      \includegraphics[width=0.48\textwidth]{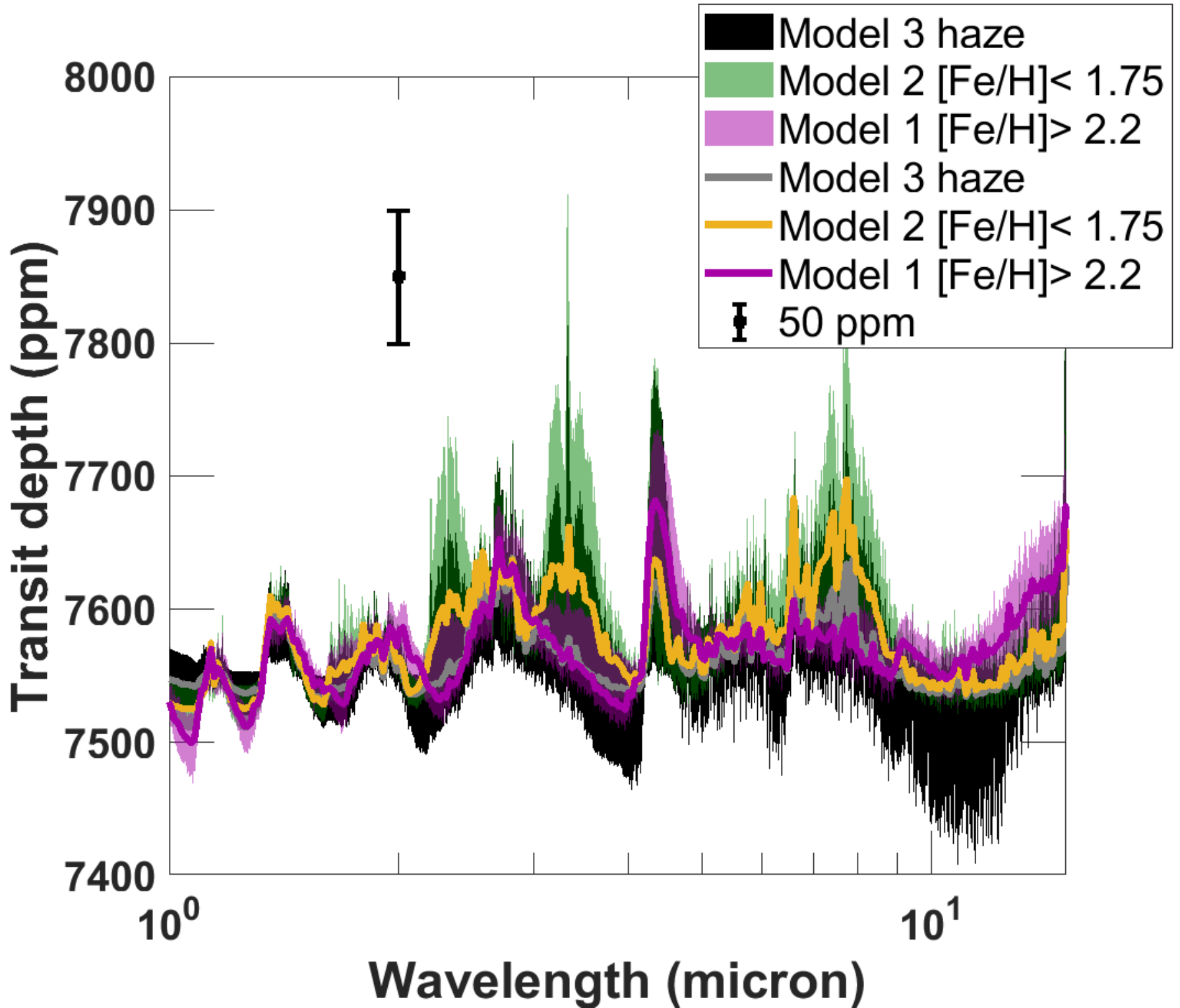}
      \caption{Synthetic spectra based on \textit{Model 1} (high metallicity, purple), \textit{Model 2} (low metallicity \& cloudy, green) and \textit{Model 3} (haze, black) based on the WASP-117b spectrum reduced with the nominal pipeline (left panels), and the CASCADE pipeline (right panels). Colored areas give the $1 \sigma$ envelope for the solutions for each model. Solid lines in grey, yellow and light-pink give the best fit for \textit{Model 1}, \textit{Model 2} and \textit{Model 3}, respectively. The spectra are shown for the wavelength range: 0.3 -0.9 $\mu m$ (upper panels) and 1 - 15 $\mu$m (lower panels). In addition, a hypothetical data point with 50 ppm accuracy is shown for guidance.          }
         \label{fig: HST_JWST_pred}
   \end{figure*}

Synthetic spectra based on the results of \textit{Model 1} (high metallicity), \textit{Model 2} (low metallicity \& cloudy) and \textit{Model 3} (hazy) (see Section~\ref{sec: atmosphere models}, Table~\ref{table: HST_retrieval}) are shown in Figure~\ref{fig: HST_JWST_pred}. 

The possible atmospheric compositions retrieved for WASP-117b produce large differences in wavelength ranges between $0.3 - 0.5 \mu$m (Figure~\ref{fig: HST_JWST_pred}, top panels). An accuracy of 200~ppm could already be sufficient to distinguish between very hazy (extreme hazy solutions of \textit{Model 3}, based on the CASCADE spectrum) and clear sky, heavy metal atmospheric composition (\textit{Model 1}), based on the HST/WFC3 G141 grism spectrum reduced with the CASCADE pipeline (Figure~\ref{fig: HST_JWST_pred}, top right). Accuracy of 100 ppm and better would be needed between 0.3 and 0.5 $\mu$m to distinguish between the medians of the clear sky, heavy metal, the grey cloudy solutions with lower metal-content and the hazy solutions, irrespective of data pipeline used (Figure~\ref{fig: HST_JWST_pred}, top panels). 

\citet{Wakeford2020} recently showed that it is, in principle, possible to accurately measure a transmission spectrum of a transiting exoplanet in this wavelength range using WFC3/UVIS G280 grism. The authors report an accuracy of 29 to 33~ppm for the broad band depth between 0.3 to 0.8 $\mu$m and 200 ppm in 10~nm spectroscopic bins. \citet{Wakeford2020} observed two transits of the hot Jupiter HAT-P-41b around a quiet F star with magnitude $m_V=11.08$. Our target WASP-117b likewise orbits a quiet F star that is even brighter than the host star of HAT-P-42b. WASP-117 has a magnitude of $m_V=10.15$.

Thus, we estimate that two transit observations of WASP-117b could be sufficient to constrain a haze layer and potentially also the atmospheric metallicity using WFC3/UVIS. We did not consider here the optical transits in particular the TESS transit depth $7200 \pm 150$~ppm. All synthetic spectra have transit depths larger than 7400~ppm in the 0.8-1 $\mu$m range and do not fit within $1 \sigma$ the TESS data (see Section~\ref{sec: TESS}).

In the JWST wavelength range with a single transit in each instrument, we expect clear \ce{H2O} features for the exo-Saturn WASP-117b in the JWST wavelength range (Figure~\ref{fig: HST_JWST_pred}, bottom panels). Furthermore, some models with low metallicity solutions (\textit{Model 2} and \textit{Model 3} framework) yield $\sim$~100 ppm strong \ce{CH4} features at 3.4 $\mu$m. For example, the best fit model for \textit{Model 2} based on the CASCADE spectrum predicts a clear \ce{CH4} signal at 3.4 $\mu$m (Figure~\ref{fig: HST_JWST_pred}, bottom, left panel). This would correspond to a \ce{CH4} abundance of $1.8\times 10^{-5}$ volume fraction (Table~\ref{table: HST_retrieval}). 

Observing WASP-117b with JWST/NIRSpec should thus at the very least impose constraints on the \ce{CH4} and \ce{H2O} content in the atmosphere. Observations of relatively high \ce{CH4} abundances of $10^{-5}$ volume fraction could also disprove high metallicity via chemistry constraints as shown in Section~\ref{sec: Disequi} (Figure~\ref{fig:diseq_bestfit_retrievals}). Non-observations of \ce{CH4} could, on the other hand, indicate either high metallicity, high atmospheric equilibrium temperatures ($T_{eq}\gtrsim$ 1000~K) during transit, or disequilibrium chemistry. In the latter case, we could also constrain vertical mixing ($K_{zz}$). It is also evident from Figure~\ref{fig: HST_JWST_pred} that CO$_2$ is a useful probe of metallicity, with the high metallicity cases showing increased \ce{CO2} absorption at 4.3 $\mu$m and at the red edge of the spectra, towards 15~$\mu$m.

\begin{figure*}
   \centering
   \textbf{Nominal}  \hspace{7.5 cm}  \textbf{CASCADE}\par
   \includegraphics[width=0.48\textwidth, height=5 cm]{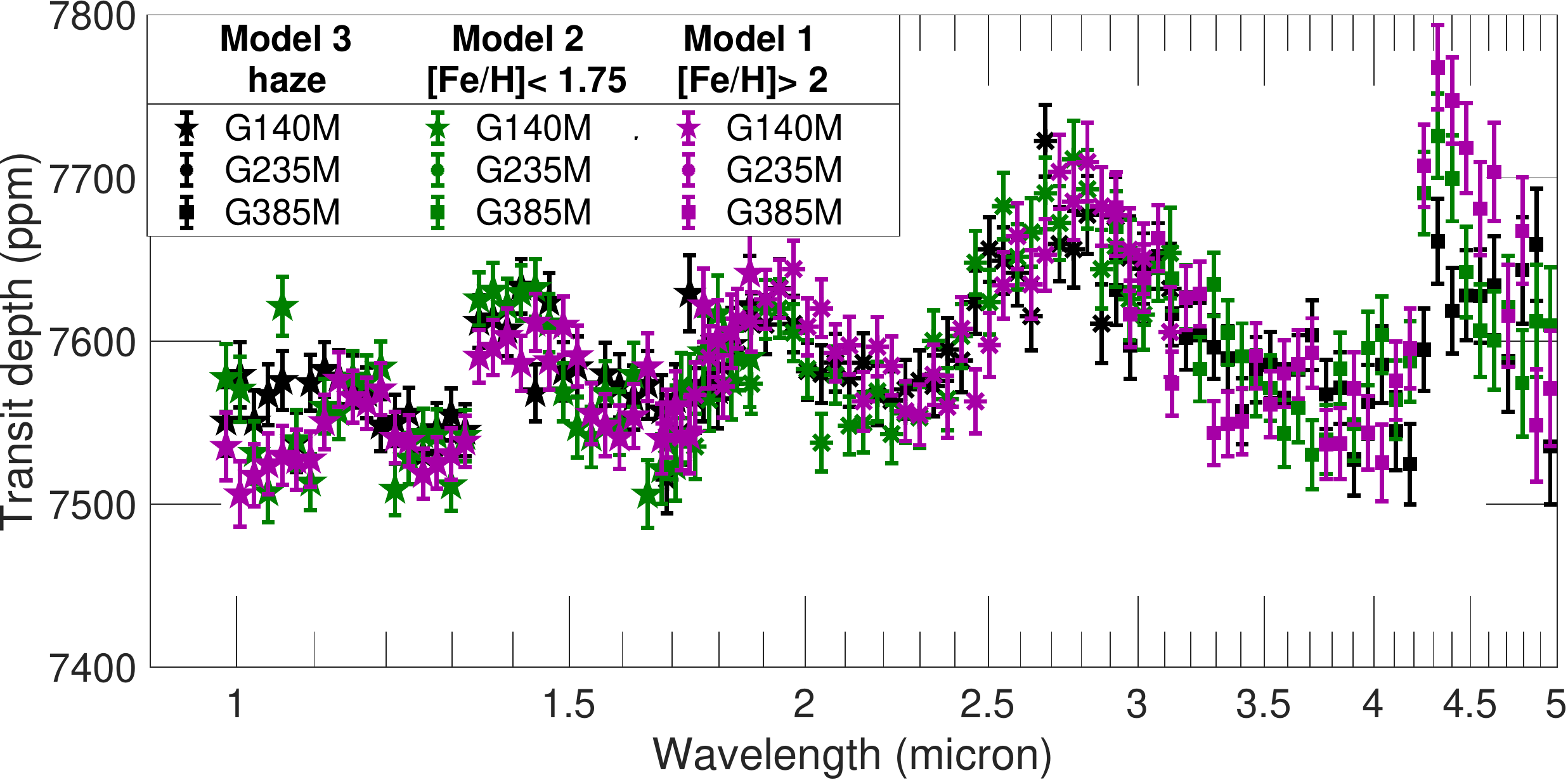}
   \includegraphics[width=0.48\textwidth, height=5 cm]{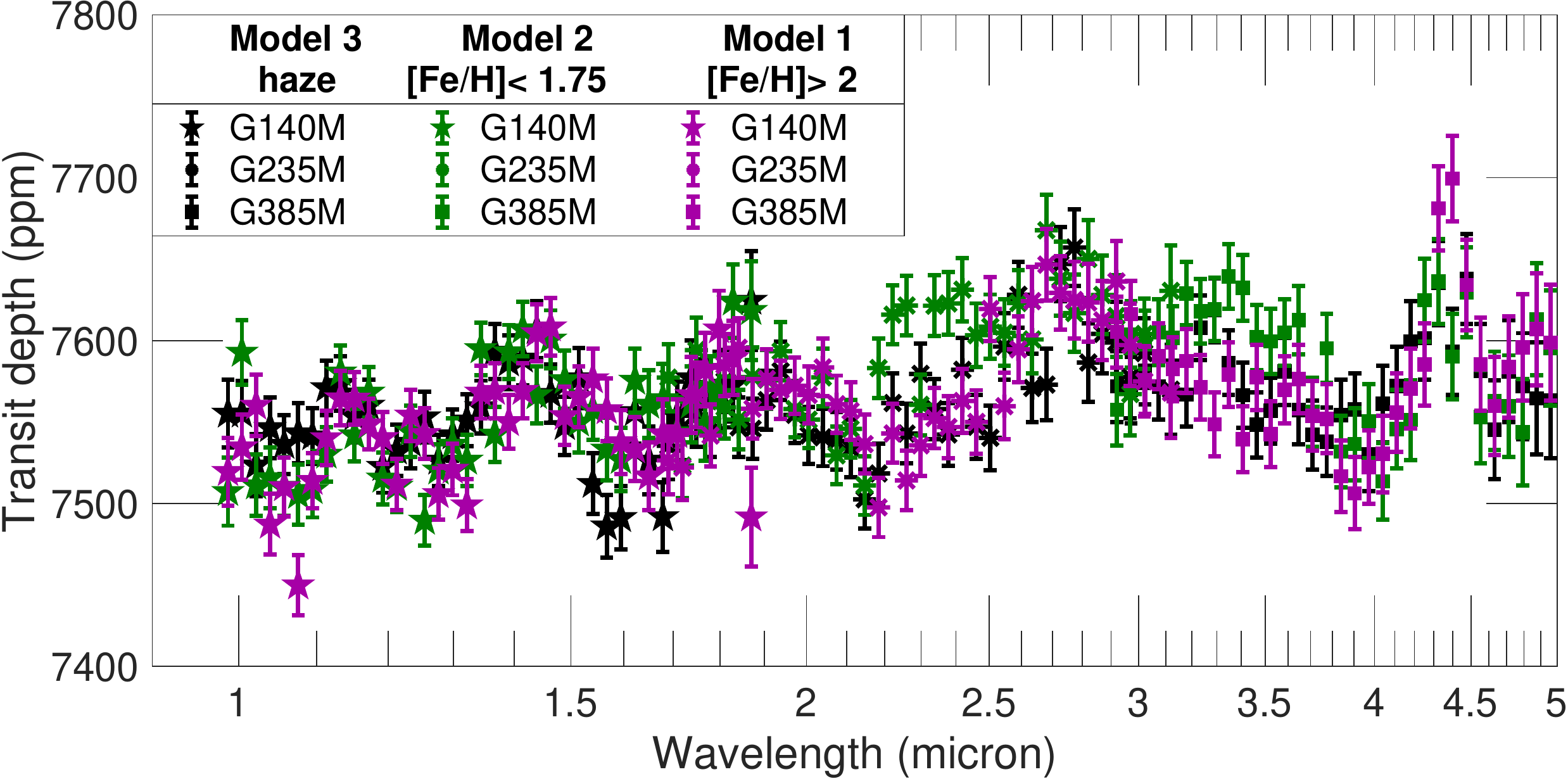}
      \caption{Predicted JWST/NIRSpec observations in the 0.8 - 5~$\mu$m wavelength range based on the median of \textit{Model 1} (high metallicity, black), \textit{Model 2} (low metallicity \& cloudy, green) and \textit{Model 3} (haze, purple) based on the WASP-117b spectrum reduced with the nominal pipeline (left panel), and the CASCADE pipeline (right panel). The model with estimated instrument performance was generated by \textit{PandExo}\citep{Batalha2017}, assuming observation of a single transit with resolution $R\sim 30$. Hexagons, stars and squares denote simulated NIRSpec G140M, G235M and G385M data, respectively. }
         \label{fig: Pandexo}
   \end{figure*}

Taking into account the expected instrument performance confirms our statement that NIRSpec should be capable, even with one single transit observation of WASP-117b, to resolve differences in \ce{CO2} at 4.3 $\mu$m (NIRSpec/G385M) that are due to  different metallicity assumptions in between atmospheric models (Figure~\ref{fig: Pandexo}) as well as constrain \ce{CH4} at both, 2.3 and 3.4 $\mu$m (NIRSpec/G235M). This is readily apparent in Figure~\ref{fig: Pandexo}, right panel for CASCADE Model 2 (green) with enhanced \ce{CH4} content  compared to other retrieved models (see Table~\ref{table: HST_retrieval}).

The comparison between spectra derived from the nominal and CASCADE pipelines (Figure~\ref{fig: Pandexo} left and right panel, respectively) also shows that the nominal \ce{H2O} features in the NIRSpec/G235M wavelength range are in the former case significantly enhanced compared to the latter, regardless of which underlying model is used. The differences in depth of the water absorption is mainly due to the differences in atmospheric temperatures assumed in the model medians. Models retrieved from the nominal HST/WFC3 spectrum are unconstrained in temperature within the prior range of 500--1500~K, so that the median model assumes 1000~K, whereas models retrieving the CASCADE HST/WFC3 spectrum retrieve lower median temperature of 800~K and cooler (Table~\ref{table: HST_retrieval}).

 The simultaneous measurement of \ce{H2O}, \ce{CH4}, \ce{CO} and \ce{CO2} with JWST thus promises to improve the constraint of the C/O ratio of the planetary atmosphere and thus either confirm or disprove the sub-solar C/O ratio that we tend to find for WASP-117b. JWST will further be extremely important to constrain temperature, metallicity, \ce{CH4} content and thus methane chemistry in the exo-Saturn WASP-117b during transit. HST/UVIS observations will be more important to constrain hazes and cloudiness, which could also constrain atmospheric metallicity if an accuracy of 100 ppm or better is reached between 0.3 and 0.6 $\mu$m.
 
\section{Discussion}
\label{sec: dicuss}

This work shows that it is possible but also very challenging to investigate the atmospheric properties of a warm gas planet with an orbital period of 10~days and longer. The longer the orbital period, the longer the transit duration that needs to be captured in its entirety to perform accurate transmission spectroscopy. For WASP-117b, the total transit duration was six hours, and a further 1--2 hours of out-of-transit observations before and after transit were needed for calibration. Thus, a relatively large time investment was needed to obtain accurate transit spectroscopy with one single transit with HST (11 consecutive orbits) and VLT/ESPRESSO (10 hour continuous observation), respectively. We also note that, in 2018, there were only two occasions to capture the full transit with VLT in its entirety from the ground. Otherwise, only partial transits were observable at night.

Atmosphere retrieval for WASP-117b favors, based on Bayesian analysis alone, a high metallicity ([Fe/H] $>2.2$), clear atmospheric composition (\textit{Model 1}) ``substantially'' ($B>4$) over the cloudy, lower metallicity solution (\textit{Model 2}) as fit to the detected $3 \sigma$ water spectrum. See Section~\ref{sec: Significance_model} for an overview of the statistical evidence for the different models. We caution, however, against using Bayesian analysis of the HST/WFC3 data alone to rule out cloudy, lower metallicity solutions, completely. We also caution that patchy clouds could mimick a high metalliticty composition \citep{lineparmentier2016}. We also note that \citet{Anisman2020} retrieved atmospheric properties of WASP-117b based on the same HST data with their own pipeline. While these authors confirm the \ce{H2O} detection, they can not place constraints on atmospheric metallicity or \ce{CH4} abundances.

VLT/ESPRESSO observations of WASP-117b in the optical could not be used to constrain the atmospheric properties of WASP-117b further. The planetary Na and K signal is too small ($2.4\sigma$). Thus, we conclude that exoplanets with long orbital periods should only be observed with ground based telescopes, after space based observations have yielded first constraints on the cloud coverage and thus the feasibility of obtaining a strong enough planetary signal in the optical. 

TESS data could also not be used for atmospheric constraints, because the TESS transit depth is abnormally shallow. We find that the TESS data is inconsistent with all atmospheric models that we explored, including cloudy and haze atmosphere models (Section~\ref{sec: W117_prediction}). Stellar activity is also unlikely to cause a 4\% transit depth differences between the TESS and the WFC3/NIR wavelength range (Figure~\ref{fig: TESS_compare}). This example shows, however, how difficult it can be to combine measurements taken at two different epochs in two different wavelengths, even when the respective atmospheric measurements are precise and the host star is quiet \citep{Rackham2019}.

\begin{figure*}
   \centering
   \includegraphics[width=0.48\textwidth, height=7 cm]{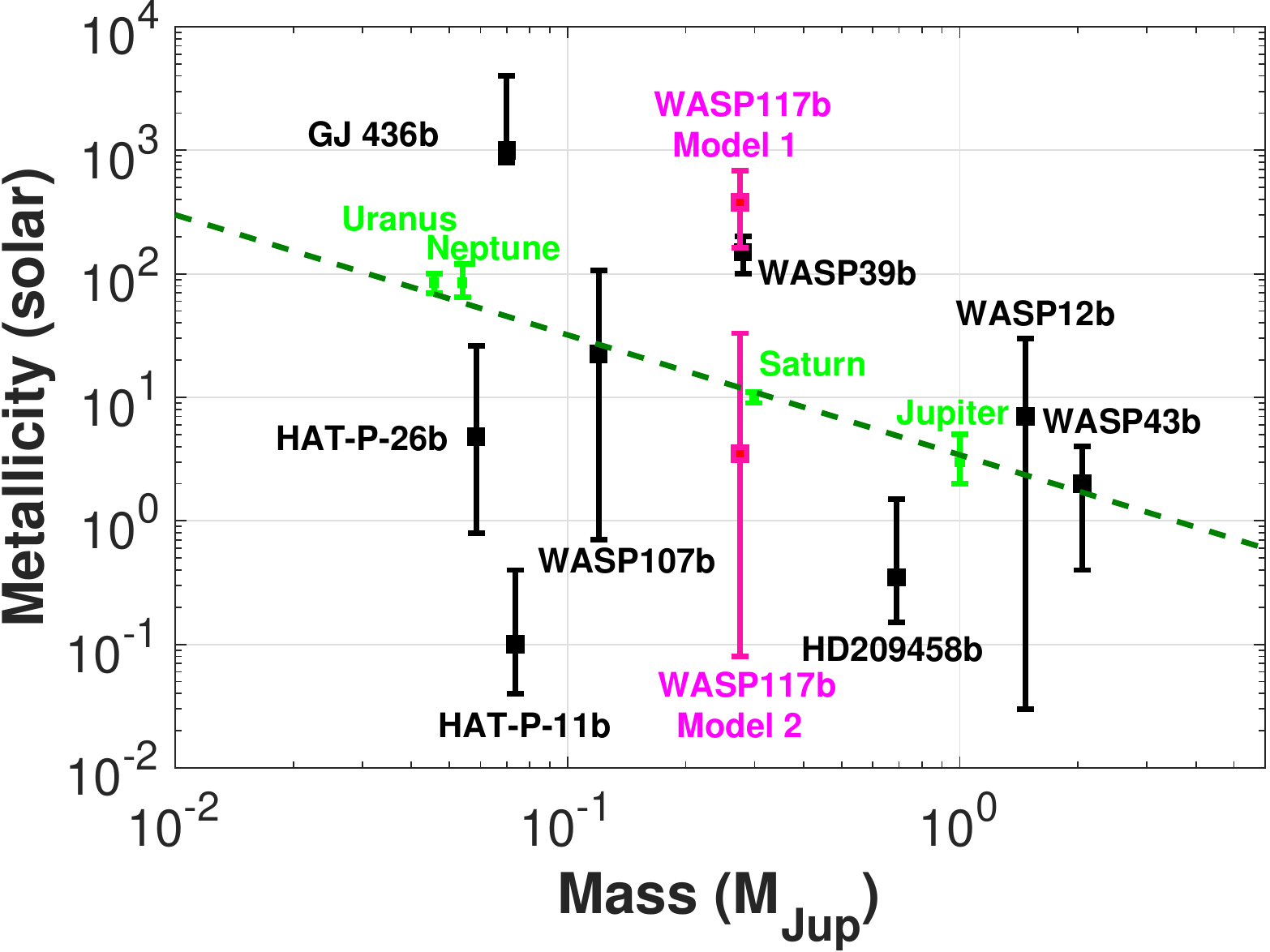}
   \includegraphics[width=0.48\textwidth, height=8.5 cm]{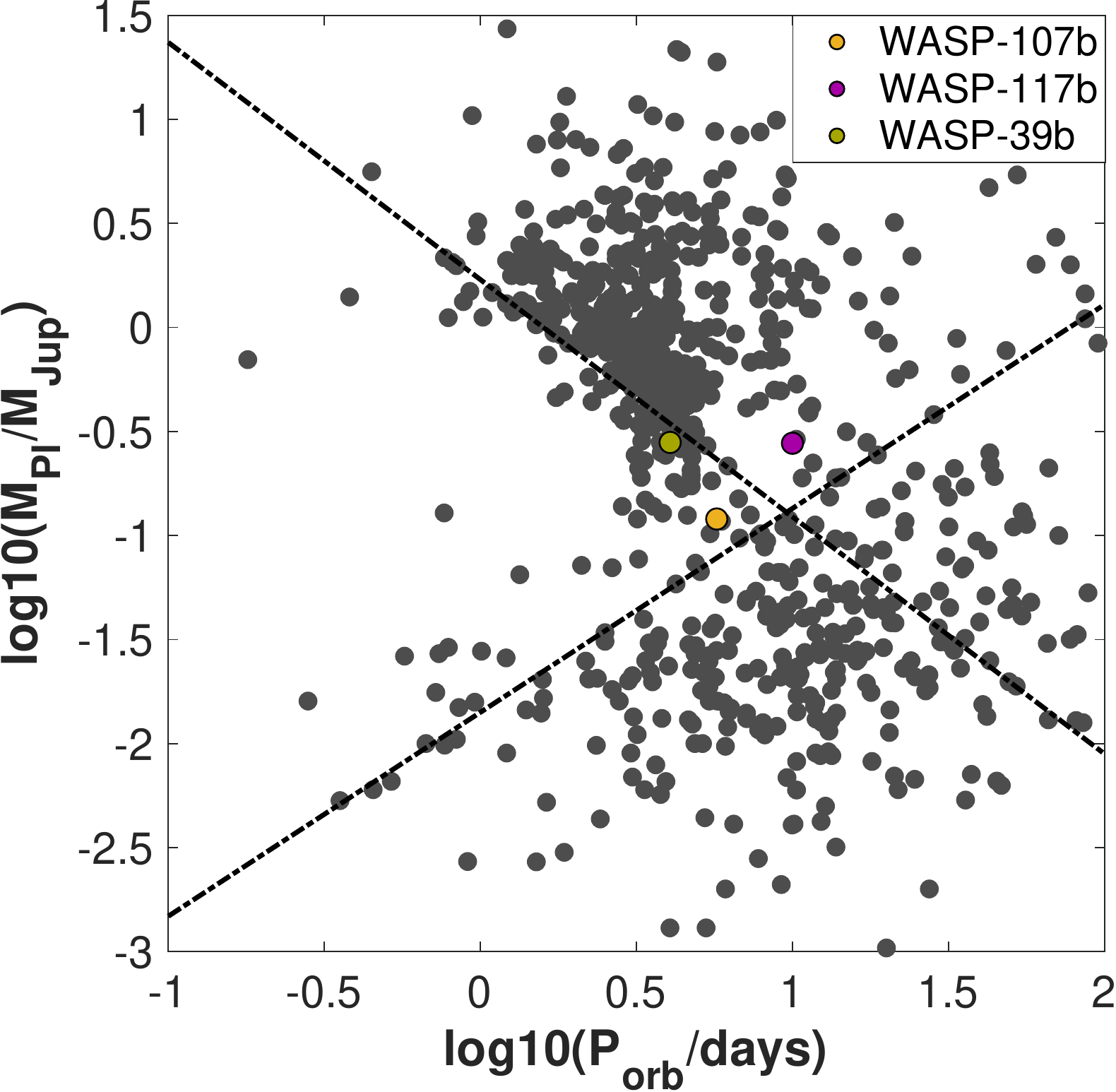}
      \caption{\textbf{Left:} Metallicity content derived for several exoplanets, adapted from \citet{Chachan2019}. WASP-117b is mainly comparable in mass and metallicity to WASP-107b \citep{Kreidberg2018} and WASP-39b \citep{Wakeford2018}. Solar System values are from \citet[e.g.][]{Wong2004, Helled2018,Chachan2019,Movshovitz2020}\newline
      \textbf{Right:} All transiting planets with orbital periods shorter than 100~days and the Neptune desert limits as proposed by \citet{Mazeh2016}. WASP-39b and WASP-107b lie clearly inside the Neptune desert are thus probably subject to erosion processes. WASP-117b lies outside the Neptune desert. (Derived from exoplanets.eu on 30.3.2020)
              }
         \label{fig: W117_metal_Neptune}
   \end{figure*}

We maintain that a better constraint of the metallicity in the atmosphere of the eccentric WASP-117b is warranted as its properties would be complementary to other exoplanets on circular orbits in the same mass range: A clear sky, high metallicity atmospheric composition would place WASP-117b significantly outside of the mass-metallicity relation inferred from the Solar System\footnote{The values for the Solar System bodies come with the caveat that even there the inferred metallicity depend on uncertainties like core size and \ce{H2O} content \citep{Thorngren2019}. (Figure~\ref{fig: W117_metal_Neptune}, left panel). The exo-Saturn WASP-117b would thus join ranks with planets like WASP-39b with comparatively high metallicity \citep{Wakeford2018}, and HAT-P-11b \citep{Chachan2019}, which also do not follow the Solar System mass-metallicity relation. In fact, a high atmospheric metallicity and its mass and radius would make WASP-117b very similar to WASP-39b. A cloudy, lower atmospheric metallicity composition (\textit{Model 2}), on the other hand, would make WASP-117b more comparable to WASP-107b, WASP-43b and WASP-12b that all follow the inferred Solar System mass-metallicity correlation \citep[see also][]{Thorngren2019}.} 

Further, it is unclear in how far erosion processes have influenced the atmospheric metallicity of extrasolar Neptune and Saturn-mass exoplanets \citep[see e.g.][]{Owen2018,DosSantos2019,Armstrong2020}. Also, here the atmospheric properties of the eccentric WASP-117b could be illuminating. Figure~\ref{fig: W117_metal_Neptune}, right panel, shows that WASP-39b and WASP-107b lie in the Neptune\footnote{ One could also call it the sub-Jovian desert \citep{Owen2018} as the mass range of planets affected by photoevaporation is not strictly constrained to Neptune mass.} desert \citep{Mazeh2016}, which is thought to reflect significant mass loss in this planetary mass and orbital period regime. The Super-Neptune WASP-107b was indeed found to experience loss of helium \citep{Spake2018,Kirk2020} that could modify atmospheric metallicity over time. The exo-Saturn WASP-117b, on the other hand, should lie at  0.1168~AU at least during transit well outside of the Neptune desert. The large distance during transit also explains the absence of deep Na and K features in the VLT/ESPRESSO data during transit. The atmospheric metallicity of WASP-117b could thus  shed light on if a high metallicity atmospheric composition (if confirmed) is a sign of atmospheric erosion processes or instead of formation processes different from the Solar System.

The atmosphere of WASP-117b could also be illustrative for the study of disequilibrium processes. We found consistently low $\ce{CH4}$ abundances ($<10^{-4}$ volume fraction), expressed by our retrievals finding sub-solar C/O ratios. Interpreting the atmospheric chemistry in more detail requires, however, the additional knowledge of the atmospheric temperature, which we could not constrain within prescribed limits of 500 to 1500~K based on the WFC3 data alone.

We maintain, however, based on simple equilibrium temperatures calculations using \citet{Mendez2017} that temperatures between 700 and 1000~K are physical during the transit of the eccentric exo-Saturn WASP-117b (Section~\ref{sec: Disequi}, Table~\ref{table: W117_temperature}). Chemical modelling for 700 K and 1000 K for high ($350 \times$ solar) and low metallicities ($5 \times$ solar) further confirm that very low $\ce{CH4}$ abundances in WASP-117b during transit are physically plausible. The 700~K models require disequilibrium chemistry to keep atmospheric \ce{CH4} abundances below $10^{-4}$ volume fraction  at $p=10^{-4}$~bar, which we identified to first order as the main region that is probed during transmission. For the colder temperature models, we thus infer a vertical eddy diffusion coefficient of $K_{zz} \geq 10^8$~cm$^2$/s to explain absence of methane via \ce{CH4} quenching. 

The eddy diffusion coefficient $K_{zz}$ is not trivially constrained for irradiated exoplanet atmospheres. Since these atmospheres tend to have deep, stably stratified radiative zones, the quenching point of most species typically does not lie within the (deeper) convective zone, and the traditional mixing-length theory breaks down. Instead, vertical mixing in irradiated exoplanets is described as being caused by medium- to large-scale atmospheric circulation, which -- when globally averaged -- may or may not act as a diffusive process \citep[see e.g.~recent discussions in ][]{Zhang2020, Showman2020}. It is possible to study these atmospheric motions in 3D using general circulation models, and from these, derive order-of-magnitude parametrizations for the 1D eddy diffusion coefficient $K_{zz}$. For instance, for the hot Jupiter HD~209458b $K_{zz}=5\cdot10^8$~cm$^2$/s -- $10^9$~cm$^2$/s is derived at 1~bar \citep{Parmentier2013, Menou2019}. We note that WASP-117b is cooler than HD~209458b, and $K_{zz}$ should decrease with decreasing equilibrium temperature \citep{Komacek2019}. However, atmospheres with higher metallicities, as we infer for WASP-117b, are also expected to have more vertical mixing than lower-metallicity atmospheres \citep{Charnay2015}. Our tentative inference of $K_{zz} \geq 10^8$~cm$^2$/s for the ($T_{\rm eq}=700$~K)-model thus shows reasonable agreement with theoretically motivated expectations. However, a more sophisticated 3D model would be needed to further strengthen the agreement.

We argue that additional transit measurements of WASP-117b with HST/UVIS and the G280 grism could yield the necessary accuracy (200-100 ppm, \citep{Wakeford2020}) to constrain a haze and cloud layer in WASP-117b. A low resolution transmission spectrum in that range would reveal cloud properties due to the different scattering slopes between 0.3 and 0.5~$\mu$m for a clear atmosphere with high metallicity, a grey cloudy and hazy atmosphere with low metallicity (Figure~\ref{sec: W117_prediction}, upper panels). HST/UVIS measurements would thus already constrain atmospheric metallicity, which is degenerate with cloud properties in the HST/WFC3 range. In addition, HST/UVIS data could also potentially shed light on the origin of the shallow TESS transit depth. Such data would also warrant the use of a patchy cloud model \citep{lineparmentier2016}.

Additional measurements with JWST/NIRSpec observations in the 0.8-5 micron region could further constrain \ce{CH4} to at least $\leq 1.8 \times 10^{-5}$ volume fraction by observing in the $3.4~\mu$m \ce{CH4} band. A further constraint on \ce{CH4} but also on \ce{CO} at $4.7~\mu$m would improve our understanding of atmosphere chemistry in WASP-117b, in particular, in conjunction with secondary eclipse observations to constrain temperature variations of WASP-117b during one orbit. The presence and absence of \ce{CH4}, or \ce{CO2} at 4.3~micron would also constrain atmospheric metallicity and together with \ce{H2O} features confirm if WASP-117b indeed has a subsolar C/O ratio. See also \citet{mesakitzmann2020} for the capabilities to use JWST/NIRSpec to characterize the atmospheric chemistry of warm Neptunes. 

\section{Conclusions}
\label{sec: Conclusion}

We  detected a $3 \sigma$ water spectrum in the HST/WFC3 near-infrared transmission spectrum of WASP-117b. Comparison to one-dimensional atmospheric models assuming vertical isothermal temperatures, a uniform grey cloud coverage and equilibrium chemistry (\texttt{petitRADTRANS}) shows that the water absorption is less pronounced than expected with a solar-type metallicity composition. Based on the preferences derived from the Bayes factor analysis alone, WASP-117b appears to have a high atmospheric metallicity (${\rm [Fe/H]}=2.58^{+0.26}_{-0.37}$ or $380^{+311}_{-217} \times$ solar, based on the nominal pipeline) with relatively clear skies. We cannot rule, however, lower atmospheric metallicity ($\rm{Fe/H]}<1.75$) with clouds and potentially an additional haze layer. However, we conclude that the data quality is too limited to warrant the application of a hazy atmosphere layer. The high metallicities that are strongly preferred by a Bayesian factor analysis are comparable to similarly high atmospheric metallicities ($151^{+46}_{-48} \times$ solar) found in WASP-39b, an exo-Saturn on a circular 4~days orbit around a cooler G type star \citep{Wakeford2018}.

While the basic atmospheric composition could be constrained, the temperature of WASP-117b during transit is unconstrained between 500--1500~K. Thus, we can not identify if the planet changes temperature during its eccentric 10~days orbit. Based on basic energy balance equations and realistic range of heat redistribution scenarios, we estimate that the planet could theoretically adopt equilibrium temperatures between 700~K and 1000~K, depending on albedo and heat adjustment.

 Because the temperature of WASP-117b during transit is not constrained, we cannot further test multiple  possible hypotheses that may explain the low abundance of \ce{CH4} in the atmosphere of WASP-117b during transit. If the atmospheric temperature is above 1000~K, then \ce{CH4} can not form and we do not expect to detect measurable levels of \ce{CH4}. If the temperature of WASP-117b is during transit 700~K, then higher \ce{CH4} abundances than $ 10^{-4}$ volume fraction would form in the atmospheric levels that we probe ($p<10^{-4}$~bar), assuming equilibrium chemistry. Thus, in the latter case, we would need to invoke \ce{CH4} quenching and we could constrain the eddy diffusion coefficient to $K_{zz} \geq 10^8$~cm${}^2$/s. If WASP-117b is cool, it would thus show methane-quenching like indicated for the super-Neptune WASP-107b~\citep{Kreidberg2018}.

 Using VLT/ESPRESSO data alone, we do not find strong indications for Na and K within $3 \sigma$. In combination with HST/WFC3 data, however, we still find ``substantial'' ($2.4 \sigma$) preference for the presence of Na and K.

 We also compared our WFC3 spectrum to TESS broad band observations of WASP-117b and found that the TESS transit depth is 4\% shallower than the HST/WFC3 transit depth. We can not explain this difference with either systematic noise, stellar activity or clouds. 

 Furthermore, with the new VLT/ESPRESSO data, we improved via the Rossiter-MacLaughlin effect measurements of the $v \sin i$ of the star and the projected spin-orbit angle to $1.46\pm 0.14$~km/s and $-46.9^{+5.5}_{-4.8}$~deg, respectively. 

\section{Outlook}
\label{sec: Outlook}

WASP-117b is one of the first warm eccentric exo-Saturns with a long orbital period ($10$~days), the atmosphere of which was characterized. While the observations proved to be challenging, they indicate that WASP-117b could have interesting properties such that further observations with JWST and HST in the optical range would be warranted. 

 WASP-117b could be an ideal target to investigate how (disequilibrium) chemistry processes are controlled by precisely known, varying heating conditions on a 10 days orbit around a quiet F main sequence star.  This studies would complement studies of eccentric exoplanets on tighter orbits \citep[e.g.][]{Lewis2013, Lewis2014,Agundez2014b}. WASP-117b is also a complementary observation target to the super-Neptunes WASP-107b and WASP-39b on circular, tighter orbits to aid our understanding on how erosion processes shape atmospheric properties such as metallicities in the Neptune-to-Saturn mass range.

As of now, several promising extrasolar exo-Saturns on eccentric orbits with $P_{orb}\geq $10~days around bright stars suitable for in-depth atmospheric characterization are known \citep{Brahm2016,Brahm2018,Jordan2019a,Brahm2019,Rodriguez2019}. One such planet, K2-287~b, orbits a G-type star with a time-averaged equilibrium temperature of 800~K \citep{Jordan2019a}, that is, it is at least 200~K cooler than WASP-117b would thus be even more ideal to study disequilibrium chemistry in warm eccentric gas planets on relatively wide orbits.

\begin{acknowledgements}
      L.C. acknowledges support by the DLR grant 50OR1804 and DFG grant~CA 1795/3. F.~Y. acknowledges the support of the DFG priority program SPP 1992 "Exploring the Diversity of Extrasolar Planets (RE 1664/16-1)". RB is a Ph.D.~fellow of the Research Foundation--Flanders (FWO). P.M. and Th.H. acknowledge support from the European Research Council under the European Union's Horizon 2020 research and innovation program under grant agreement No. 832428. B.V.R. thanks the Heising-Simons Foundation for support. A.J.\ acknowledges support from FONDECYT project 1171208 and by the Ministry for the Economy, Development, and Tourism's Programa Iniciativa Cient\'{i}fica Milenio through grant IC\,120009, awarded to the Millennium Institute of Astrophysics (MAS). Support for Programs number GO-15301 and AR-15060 were provided by NASA through a grant from the Space Telescope Science Institute, which is operated by the Association of Universities for Research in Astronomy, Incorporated, under NASA contract NAS5-26555. O.V. thanks the CNRS/INSU Programme National de Plan\'etologie (PNP) and CNES for funding support. J.B. acknowledges support from the European Research Council under the European Union's Horizon 2020 research and innovation program ExoplANETS-A under grant agreement No.~776403. We would also like to thank the reviewer Bruno Bézard for his thoughtful and thorough review which greatly improved this paper.
\end{acknowledgements}

%
%

\bibliographystyle{aa} 
\bibliography{WASP117.bib}

\begin{thebibliography}{112}
\expandafter\ifx\csname natexlab\endcsname\relax\def\natexlab#1{#1}\fi

\bibitem[{{Ag{\'u}ndez} {et~al.}(2014){Ag{\'u}ndez}, {Venot}, {Selsis}, \&
  {Iro}}]{Agundez2014b}
{Ag{\'u}ndez}, M., {Venot}, O., {Selsis}, F., \& {Iro}, N. 2014, \apj, 781, 68

\bibitem[{{Anisman} {et~al.}(2020){Anisman}, {Edwards}, {Changeat}, {Venot},
  {Al-Refaie}, {Tsiaras}, \& {Tinetti}}]{Anisman2020}
{Anisman}, L.~O., {Edwards}, B., {Changeat}, Q., {et~al.} 2020, \aj, 160, 233

\bibitem[{{Apai} {et~al.}(2017){Apai}, {Karalidi}, {Marley}, {Yang}, {Flateau},
  {Metchev}, {Cowan}, {Buenzli}, {Burgasser}, {Radigan}, {Artigau}, \&
  {Lowrance}}]{Apai2017}
{Apai}, D., {Karalidi}, T., {Marley}, M.~S., {et~al.} 2017, Science, 357, 683

\bibitem[{{Armstrong} {et~al.}(2020){Armstrong}, {Lopez}, {Adibekyan}, {Booth},
  {Bryant}, {Collins}, {Deleuil}, {Emsenhuber}, {Huang}, {King}, {Lillo-Box},
  {Lissauer}, {Matthews}, {Mousis}, {Nielsen}, {Osborn}, {Otegi}, {Santos},
  {Sousa}, {Stassun}, {Veras}, {Ziegler}, {Acton}, {Almenara}, {Anderson},
  {Barrado}, {Barros}, {Bayliss}, {Belardi}, {Bouchy}, {Brice{\~n}o}, {Brogi},
  {Brown}, {Burleigh}, {Casewell}, {Chaushev}, {Ciardi}, {Collins},
  {Col{\'o}n}, {Cooke}, {Crossfield}, {D{\'\i}az}, {Delgado Mena}, {Demangeon},
  {Dorn}, {Dumusque}, {Eigm{\"u}ller}, {Fausnaugh}, {Figueira}, {Gan},
  {Gandhi}, {Gill}, {Gonzales}, {Goad}, {G{\"u}nther}, {Helled}, {Hojjatpanah},
  {Howell}, {Jackman}, {Jenkins}, {Jenkins}, {Jensen}, {Kennedy}, {Latham},
  {Law}, {Lendl}, {Lozovsky}, {Mann}, {Moyano}, {McCormac}, {Meru},
  {Mordasini}, {Osborn}, {Pollacco}, {Queloz}, {Raynard}, {Ricker}, {Rowden},
  {Santerne}, {Schlieder}, {Seager}, {Sha}, {Tan}, {Tilbrook}, {Ting}, {Udry},
  {Vanderspek}, {Watson}, {West}, {Wilson}, {Winn}, {Wheatley}, {Villasenor},
  {Vines}, \& {Zhan}}]{Armstrong2020}
{Armstrong}, D.~J., {Lopez}, T.~A., {Adibekyan}, V., {et~al.} 2020, \nat, 583,
  39

\bibitem[{{Asplund} {et~al.}(2009){Asplund}, {Grevesse}, {Sauval}, \&
  {Scott}}]{Asplund2009}
{Asplund}, M., {Grevesse}, N., {Sauval}, A.~J., \& {Scott}, P. 2009, \araa, 47,
  481

\bibitem[{{Astropy Collaboration} {et~al.}(2013){Astropy Collaboration},
  {Robitaille}, {Tollerud}, {Greenfield}, {Droettboom}, {Bray}, {Aldcroft},
  {Davis}, {Ginsburg}, {Price-Whelan}, {Kerzendorf}, {Conley}, {Crighton},
  {Barbary}, {Muna}, {Ferguson}, {Grollier}, {Parikh}, {Nair}, {Unther},
  {Deil}, {Woillez}, {Conseil}, {Kramer}, {Turner}, {Singer}, {Fox}, {Weaver},
  {Zabalza}, {Edwards}, {Azalee Bostroem}, {Burke}, {Casey}, {Crawford},
  {Dencheva}, {Ely}, {Jenness}, {Labrie}, {Lim}, {Pierfederici}, {Pontzen},
  {Ptak}, {Refsdal}, {Servillat}, \& {Streicher}}]{astropy:2013}
{Astropy Collaboration}, {Robitaille}, T.~P., {Tollerud}, E.~J., {et~al.} 2013,
  \aap, 558, A33

\bibitem[{{Barman} {et~al.}(2011{\natexlab{a}}){Barman}, {Macintosh},
  {Konopacky}, \& {Marois}}]{Barman2011b}
{Barman}, T.~S., {Macintosh}, B., {Konopacky}, Q.~M., \& {Marois}, C.
  2011{\natexlab{a}}, \apj, 733, 65

\bibitem[{{Barman} {et~al.}(2011{\natexlab{b}}){Barman}, {Macintosh},
  {Konopacky}, \& {Marois}}]{Barman2011}
{Barman}, T.~S., {Macintosh}, B., {Konopacky}, Q.~M., \& {Marois}, C.
  2011{\natexlab{b}}, \apjl, 735, L39

\bibitem[{{Barstow}(2020)}]{barstow2020}
{Barstow}, J.~K. 2020, arXiv e-prints, arXiv:2002.02945

\bibitem[{{Barstow} {et~al.}(2013){Barstow}, {Aigrain}, {Irwin}, {Bowles},
  {Fletcher}, \& {Lee}}]{barstowaigrain2013}
{Barstow}, J.~K., {Aigrain}, S., {Irwin}, P.~G.~J., {et~al.} 2013, \mnras, 430,
  1188

\bibitem[{{Batalha} {et~al.}(2017){Batalha}, {Mandell}, {Pontoppidan},
  {Stevenson}, {Lewis}, {Kalirai}, {Earl}, {Greene}, {Albert}, \&
  {Nielsen}}]{Batalha2017}
{Batalha}, N.~E., {Mandell}, A., {Pontoppidan}, K., {et~al.} 2017, \pasp, 129,
  064501

\bibitem[{{Benneke} {et~al.}(2019){Benneke}, {Knutson}, {Lothringer},
  {Crossfield}, {Moses}, {Morley}, {Kreidberg}, {Fulton}, {Dragomir}, {Howard},
  {Wong}, {D{\'e}sert}, {McCullough}, {Kempton}, {Fortney}, {Gilliland },
  {Deming}, \& {Kammer}}]{Benneke2019}
{Benneke}, B., {Knutson}, H.~A., {Lothringer}, J., {et~al.} 2019, Nature
  Astronomy, 3, 813

\bibitem[{{Benneke} \& {Seager}(2013)}]{Benneke2013}
{Benneke}, B. \& {Seager}, S. 2013, \apj, 778, 153

\bibitem[{{Brahm} {et~al.}(2019){Brahm}, {Espinoza}, {Jord{\'a}n}, {Henning},
  {Sarkis}, {Jones}, {D{\'\i}az}, {Jenkins}, {Vanzi}, {Zapata}, {Petrovich},
  {Kossakowski}, {Rabus}, {Rojas}, \& {Torres}}]{Brahm2019}
{Brahm}, R., {Espinoza}, N., {Jord{\'a}n}, A., {et~al.} 2019, \aj, 158, 45

\bibitem[{{Brahm} {et~al.}(2018){Brahm}, {Espinoza}, {Jord{\'a}n}, {Rojas},
  {Sarkis}, {D{\'\i}az}, {Rabus}, {Drass}, {Lachaume}, {Soto}, {Jenkins},
  {Jones}, {Henning}, {Pantoja}, \& {Vu{\v{c}}kovi{\'c}}}]{Brahm2018}
{Brahm}, R., {Espinoza}, N., {Jord{\'a}n}, A., {et~al.} 2018, \mnras, 477, 2572

\bibitem[{{Brahm} {et~al.}(2016){Brahm}, {Jord{\'a}n}, {Bakos}, {Penev},
  {Espinoza}, {Rabus}, {Hartman}, {Bayliss}, {Ciceri}, {Zhou}, {Mancini},
  {Tan}, {de Val-Borro}, {Bhatti}, {Csubry}, {Bento}, {Henning}, {Schmidt},
  {Rojas}, {Suc}, {L{\'a}z{\'a}r}, {Papp}, \& {S{\'a}ri}}]{Brahm2016}
{Brahm}, R., {Jord{\'a}n}, A., {Bakos}, G.~{\'A}., {et~al.} 2016, \aj, 151, 89

\bibitem[{{Brammer} {et~al.}(2015){Brammer}, {Ryan}, \&
  {Pirzkal}}]{Brammer2015wfc}
{Brammer}, G., {Ryan}, R., \& {Pirzkal}, N. 2015, {Source-dependent master sky
  images for the WFC3/IR grisms}, Space Telescope WFC Instrument Science Report

\bibitem[{{Buchner}(2014)}]{Buchner2014}
{Buchner}, J. 2014, arXiv e-prints, arXiv:1407.5459

\bibitem[{{Caldas} {et~al.}(2019){Caldas}, {Leconte}, {Selsis}, {Waldmann},
  {Bord{\'e}}, {Rocchetto}, \& {Charnay}}]{caldasleconte2019}
{Caldas}, A., {Leconte}, J., {Selsis}, F., {et~al.} 2019, \aap, 623, A161

\bibitem[{{Carnall}(2017)}]{SpectRes2017}
{Carnall}, A.~C. 2017, arXiv e-prints, arXiv:1705.05165

\bibitem[{{Castelli} \& {Kurucz}(2003)}]{Castelli2003}
{Castelli}, F. \& {Kurucz}, R.~L. 2003, in IAU Symposium, Vol. 210, Modelling
  of Stellar Atmospheres, ed. N.~{Piskunov}, W.~W. {Weiss}, \& D.~F. {Gray},
  A20

\bibitem[{{Chachan} {et~al.}(2019){Chachan}, {Knutson}, {Gao}, {Kataria},
  {Wong}, {Henry}, {Benneke}, {Zhang}, {Barstow}, {Bean}, {Mikal-Evans},
  {Lewis}, {Mansfield}, {L{\'o}pez-Morales}, {Nikolov}, {Sing}, \&
  {Wakeford}}]{Chachan2019}
{Chachan}, Y., {Knutson}, H.~A., {Gao}, P., {et~al.} 2019, \aj, 158, 244

\bibitem[{{Charnay} {et~al.}(2015){Charnay}, {Meadows}, \&
  {Leconte}}]{Charnay2015}
{Charnay}, B., {Meadows}, V., \& {Leconte}, J. 2015, \apj, 813, 15

\bibitem[{{Crossfield}(2015)}]{Crossfield2015}
{Crossfield}, I. J.~M. 2015, \pasp, 127, 941

\bibitem[{{Czesla} {et~al.}(2015){Czesla}, {Klocov{\'a}}, {Khalafinejad},
  {Wolter}, \& {Schmitt}}]{Czesla2015}
{Czesla}, S., {Klocov{\'a}}, T., {Khalafinejad}, S., {Wolter}, U., \&
  {Schmitt}, J.~H.~M.~M. 2015, \aap, 582, A51

\bibitem[{{Czesla} {et~al.}(2019){Czesla}, {Schr{\"o}ter}, {Schneider},
  {Huber}, {Pfeifer}, {Andreasen}, \& {Zechmeister}}]{Czesla2019}
{Czesla}, S., {Schr{\"o}ter}, S., {Schneider}, C.~P., {et~al.} 2019, {PyA:
  Python astronomy-related packages}

\bibitem[{{Deming} {et~al.}(2015){Deming}, {Knutson}, {Kammer}, {Fulton},
  {Ingalls}, {Carey}, {Burrows}, {Fortney}, {Todorov}, {Agol}, {Cowan},
  {Desert}, {Fraine}, {Langton}, {Morley}, \& {Showman}}]{pld}
{Deming}, D., {Knutson}, H., {Kammer}, J., {et~al.} 2015, \apj, 805, 132

\bibitem[{{Deming} {et~al.}(2013){Deming}, {Wilkins}, {McCullough}, {Burrows},
  {Fortney}, {Agol}, {Dobbs-Dixon}, {Madhusudhan}, {Crouzet}, {Desert},
  {Gilliland}, {Haynes}, {Knutson}, {Line}, {Magic}, {Mand ell}, {Ranjan},
  {Charbonneau}, {Clampin}, {Seager}, \& {Showman}}]{Deming2013}
{Deming}, D., {Wilkins}, A., {McCullough}, P., {et~al.} 2013, \apj, 774, 95

\bibitem[{{dos Santos} {et~al.}(2019){dos Santos}, {Ehrenreich}, {Bourrier},
  {Lecavelier des Etangs}, {L{\'o}pez-Morales}, {Sing}, {Ballester},
  {Ben-Jaffel}, {Buchhave}, {Garc{\'\i}a Mu{\~n}oz}, {Henry}, {Kataria},
  {Lavie}, {Lavvas}, {Lewis}, {Mikal-Evans}, {Sanz-Forcada}, \&
  {Wakeford}}]{DosSantos2019}
{dos Santos}, L.~A., {Ehrenreich}, D., {Bourrier}, V., {et~al.} 2019, \aap,
  629, A47

\bibitem[{{Espinoza}(2018)}]{Espinoza2018}
{Espinoza}, N. 2018, Research Notes of the American Astronomical Society, 2,
  209

\bibitem[{{Espinoza} {et~al.}(2019){Espinoza}, {Kossakowski}, \&
  {Brahm}}]{Espinoza2019}
{Espinoza}, N., {Kossakowski}, D., \& {Brahm}, R. 2019, \mnras, 490, 2262

\bibitem[{{Foreman-Mackey} {et~al.}(2013){Foreman-Mackey}, {Hogg}, {Lang}, \&
  {Goodman}}]{Mackey2013}
{Foreman-Mackey}, D., {Hogg}, D.~W., {Lang}, D., \& {Goodman}, J. 2013, \pasp,
  125, 306

\bibitem[{Golub {et~al.}(1979)Golub, Heath, \& Wahba}]{Golub1979}
Golub, G.~H., Heath, M., \& Wahba, G. 1979, Technometrics, 21, 215

\bibitem[{{Guilluy} {et~al.}(2019){Guilluy}, {Sozzetti}, {Brogi}, {Bonomo},
  {Giacobbe}, {Claudi}, \& {Benatti}}]{Guilluy2019}
{Guilluy}, G., {Sozzetti}, A., {Brogi}, M., {et~al.} 2019, \aap, 625, A107

\bibitem[{{Guzm{\'a}n Mesa} {et~al.}(2020){Guzm{\'a}n Mesa}, {Kitzmann},
  {Fisher}, {Burgasser}, {Hoeijmakers}, {M{\'a}rquez-Neila}, {Grimm},
  {Mandell}, {Sznitman}, \& {Heng}}]{mesakitzmann2020}
{Guzm{\'a}n Mesa}, A., {Kitzmann}, D., {Fisher}, C., {et~al.} 2020, arXiv
  e-prints, arXiv:2004.10106

\bibitem[{{Helled} \& {Guillot}(2018)}]{Helled2018}
{Helled}, R. \& {Guillot}, T. 2018, {Internal Structure of Giant and Icy
  Planets: Importance of Heavy Elements and Mixing}, 44

\bibitem[{Hoerl \& Kennard(1970)}]{Hoerl1970}
Hoerl, A.~E. \& Kennard, R.~W. 1970, Technometrics, 12, 55

\bibitem[{{Horne}(1986)}]{Horne1986}
{Horne}, K. 1986, \pasp, 98, 609

\bibitem[{{Janson} {et~al.}(2013){Janson}, {Brandt}, {Kuzuhara}, {Spiegel},
  {Thalmann}, {Currie}, {Bonnefoy}, {Zimmerman}, {Sorahana}, {Kotani},
  {Schlieder}, {Hashimoto}, {Kudo}, {Kusakabe}, {Abe}, {Brand ner}, {Carson},
  {Egner}, {Feldt}, {Goto}, {Grady}, {Guyon}, {Hayano}, {Hayashi}, {Hayashi},
  {Henning}, {Hodapp}, {Ishii}, {Iye}, {Kandori}, {Knapp}, {Kwon}, {Matsuo},
  {McElwain}, {Mede}, {Miyama}, {Morino}, {Moro-Mart{\'\i}n}, {Nakagawa},
  {Nishimura}, {Pyo}, {Serabyn}, {Suenaga}, {Suto}, {Suzuki}, {Takahashi},
  {Takami}, {Takato}, {Terada}, {Tomono}, {Turner}, {Watanabe}, {Wisniewski},
  {Yamada}, {Takami}, {Usuda}, \& {Tamura}}]{Janson2013}
{Janson}, M., {Brandt}, T.~D., {Kuzuhara}, M., {et~al.} 2013, \apjl, 778, L4

\bibitem[{{Jord{\'a}n} {et~al.}(2019){Jord{\'a}n}, {Brahm}, {Espinoza},
  {Cort{\'e}s}, {D{\'\i}az}, {Drass}, {Henning}, {Jenkins}, {Jones}, {Rabus},
  {Rojas}, {Sarkis}, {Vu{\v{c}}kovi{\'c}}, {Zapata}, {Soto}, {Bakos},
  {Bayliss}, {Bhatti}, {Csubry}, {Lachaume}, {Moraga}, {Pantoja}, {Osip},
  {Shporer}, {Suc}, \& {V{\'a}squez}}]{Jordan2019a}
{Jord{\'a}n}, A., {Brahm}, R., {Espinoza}, N., {et~al.} 2019, \aj, 157, 100

\bibitem[{{Jord{\'a}n} {et~al.}(2020){Jord{\'a}n}, {Brahm}, {Espinoza},
  {Henning}, {Jones}, {Kossakowski}, {Sarkis}, {Trifonov}, {Rojas}, {Torres},
  {Drass}, {Nandakumar}, {Barbieri}, {Davis}, {Wang}, {Bayliss}, {Bouma},
  {Dragomir}, {Eastman}, {Daylan}, {Guerrero}, {Barclay}, {Ting}, {Henze},
  {Ricker}, {Vanderspek}, {Latham}, {Seager}, {Winn}, {Jenkins}, {Wittenmyer},
  {Bowler}, {Crossfield}, {Horner}, {Kane}, {Kielkopf}, {Morton}, {Plavchan},
  {Tinney}, {Addison}, {Mengel}, {Okumura}, {Shahaf}, {Mazeh}, {Rabus},
  {Shporer}, {Ziegler}, {Mann}, \& {Hart}}]{Jordan2020}
{Jord{\'a}n}, A., {Brahm}, R., {Espinoza}, N., {et~al.} 2020, \aj, 159, 145

\bibitem[{Kass \& Raftery(1995)}]{Kass1995}
Kass, R.~E. \& Raftery, A.~E. 1995, Journal of the American Statistical
  Association, 90, 773

\bibitem[{{Kipping}(2013)}]{kipping}
{Kipping}, D.~M. 2013, \mnras, 435, 2152

\bibitem[{{Kirk} {et~al.}(2020){Kirk}, {Alam}, {L{\'o}pez-Morales}, \&
  {Zeng}}]{Kirk2020}
{Kirk}, J., {Alam}, M.~K., {L{\'o}pez-Morales}, M., \& {Zeng}, L. 2020, \aj,
  159, 115

\bibitem[{{Kochanek} {et~al.}(2017){Kochanek}, {Shappee}, {Stanek}, {Holoien},
  {Thompson}, {Prieto}, {Dong}, {Shields}, {Will}, {Britt}, {Perzanowski}, \&
  {Pojma{\'n}ski}}]{Kochanek2017}
{Kochanek}, C.~S., {Shappee}, B.~J., {Stanek}, K.~Z., {et~al.} 2017, \pasp,
  129, 104502

\bibitem[{{Komacek} {et~al.}(2019){Komacek}, {Showman}, \&
  {Parmentier}}]{Komacek2019}
{Komacek}, T.~D., {Showman}, A.~P., \& {Parmentier}, V. 2019, \apj, 881, 152

\bibitem[{{Kreidberg}(2015)}]{Kreidberg2015}
{Kreidberg}, L. 2015, \pasp, 127, 1161

\bibitem[{{Kreidberg} {et~al.}(2014){Kreidberg}, {Bean}, {D{\'e}sert}, {Line},
  {Fortney}, {Madhusudhan}, {Stevenson}, {Showman}, {Charbonneau},
  {McCullough}, {Seager}, {Burrows}, {Henry}, {Williamson}, {Kataria}, \&
  {Homeier}}]{Kreidberg2014}
{Kreidberg}, L., {Bean}, J.~L., {D{\'e}sert}, J.-M., {et~al.} 2014, \apjl, 793,
  L27

\bibitem[{{Kreidberg} {et~al.}(2018){Kreidberg}, {Line}, {Thorngren}, {Morley},
  \& {Stevenson}}]{Kreidberg2018}
{Kreidberg}, L., {Line}, M.~R., {Thorngren}, D., {Morley}, C.~V., \&
  {Stevenson}, K.~B. 2018, \apjl, 858, L6

\bibitem[{{Kurucz}(1993)}]{Kurucz1993}
{Kurucz}, R.~L. 1993, Astronomical Society of the Pacific Conference Series,
  Vol.~44, {A New Opacity-Sampling Model Atmosphere Program for Arbitrary
  Abundances}, ed. M.~M. {Dworetsky}, F.~{Castelli}, \& R.~{Faraggiana}, 87

\bibitem[{{Lendl} {et~al.}(2014){Lendl}, {Triaud}, {Anderson}, {Collier
  Cameron}, {Delrez}, {Doyle}, {Gillon}, {Hellier}, {Jehin}, {Maxted},
  {Neveu-VanMalle}, {Pepe}, {Pollacco}, {Queloz}, {S{\'e}gransan}, {Smalley},
  {Smith}, {Udry}, {Van Grootel}, \& {West}}]{Lendl2014}
{Lendl}, M., {Triaud}, A.~H.~M.~J., {Anderson}, D.~R., {et~al.} 2014, \aap,
  568, A81

\bibitem[{{Lewis} {et~al.}(2013){Lewis}, {Knutson}, {Showman}, {Cowan},
  {Laughlin}, {Burrows}, {Deming}, {Crepp}, {Mighell}, {Agol}, {Bakos},
  {Charbonneau}, {D{\'e}sert}, {Fischer}, {Fortney}, {Hartman}, {Hinkley},
  {Howard}, {Johnson}, {Kao}, {Langton}, \& {Marcy}}]{Lewis2013}
{Lewis}, N.~K., {Knutson}, H.~A., {Showman}, A.~P., {et~al.} 2013, \apj, 766,
  95

\bibitem[{{Lewis} {et~al.}(2014){Lewis}, {Showman}, {Fortney}, {Knutson}, \&
  {Marley}}]{Lewis2014}
{Lewis}, N.~K., {Showman}, A.~P., {Fortney}, J.~J., {Knutson}, H.~A., \&
  {Marley}, M.~S. 2014, \apj, 795, 150

\bibitem[{{Line} \& {Parmentier}(2016)}]{lineparmentier2016}
{Line}, M.~R. \& {Parmentier}, V. 2016, \apj, 820, 78

\bibitem[{{Lodders} \& {Fegley}(2002)}]{Lodders2002}
{Lodders}, K. \& {Fegley}, B. 2002, \icarus, 155, 393

\bibitem[{{Lomb}(1976)}]{Lomb1976}
{Lomb}, N.~R. 1976, \apss, 39, 447

\bibitem[{{Luger} {et~al.}(2016){Luger}, {Agol}, {Kruse}, {Barnes}, {Becker},
  {Foreman-Mackey}, \& {Deming}}]{pldk2}
{Luger}, R., {Agol}, E., {Kruse}, E., {et~al.} 2016, \aj, 152, 100

\bibitem[{{MacDonald} {et~al.}(2020){MacDonald}, {Goyal}, \&
  {Lewis}}]{macdonaldgoyal2020}
{MacDonald}, R.~J., {Goyal}, J.~M., \& {Lewis}, N.~K. 2020, arXiv e-prints,
  arXiv:2003.11548

\bibitem[{{MacDonald} \& {Madhusudhan}(2017)}]{macdonaldmadhusudhan2017}
{MacDonald}, R.~J. \& {Madhusudhan}, N. 2017, \mnras, 469, 1979

\bibitem[{{Madhusudhan}(2012)}]{Maddu2012}
{Madhusudhan}, N. 2012, \apj, 758, 36

\bibitem[{{Mallonn} {et~al.}(2019){Mallonn}, {von Essen}, {Herrero},
  {Alexoudi}, {Granzer}, {Sosa}, {Strassmeier}, {Bakos}, {Bayliss}, {Brahm},
  {Bretton}, {Campos}, {Carone}, {Col{\'o}n}, {Dale}, {Dragomir}, {Espinoza},
  {Evans}, {Garcia}, {Gu}, {Guerra}, {Jongen}, {Jord{\'a}n}, {Kang}, {Keles},
  {Kim}, {Lendl}, {Molina}, {Salisbury}, {Scaggiante}, {Shporer}, {Siverd},
  {Sokov}, {Sokova}, \& {W{\"u}nsche}}]{Mallonn2019}
{Mallonn}, M., {von Essen}, C., {Herrero}, E., {et~al.} 2019, \aap, 622, A81

\bibitem[{{Mazeh} {et~al.}(2016){Mazeh}, {Holczer}, \& {Faigler}}]{Mazeh2016}
{Mazeh}, T., {Holczer}, T., \& {Faigler}, S. 2016, \aap, 589, A75

\bibitem[{{M{\'e}ndez} \& {Rivera-Valent{\'\i}n}(2017)}]{Mendez2017}
{M{\'e}ndez}, A. \& {Rivera-Valent{\'\i}n}, E.~G. 2017, \apjl, 837, L1

\bibitem[{{Menou}(2019)}]{Menou2019}
{Menou}, K. 2019, \mnras, 485, L98

\bibitem[{{Miles} {et~al.}(2018){Miles}, {Skemer}, {Barman}, {Allers}, \&
  {Stone}}]{Miles2018}
{Miles}, B.~E., {Skemer}, A.~J., {Barman}, T.~S., {Allers}, K.~N., \& {Stone},
  J.~M. 2018, \apj, 869, 18

\bibitem[{{Molli{\`e}re} {et~al.}(2017){Molli{\`e}re}, {van Boekel}, {Bouwman},
  {Henning}, {Lagage}, \& {Min}}]{Molliere2017}
{Molli{\`e}re}, P., {van Boekel}, R., {Bouwman}, J., {et~al.} 2017, \aap, 600,
  A10

\bibitem[{{Molli{\`e}re} {et~al.}(2015){Molli{\`e}re}, {van Boekel},
  {Dullemond}, {Henning}, \& {Mordasini}}]{Molliere2015}
{Molli{\`e}re}, P., {van Boekel}, R., {Dullemond}, C., {Henning}, T., \&
  {Mordasini}, C. 2015, \apj, 813, 47

\bibitem[{{Molli{\`e}re} {et~al.}(2019){Molli{\`e}re}, {Wardenier}, {van
  Boekel}, {Henning}, {Molaverdikhani}, \& {Snellen}}]{Molliere2019}
{Molli{\`e}re}, P., {Wardenier}, J.~P., {van Boekel}, R., {et~al.} 2019, \aap,
  627, A67

\bibitem[{{Moses} {et~al.}(2016){Moses}, {Marley}, {Zahnle}, {Line}, {Fortney},
  {Barman}, {Visscher}, {Lewis}, \& {Wolff}}]{Moses2016}
{Moses}, J.~I., {Marley}, M.~S., {Zahnle}, K., {et~al.} 2016, \apj, 829, 66

\bibitem[{{Movshovitz} {et~al.}(2020){Movshovitz}, {Fortney}, {Mankovich},
  {Thorngren}, \& {Helled}}]{Movshovitz2020}
{Movshovitz}, N., {Fortney}, J.~J., {Mankovich}, C., {Thorngren}, D., \&
  {Helled}, R. 2020, \apj, 891, 109

\bibitem[{Nagao \& Matsuyama(1979)}]{NAGAO1979394}
Nagao, M. \& Matsuyama, T. 1979, Computer Graphics and Image Processing, 9, 394

\bibitem[{{Owen} \& {Lai}(2018)}]{Owen2018}
{Owen}, J.~E. \& {Lai}, D. 2018, \mnras, 479, 5012

\bibitem[{{Parmentier} {et~al.}(2016){Parmentier}, {Fortney}, {Showman},
  {Morley}, \& {Marley}}]{Parmentier2016}
{Parmentier}, V., {Fortney}, J.~J., {Showman}, A.~P., {Morley}, C., \&
  {Marley}, M.~S. 2016, \apj, 828, 22

\bibitem[{{Parmentier} {et~al.}(2013){Parmentier}, {Showman}, \&
  {Lian}}]{Parmentier2013}
{Parmentier}, V., {Showman}, A.~P., \& {Lian}, Y. 2013, \aap, 558, A91

\bibitem[{{Pepe} {et~al.}(2010){Pepe}, {Cristiani}, {Rebolo Lopez}, {Santos},
  {Amorim}, {Avila}, {Benz}, {Bonifacio}, {Cabral}, {Carvas}, {Cirami},
  {Coelho}, {Comari}, {Coretti}, {De Caprio}, {Dekker}, {Delabre}, {Di
  Marcantonio}, {D'Odorico}, {Fleury}, {Garc{\'\i}a}, {Herreros Linares},
  {Hughes}, {Iwert}, {Lima}, {Lizon}, {Lo Curto}, {Lovis}, {Manescau},
  {Martins}, {M{\'e}gevand}, {Moitinho}, {Molaro}, {Monteiro}, {Monteiro},
  {Pasquini}, {Mordasini}, {Queloz}, {Rasilla}, {Rebord{\~a}o}, {Santana
  Tschudi}, {Santin}, {Sosnowska}, {Span{\`o}}, {Tenegi}, {Udry}, {Vanzella},
  {Viel}, {Zapatero Osorio}, \& {Zerbi}}]{Pepe2010}
{Pepe}, F.~A., {Cristiani}, S., {Rebolo Lopez}, R., {et~al.} 2010, Society of
  Photo-Optical Instrumentation Engineers (SPIE) Conference Series, Vol. 7735,
  {ESPRESSO: the Echelle spectrograph for rocky exoplanets and stable
  spectroscopic observations}, 77350F

\bibitem[{{Pinhas} \& {Madhusudhan}(2017)}]{pinhasmadhu2017}
{Pinhas}, A. \& {Madhusudhan}, N. 2017, \mnras, 471, 4355

\bibitem[{{Pirzkal} {et~al.}(2016){Pirzkal}, {Ryan}, \&
  {Brammer}}]{Pirzkal2016wfc}
{Pirzkal}, N., {Ryan}, R., \& {Brammer}, G. 2016, {Trace and Wavelength
  Calibrations of the WFC3 G102 and G141 IR Grisms}, Space Telescope WFC
  Instrument Science Report

\bibitem[{{Piskunov} \& {Valenti}(2017)}]{Piskunov2016}
{Piskunov}, N. \& {Valenti}, J.~A. 2017, \aap, 597, A16

\bibitem[{{Pluriel} {et~al.}(2020){Pluriel}, {Zingales}, {Leconte}, \&
  {Parmentier}}]{plurielzingales2020}
{Pluriel}, W., {Zingales}, T., {Leconte}, J., \& {Parmentier}, V. 2020, \aap,
  636, A66

\bibitem[{{Price-Whelan} {et~al.}(2018){Price-Whelan}, {Sip{\H{o}}cz},
  {G{\"u}nther}, {Lim}, {Crawford}, {Conseil}, {Shupe}, {Craig}, {Dencheva},
  {Ginsburg}, {VanderPlas}, {Bradley}, {P{\'e}rez-Su{\'a}rez}, {de Val-Borro},
  {Paper Contributors}, {Aldcroft}, {Cruz}, {Robitaille}, {Tollerud},
  {Coordination Committee}, {Ardelean}, {Babej}, {Bach}, {Bachetti}, {Bakanov},
  {Bamford}, {Barentsen}, {Barmby}, {Baumbach}, {Berry}, {Biscani}, {Boquien},
  {Bostroem}, {Bouma}, {Brammer}, {Bray}, {Breytenbach}, {Buddelmeijer},
  {Burke}, {Calderone}, {Cano Rodr{\'\i}guez}, {Cara}, {Cardoso}, {Cheedella},
  {Copin}, {Corrales}, {Crichton}, {D{\textquoteright}Avella}, {Deil},
  {Depagne}, {Dietrich}, {Donath}, {Droettboom}, {Earl}, {Erben}, {Fabbro},
  {Ferreira}, {Finethy}, {Fox}, {Garrison}, {Gibbons}, {Goldstein}, {Gommers},
  {Greco}, {Greenfield}, {Groener}, {Grollier}, {Hagen}, {Hirst}, {Homeier},
  {Horton}, {Hosseinzadeh}, {Hu}, {Hunkeler}, {Ivezi{\'c}}, {Jain}, {Jenness},
  {Kanarek}, {Kendrew}, {Kern}, {Kerzendorf}, {Khvalko}, {King}, {Kirkby},
  {Kulkarni}, {Kumar}, {Lee}, {Lenz}, {Littlefair}, {Ma}, {Macleod},
  {Mastropietro}, {McCully}, {Montagnac}, {Morris}, {Mueller}, {Mumford},
  {Muna}, {Murphy}, {Nelson}, {Nguyen}, {Ninan}, {N{\"o}the}, {Ogaz}, {Oh},
  {Parejko}, {Parley}, {Pascual}, {Patil}, {Patil}, {Plunkett}, {Prochaska},
  {Rastogi}, {Reddy Janga}, {Sabater}, {Sakurikar}, {Seifert}, {Sherbert},
  {Sherwood-Taylor}, {Shih}, {Sick}, {Silbiger}, {Singanamalla}, {Singer},
  {Sladen}, {Sooley}, {Sornarajah}, {Streicher}, {Teuben}, {Thomas},
  {Tremblay}, {Turner}, {Terr{\'o}n}, {van Kerkwijk}, {de la Vega}, {Watkins},
  {Weaver}, {Whitmore}, {Woillez}, {Zabalza}, \& {Contributors}}]{astropy:2018}
{Price-Whelan}, A.~M., {Sip{\H{o}}cz}, B.~M., {G{\"u}nther}, H.~M., {et~al.}
  2018, \aj, 156, 123

\bibitem[{{Queloz} {et~al.}(2000){Queloz}, {Eggenberger}, {Mayor}, {Perrier},
  {Beuzit}, {Naef}, {Sivan}, \& {Udry}}]{Queloz2000}
{Queloz}, D., {Eggenberger}, A., {Mayor}, M., {et~al.} 2000, \aap, 359, L13

\bibitem[{{Rackham} {et~al.}(2018){Rackham}, {Apai}, \&
  {Giampapa}}]{Rackham2018}
{Rackham}, B.~V., {Apai}, D., \& {Giampapa}, M.~S. 2018, \apj, 853, 122

\bibitem[{{Rackham} {et~al.}(2019){Rackham}, {Apai}, \&
  {Giampapa}}]{Rackham2019}
{Rackham}, B.~V., {Apai}, D., \& {Giampapa}, M.~S. 2019, \aj, 157, 96

\bibitem[{{Rocchetto} {et~al.}(2016){Rocchetto}, {Waldmann}, {Venot}, {Lagage},
  \& {Tinetti}}]{rocchettowaldmann2016}
{Rocchetto}, M., {Waldmann}, I.~P., {Venot}, O., {Lagage}, P.~O., \& {Tinetti},
  G. 2016, \apj, 833, 120

\bibitem[{{Rodriguez} {et~al.}(2019){Rodriguez}, {Quinn}, {Huang},
  {Vanderburg}, {Penev}, {Brahm}, {Jord{\'a}n}, {Ikwut-Ukwa}, {Tsirulik},
  {Latham}, {Stassun}, {Shporer}, {Ziegler}, {Matthews}, {Eastman}, {Gaudi},
  {Collins}, {Guerrero}, {Relles}, {Barclay}, {Batalha}, {Berlind}, {Bieryla},
  {Bouma}, {Boyd}, {Burt}, {Calkins}, {Christiansen}, {Ciardi}, {Col{\'o}n},
  {Conti}, {Crossfield}, {Daylan}, {Dittmann}, {Dragomir}, {Dynes}, {Espinoza},
  {Esquerdo}, {Essack}, {Garcia Soto}, {Glidden}, {G{\"u}nther}, {Henning},
  {Jenkins}, {Kielkopf}, {Krishnamurthy}, {Law}, {Levine}, {Lewin}, {Mann},
  {Morgan}, {Morris}, {Oelkers}, {Paegert}, {Pepper}, {Quintana}, {Ricker},
  {Rowden}, {Seager}, {Sarkis}, {Schlieder}, {Sha}, {Tokovinin}, {Torres},
  {Vand erspek}, {Villanueva}, {Villase{\~n}or}, {Winn}, {Wohler}, {Wong},
  {Yahalomi}, {Yu}, {Zhan}, \& {Zhou}}]{Rodriguez2019}
{Rodriguez}, J.~E., {Quinn}, S.~N., {Huang}, C.~X., {et~al.} 2019, \aj, 157,
  191

\bibitem[{{Samland} {et~al.}(2020){Samland}, {Bouwman}, {Hogg}, {Brandner}, \&
  {Henning}}]{Samland2020}
{Samland}, M., {Bouwman}, J., {Hogg}, D.~W., {Brandner}, W., \& {Henning}, T.
  2020, \aap

\bibitem[{{Scargle}(1982)}]{Scargle1982}
{Scargle}, J.~D. 1982, \apj, 263, 835

\bibitem[{Sch{\"o}lkopf {et~al.}(2016)Sch{\"o}lkopf, Hogg, Wang,
  Foreman-Mackey, Janzing, Simon-Gabriel, \& Peters}]{Schoelkopf7391}
Sch{\"o}lkopf, B., Hogg, D.~W., Wang, D., {et~al.} 2016, Proceedings of the
  National Academy of Sciences, 113, 7391

\bibitem[{{Shapiro} {et~al.}(2014){Shapiro}, {Solanki}, {Krivova}, {Schmutz},
  {Ball}, {Knaack}, {Rozanov}, \& {Unruh}}]{Shapiro2014}
{Shapiro}, A.~I., {Solanki}, S.~K., {Krivova}, N.~A., {et~al.} 2014, \aap, 569,
  A38

\bibitem[{{Shappee} {et~al.}(2014){Shappee}, {Prieto}, {Grupe}, {Kochanek},
  {Stanek}, {De Rosa}, {Mathur}, {Zu}, {Peterson}, {Pogge}, {Komossa}, {Im},
  {Jencson}, {Holoien}, {Basu}, {Beacom}, {Szczygie{\l}}, {Brimacombe},
  {Adams}, {Campillay}, {Choi}, {Contreras}, {Dietrich}, {Dubberley},
  {Elphick}, {Foale}, {Giustini}, {Gonzalez}, {Hawkins}, {Howell}, {Hsiao},
  {Koss}, {Leighly}, {Morrell}, {Mudd}, {Mullins}, {Nugent}, {Parrent},
  {Phillips}, {Pojmanski}, {Rosing}, {Ross}, {Sand}, {Terndrup}, {Valenti},
  {Walker}, \& {Yoon}}]{Shappee2014}
{Shappee}, B.~J., {Prieto}, J.~L., {Grupe}, D., {et~al.} 2014, \apj, 788, 48

\bibitem[{{Showman} {et~al.}(2014){Showman}, {Lewis}, \&
  {Fortney}}]{Showman2014}
{Showman}, A.~P., {Lewis}, N.~K., \& {Fortney}, J.~J. 2014, arXiv e-prints,
  arXiv:1411.4731

\bibitem[{{Showman} {et~al.}(2020){Showman}, {Tan}, \&
  {Parmentier}}]{Showman2020}
{Showman}, A.~P., {Tan}, X., \& {Parmentier}, V. 2020, \ssr, 216, 139

\bibitem[{{Showman} {et~al.}(2019){Showman}, {Tan}, \& {Zhang}}]{Showman2019}
{Showman}, A.~P., {Tan}, X., \& {Zhang}, X. 2019, \apj, 883, 4

\bibitem[{{Spake} {et~al.}(2018){Spake}, {Sing}, {Evans}, {Oklop{\v{c}}i{\'c}},
  {}, {Bourrier}, {Kreidberg}, {Rackham}, {Irwin}, {Ehrenreich}, {Wyttenbach},
  {Wakeford}, {Zhou}, {Chubb}, {Nikolov}, {Goyal}, {Henry}, {Williamson},
  {Blumenthal}, {Anderson}, {Hellier}, {Charbonneau}, {Udry}, \&
  {Madhusudhan}}]{Spake2018}
{Spake}, J.~J., {Sing}, D.~K., {Evans}, T.~M., {et~al.} 2018, \nat, 557, 68

\bibitem[{{Thorngren} \& {Fortney}(2019)}]{Thorngren2019}
{Thorngren}, D. \& {Fortney}, J.~J. 2019, \apjl, 874, L31

\bibitem[{{Thorngren} {et~al.}(2019){Thorngren}, {Gao}, \&
  {Fortney}}]{Thorngren2019_tint_and_RCB_depth}
{Thorngren}, D., {Gao}, P., \& {Fortney}, J.~J. 2019, \apjl, 884, L6

\bibitem[{{Thuillier} {et~al.}(2004){Thuillier}, {Floyd}, {Woods}, {Cebula},
  {Hilsenrath}, {Hers{\'e}}, \& {Labs}}]{Thuillier2004}
{Thuillier}, G., {Floyd}, L., {Woods}, T.~N., {et~al.} 2004, Advances in Space
  Research, 34, 256

\bibitem[{{Trotta}(2008)}]{Trotta2008}
{Trotta}, R. 2008, Contemporary Physics, 49, 71

\bibitem[{van~der Walt {et~al.}(2014)van~der Walt, {S}ch\"onberger,
  {Nunez-Iglesias}, {B}oulogne, {W}arner, {Y}ager, {G}ouillart, {Y}u, \& the
  scikit-image contributors}]{scikit-image}
van~der Walt, S., {S}ch\"onberger, J.~L., {Nunez-Iglesias}, J., {et~al.} 2014,
  PeerJ, 2, e453

\bibitem[{{Venot} {et~al.}(2012){Venot}, {H{\'e}brard}, {Ag{\'u}ndez},
  {Dobrijevic}, {Selsis}, {Hersant}, {Iro}, \& {Bounaceur}}]{Venot2012}
{Venot}, O., {H{\'e}brard}, E., {Ag{\'u}ndez}, M., {et~al.} 2012, \aap, 546,
  A43

\bibitem[{{Wakeford} {et~al.}(2018){Wakeford}, {Sing}, {Deming}, {Lewis},
  {Goyal}, {Wilson}, {Barstow}, {Kataria}, {Drummond}, {Evans}, {Carter},
  {Nikolov}, {Knutson}, {Ballester}, \& {Mand ell}}]{Wakeford2018}
{Wakeford}, H.~R., {Sing}, D.~K., {Deming}, D., {et~al.} 2018, \aj, 155, 29

\bibitem[{{Wakeford} {et~al.}(2017){Wakeford}, {Sing}, {Kataria}, {Deming},
  {Nikolov}, {Lopez}, {Tremblin}, {Amundsen}, {Lewis}, {Mandell}, {Fortney},
  {Knutson}, {Benneke}, \& {Evans}}]{Wakeford2017}
{Wakeford}, H.~R., {Sing}, D.~K., {Kataria}, T., {et~al.} 2017, Science, 356,
  628

\bibitem[{{Wakeford} {et~al.}(2020){Wakeford}, {Sing}, {Stevenson}, {Lewis},
  {Pirzkal}, {Wilson}, {Goyal}, {Kataria}, {Mikal-Evans}, {Nikolov}, \&
  {Spake}}]{Wakeford2020}
{Wakeford}, H.~R., {Sing}, D.~K., {Stevenson}, K.~B., {et~al.} 2020, \aj, 159,
  204

\bibitem[{{Wang} {et~al.}(2016){Wang}, {Hogg}, {Foreman-Mackey}, \&
  {Sch{\"o}lkopf}}]{Wang2016PASP}
{Wang}, D., {Hogg}, D.~W., {Foreman-Mackey}, D., \& {Sch{\"o}lkopf}, B. 2016,
  \pasp, 128, 094503

\bibitem[{{Wong} {et~al.}(2004){Wong}, {Mahaffy}, {Atreya}, {Niemann}, \&
  {Owen}}]{Wong2004}
{Wong}, M.~H., {Mahaffy}, P.~R., {Atreya}, S.~K., {Niemann}, H.~B., \& {Owen},
  T.~C. 2004, \icarus, 171, 153

\bibitem[{{Yan} {et~al.}(2019){Yan}, {Casasayas-Barris}, {Molaverdikhani},
  {Alonso-Floriano}, {Reiners}, {Pall{\'e}}, {Henning}, {Molli{\`e}re}, {Chen},
  {Nortmann}, {Snellen}, {Ribas}, {Quirrenbach}, {Caballero}, {Amado},
  {Azzaro}, {Bauer}, {Cort{\'e}s Contreras}, {Czesla}, {Khalafinejad}, {Lara},
  {L{\'o}pez-Puertas}, {Montes}, {Nagel}, {Oshagh}, {S{\'a}nchez-L{\'o}pez},
  {Stangret}, \& {Zechmeister}}]{Yan2019}
{Yan}, F., {Casasayas-Barris}, N., {Molaverdikhani}, K., {et~al.} 2019, \aap,
  632, A69

\bibitem[{{Yan} {et~al.}(2015){Yan}, {Fosbury}, {Petr-Gotzens}, {Zhao}, \&
  {Pall{\'e}}}]{Yan2015a}
{Yan}, F., {Fosbury}, R.~A.~E., {Petr-Gotzens}, M.~G., {Zhao}, G., \&
  {Pall{\'e}}, E. 2015, \aap, 574, A94

\bibitem[{Yan {et~al.}(2015)Yan, Fosbury, Petr-Gotzens, Zhao, Wang, Wang, Liu,
  \& Palll\'{e}}]{Yan2015b}
Yan, F., Fosbury, R. A.~E., Petr-Gotzens, M.~G., {et~al.} 2015, International
  Journal of Astrobiology, 14, 255

\bibitem[{{Yan} \& {Henning}(2018)}]{Yan2018}
{Yan}, F. \& {Henning}, T. 2018, Nature Astronomy, 2, 714

\bibitem[{{Yan} {et~al.}(2017){Yan}, {Pall{\'e}}, {Fosbury}, {Petr-Gotzens}, \&
  {Henning}}]{Yan2017}
{Yan}, F., {Pall{\'e}}, E., {Fosbury}, R.~A.~E., {Petr-Gotzens}, M.~G., \&
  {Henning}, T. 2017, \aap, 603, A73

\bibitem[{{Zhang}(2020)}]{Zhang2020}
{Zhang}, X. 2020, Research in Astronomy and Astrophysics, 20, 099

\bibitem[{{Zhou} {et~al.}(2017){Zhou}, {Apai}, {Lew}, \&
  {Schneider}}]{Zhou2017}
{Zhou}, Y., {Apai}, D., {Lew}, B. W.~P., \& {Schneider}, G. 2017, \aj, 153, 243

\end{thebibliography}

\begin{appendix} 

\section{HST data reduction pipeline description}
\label{sec: HST_pipelines}

In the following, the two data reduction pipelines used in this work are described in more details.

\subsection{Nominal pipeline}
\label{sec: Zhou}

\begin{figure}[!ht]
  \centering
  \includegraphics[width=\columnwidth]{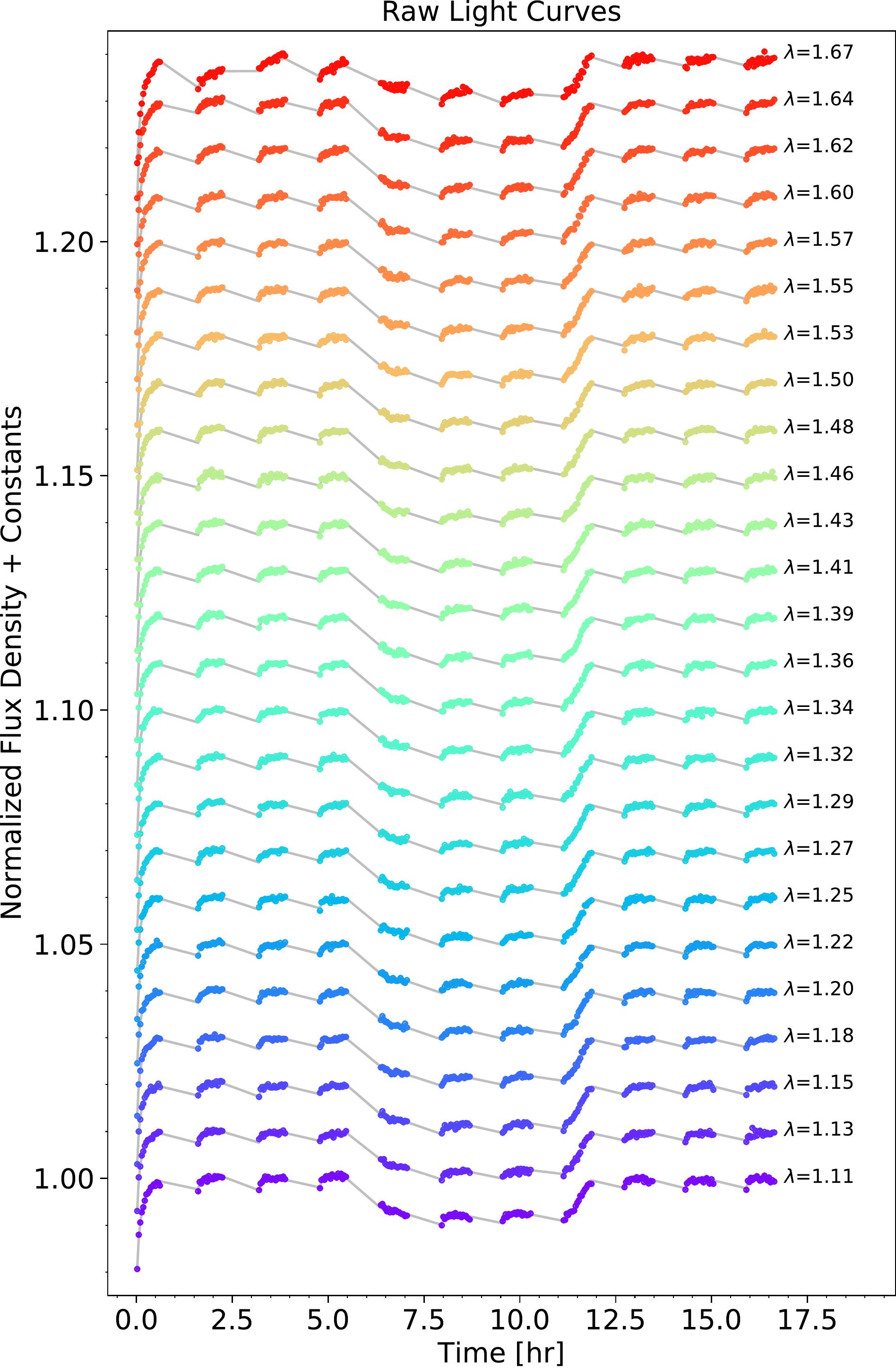}
  \caption{The raw light curves extracted from HST/WFC3 observations of WASP-117b. Each light curve is constructed by the integration of a 5-pixel-wide wavelength bin. Colors of the light curve represent their central wavelengths (using a rainbow color scheme, from red for longest wavelength ($1.67~\mu$m) to blue for the shortest ($1.11~\mu$m)). Light curve systematics are clearly visible, particularly at the first orbit of the observations. The gray curves show the best-fitting RECTE+transit profile models.}
  \label{fig:rawlc}
\end{figure}

We started our data reduction with the \texttt{ima} files, which were produced by the \texttt{CALWFC3} data reduction pipeline. The \texttt{ima} files contain all calibrated non-destructive reads. Before extracting the light curves, we first measure the sub-pixel-level telescope pointing drifts using the cross-correlation method and correct the drift-induced wavelength calibration error by aligning images in the $x$-direction (dispersion direction) using linear interpolation shift. We then followed the standard procedures \citep{Deming2013} to extract light curves from the \texttt{ima} files. First, differential images, which represent the spectrum collected in each non-destructive read, were obtained by subtracting each \texttt{ima} frame from its previous read. Second,  we identified and corrected bad pixels and cosmic rays. We first marked pixels with data quality flags of 4, 16, 32, 256 as bad pixels. We then searched for spurious pixels caused by cosmic ray hits by comparing the differential images with their median-filtered images and selected the $5\sigma$ outliers as cosmic ray pixels. Pixels identified as either bad or cosmic rays were replaced by linear interpolations of their neighboring pixels. Third, we subtracted the sky background from each differential image. The sky background was estimated by taking the median values of pixels that were neither illuminated by any astrophysical sources or bad pixels. To select the background region, we manually constructed pixel masks to exclude all visible spectral traces. Besides, we applied $5\sigma$, 10-iteration sigma-clipping on the unmasked images to remove remaining bright pixels from background estimation. Finally, we measured the raw light curve of each column by taking a $R=$ 45 pixels aperture photometry. We integrated spectroscopic light curves using a 5-pixel wide bin. The raw light curves are presented in Figure~\ref{fig:rawlc}.

The raw light curves demonstrate systematic trends that are typical to WFC3/IR observations (the ``ramp effect''). \citet{Zhou2017} introduced a physically-motivated model for this effect (RECTE) by assuming that the rise of the light curve at the beginning of each orbit is introduced by stimuli gradually filling charge traps that are caused by detector defects. In the RECTE model, there are two populations of charge traps differing in their trapping efficiencies and sustaining lifetime. We corrected the systematics and fit the transit profile simultaneously. We used RECTE to model the systematics. For the transit profile, we adopted the \texttt{batman} \citep{Kreidberg2015} code and assumed the linear limb-darkening formula. The parameters related to the properties of the charge trap, including the numbers, trapping efficiencies, and trapping lifetimes were pre-determined. There were four free parameters in the RECTE model. They were the initially trapped charges and the amounts of trapped charges added during earth occultations for both trap populations. We also included linear trends (two free parameters) for light curves in both forward and backward scanning directions as part of the systematic model. We noted that the light curves demonstrated correlated changes with the telescope pointing variations in the $y$-direction (cross-dispersion direction), which could be adequately modeled by a linear function of $\Delta y$. We added this $\Delta y$ linear term as part of the systematics model as well.

\begin{figure}[!h]
  \centering
  \includegraphics[width=\columnwidth]{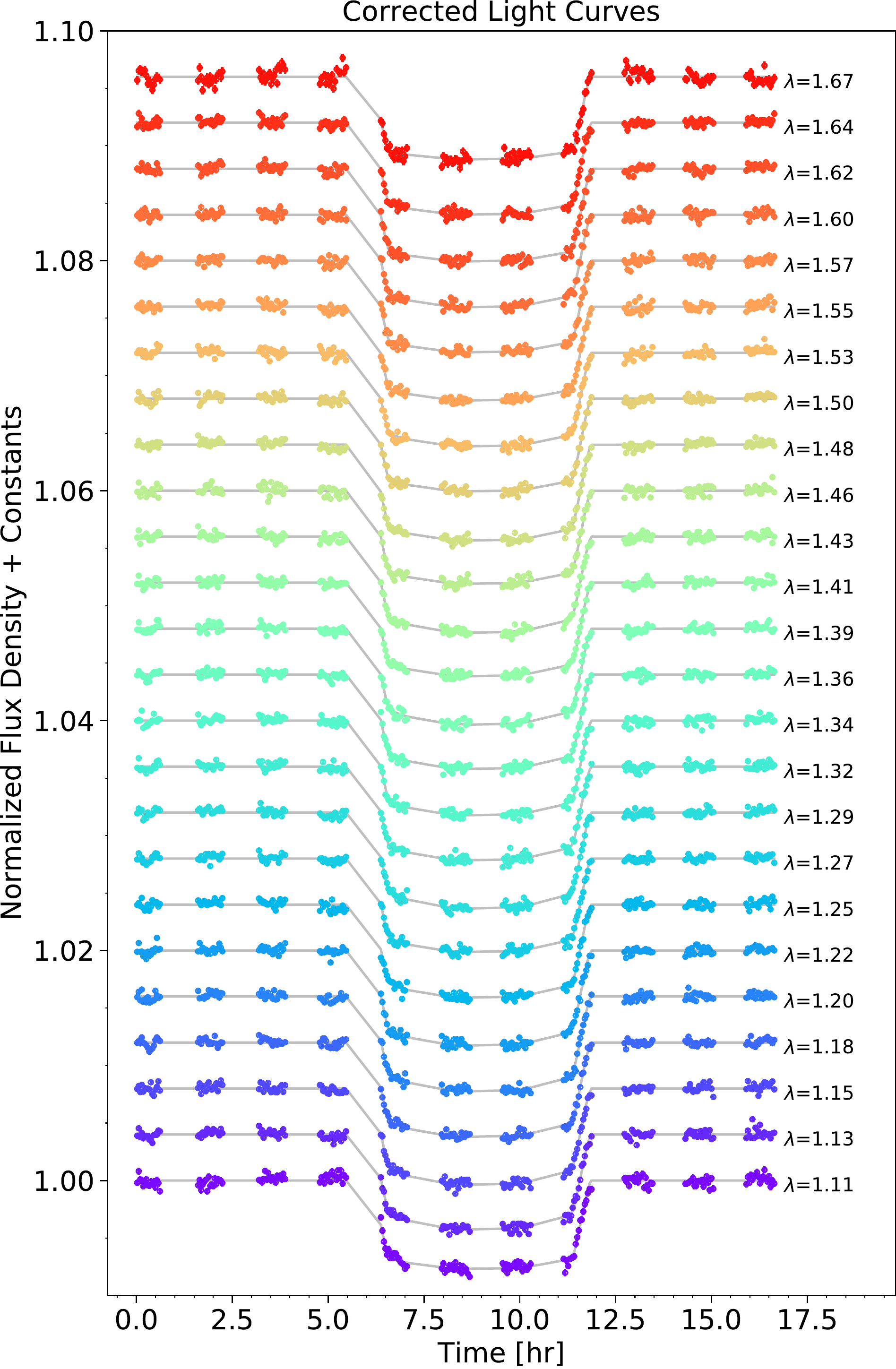}
  \caption{Systematics-corrected light curves of WASP-117b. Compared to the ones shown in Figure~\ref{fig:rawlc}, the systematics are divided-out of the light curves. The light curves are color-coded identically to Figure~\ref{fig:rawlc}. The gray curves represent the best-fitting transit profile models.}
  \label{fig:lc}
\end{figure}

We first conducted the joint transit+systematics fit on the 1.10 to 1.70 $\mu$m broadband light curve. In addition to the parameters for the systematics model described above, parameters of the transit mid-time, the transit depth, and the limb-darkening coefficient were optimized using the \texttt{emcee} code \citep{Mackey2013}. The other planetary-system related parameters, including the orbital period, system semi-major axis, eccentricity, inclination, and the longitude of periastron, were fixed using values from the literature \citep{Lendl2014, Mallonn2019}, which were already good enough for the fit. We then conducted the fit to spectroscopic light curves using the same method except fixing the transit mid-time using the best-fitting value from the broadband fit. The corrected light curves and their best-fitting transit profile models are presented in Figure~\ref{fig:lc}.

\subsection{CASCADE}
\label{sec: CASCADE}
In addition to the nominal pipeline we also used the Calibration of trAnsit Spectroscopy using CAusal Data (CASCADe\footnote{\url{https://jbouwman.gitlab.io/CASCADe/}}) data reduction package developed with the  \emph{Exoplanet Atmosphere New Emission Transmission Spectra Analysis} (ExoplANETS-A\footnote{\url{https://cordis.europa.eu/project/rcn/212911/factsheet/en}}) Horizon-2020 programme.
The CASCADe pipeline also starts the data reduction with the \texttt{ima} intermediate data product, which were produced by the \texttt{CALWFC3} data reduction pipeline. As the data was observed using the \texttt{SPARS10} readout mode with \texttt{NSAMP = 15}, the first non-destructive readout of the 16 samples of the detector ramp stored in the \texttt{ima} data product was removed as it has a much shorter readout time than the other readouts. For the remaining readouts, we constructed pairwise differences between consecutive readouts resulting in 14 signal measurements per detector ramp.

To subtract the background signal from our target signal we used the background subtraction procedure outlined in \cite{Brammer2015wfc}. This entails fitting  and subtracting ``master sky'' images for the zodiacal light and a 1.083 $\mu$m emission line from the Earth’s upper atmosphere appearing in exposures outside of the Earth’s shadow. For details on the exact procedure we refer to section~6 of \citep{Brammer2015wfc}.

To identify and clean bad pixels we first masked all pixels with quality flag other than zero. We then employed an edge-preserving filtering technique similar to  \cite{NAGAO1979394}. We refer to this paper for details on the method. This procedure ensures that the spatial profile of the dispersed light is preserved. As directional filters we use a kernel defined by a two dimensional Gausian kernel who's width is defined by the two standard deviations $\sigma_x$ and $\sigma_y$ of 0.1, respectively 3 pixels which we rotated to obtain a series of different orientations. We flag all pixels deviating more than 4 sigma from the mean determined by the filter profile. After the filtering, we have three data products: one where all bad pixels are masked; a cleaned data set where all bad pixels are replaced by a mean value determined by the optimal filter for those pixels, which we will use to determine the relative telescope movements; and a smoothed data product on which we base our spectral extraction profile.

\begin{figure}[!ht]
  \centering
  \includegraphics[width=\columnwidth]{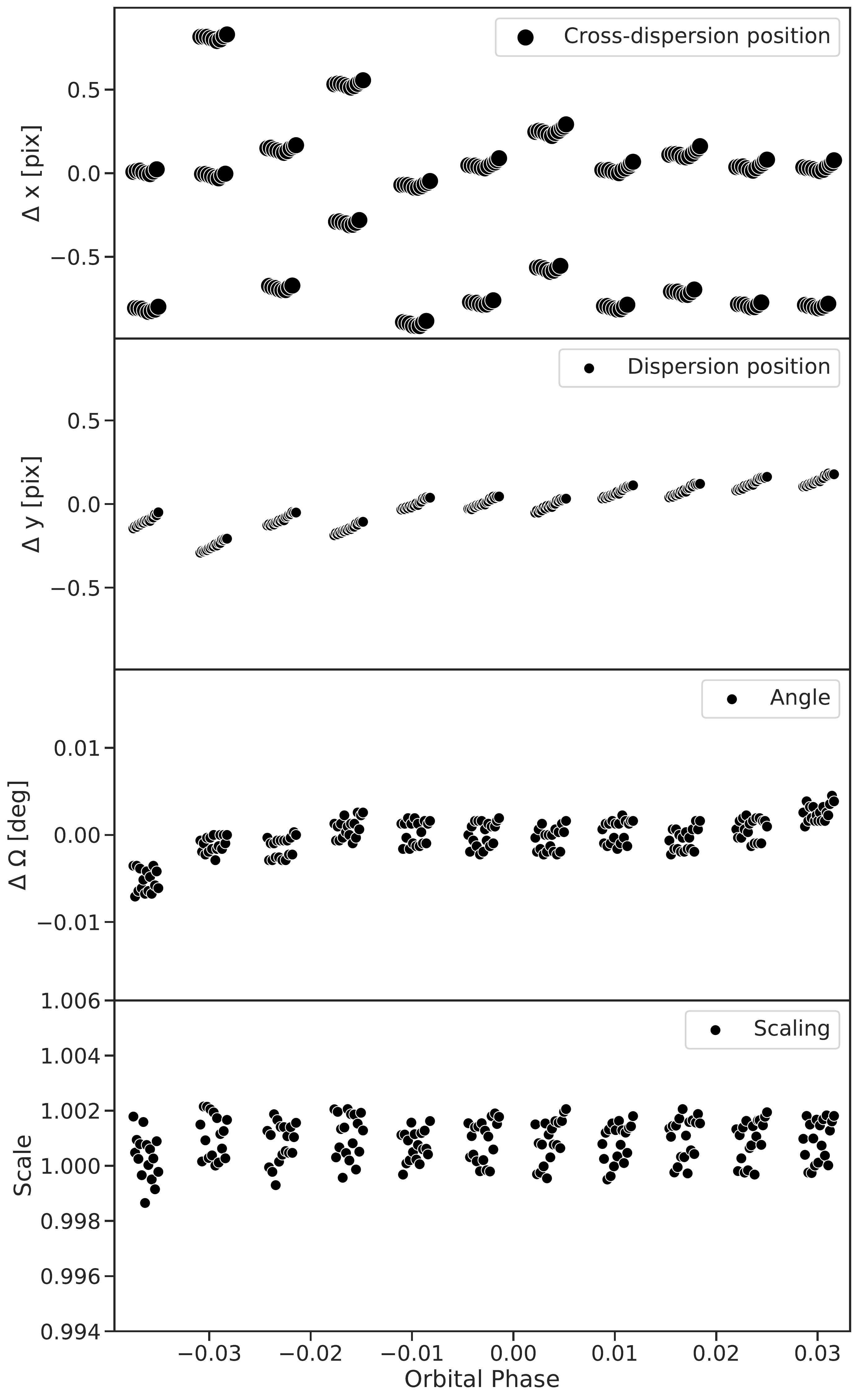}
  \caption{Derived pointing jitter and image distortion during the transit observation. The top panel shows the relative pointing movement in pixels for each spatial scan in the cross-dispersion direction. The second panel shows the relative pointing variations in the dispersion direction. The third panel shows the relative change in the rotation angle and the bottom panel the change in size.}
  \label{fig:cascade_tel_move}
\end{figure}

We use the image of the target with the F126N filter taken at the start of the first orbit to determine the initial wavelength solution and position of the time-series of spectral images taken with the \texttt{G141} grism. We implemented the method outlined in \cite{Pirzkal2016wfc} to determine wavelength calibration and spectral trace from the acquisition image, using the \emph{G141.F126N.V4.32.conf} configuration file for the correct polynomial solution. The initial wavelength and trace solution is expected to have an 0.1 pixel accuracy for the trace description and better than 0.5 pixel ($\sim20$~nanometers) accuracy for the overall wavelength solution. As our observations are performed using the spatial-scan mode, and thus move back and forth around the initial detector position of the target system, we determined the center-of-light position of the first signal measurement to determine the absolute offset to the initial trace position. We then employ a cross-correlation technique in Fourier space, using the register{\textunderscore}translation and  warp{\textunderscore}polar functions in the 
scikit-image package \citep{scikit-image} version 16.2, to determine the relative positional shifts ($\Delta x, \Delta y$), rotation ($\Delta \Omega$) and scale changes of all subsequent exposures. The average change per scan can be seen in Figure~\ref{fig:cascade_tel_move}. We use these relative movements and rotations to update our wavelength solution for the spectral images at each time step. 

\begin{figure}[!ht]
  \centering
  \includegraphics[width=\columnwidth]{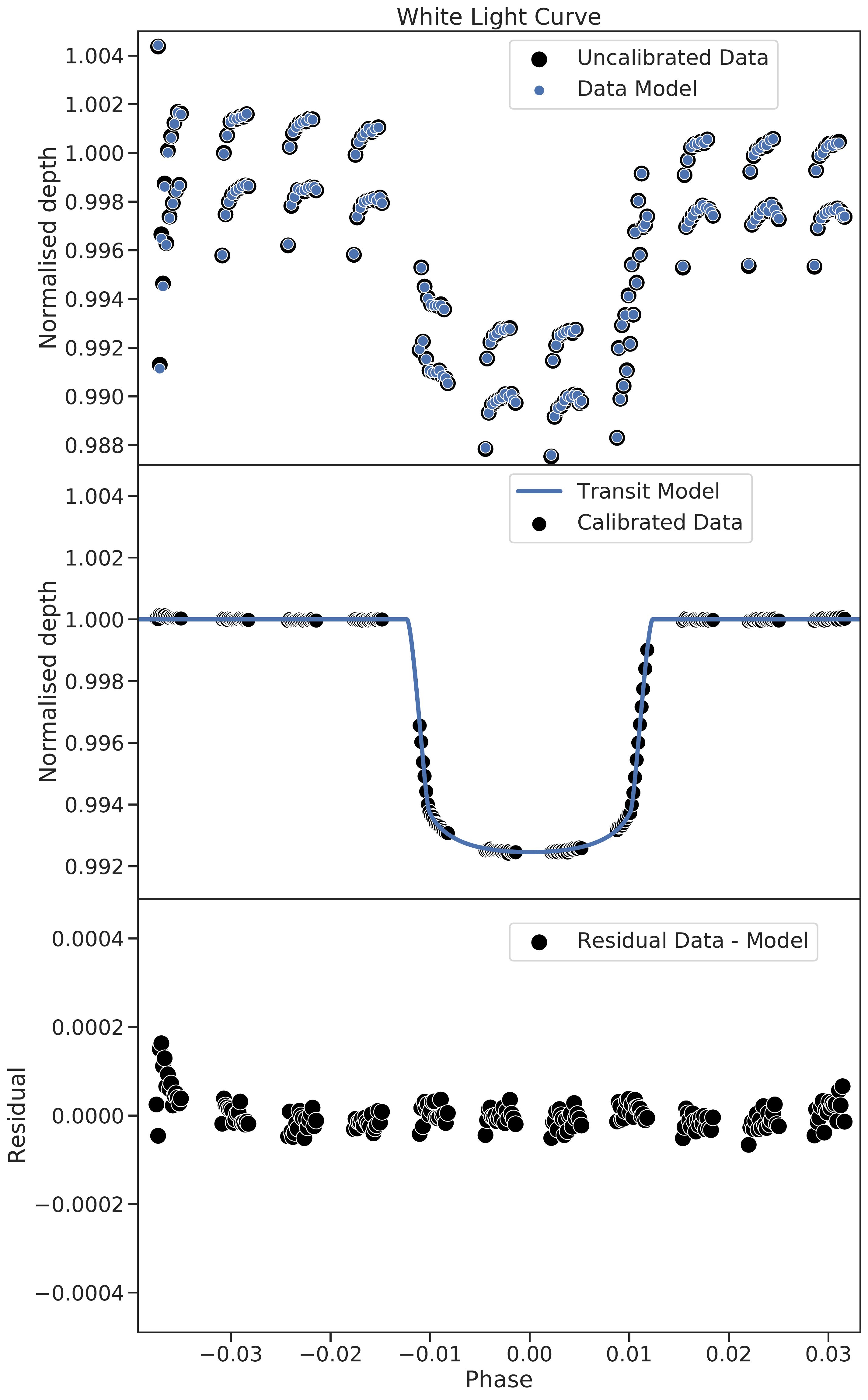}
  \caption{Calibration results of the spectral lightcurves with CASCADe. The top panel shows the wavelength averaged uncalibrated lightcurves for each spatial scan and the wavelength averaged fitted regression model. The bottom panel shows the fit residuals. The middle panel shows the derived calibrated lightcurve and transit model.}
  \label{fig:cascade_lcfit}
\end{figure}

We then optimally extract \citep[see][]{Horne1986} the 1 dimensional spectra from the spectral images using as an extraction profile for each time step the normalized filtered and smoothed spectral images. After the spectral extraction step we rebin the spectra to a uniform wavelength grid using the method outlined in \cite{SpectRes2017} to determine the corresponding flux values and error estimates, after which we create an average spectra per spectral scan. The resulting white light curve of the uncalibrated time series data can be seen in the top panel of Figure~\ref{fig:cascade_lcfit}.

To calibrate the extracted spectral timeseries data and to extract the transmission spectrum of WASP~117b we make use of the \emph{half-sibling-regression} methodology developed by \cite{Schoelkopf7391} using causal connections within the dataset to model both transit signal and any systematics, and which has been successfully applied to transit observations from the Kepler mission \citep{Wang2016PASP} and field-stabilized imaging data \citep{Samland2020}. For a detailed discussion and the mathematical proof of this method we refer to the \cite{Schoelkopf7391} paper. Here we briefly discuss the main concepts and our implementation of the method for the HST spectroscopic data. 

Let's assume we have a time series of (spectral) image data $D_{i,k} \in \mathbb{R}^{M \times T}$, where $M$ is the number of detector pixels and $T$ the number of discrete measurement times in the time series. The main idea is that all data $D_{i,k}$ as a function of detector position $x_i$ at time $t_k$ can be described by an additive noise model:
\begin{equation}
    \begin{split}
    & D(x_i, t_k \,|\, N) =  S(x_i, t_k) + g(x_i, t_k \,|\, N) + \epsilon(x_i, t_k) \\
    & \quad \forall i \in \{1, 2, \dots, M\} \quad \forall k \in \{1, 2, \dots, T\}
    \end{split}
    \label{Eq:additative}
\end{equation}
where $S$ (transit) signal contribution we which to measure, $g$ denotes the functional form of the systematics affecting this measurement, and $\epsilon$ is the stochastic noise (e.g., photon noise). $N$ here is a stand-in for all hidden parameters that may be responsible for causing the systematics. We can now split the dataset into two subsets:
\begin{equation}
    \begin{split}
    \setPs &= \left\{ x_i  \,|\, S(x_i, t_k) > 0 \right\}, \quad \forall k \in \{1, 2, \dots, T\}, \quad i \in \{1, 2, \dots, N\}\\
    \setPc &= \left\{ x_i \,|\, S(x_i, t_k) = 0 \right\}
    \end{split}
    \label{Eq:subsets}
\end{equation}
where the set $\setPs$ consists of all detector data which "see" the signal $S$ and $\setPc$ all other detector data. In practice one assigns those detector data to $\setPc$ which is far enough removed from the position of the source $S$ on the detector such that the calibrator signal is independent of the signal of interest. This is typically a few times the full width half max of the point spread function. Further, as some of the systematics are unique to point sources (e.g. intra-pixel sensitivity variations) one limits $\setPc$ to those pixels seeing other point sources in case of multi object observations \citep[see also][]{Wang2016PASP}. An estimate of the signal $S$ can be made as follows:
\begin{equation}
 \begin{split}
& \hat{S}(x_i, t_k) = D(x_i, t_k \,|\, N) - \E[D(x_i,t_k) \,|\, D(x_j,t_k), \Theta(t_k)], \\
& \quad \forall x_i \in \setPs, \quad \forall x_j \in \setPc, \quad \forall k \in \{1, 2, \dots, T\}
 \end{split}
\end{equation}
where $\E$ is the expectation value i.e. regression model for the dataset $D_{i,k}$ given the other data $D_{j,k}$ and any auxilary data $\Theta_k$, such as derived telescope movement as seen in Figure~\ref{fig:cascade_tel_move} correlated with the systematics present in the data. 

In the case of single object spectra, in contrast to the multi-object photometry like the Kepler data, rather than using the subset of data which contain the signals of other stars as regressors, we make use of the dispersed light, calibrating the observed signal at a particular wavelength with the  data at other wavelengths. For the extracted HST spectra, we can rewrite Equation~\ref{Eq:additative} as:
\begin{equation}
    \begin{split}
    & D(\lambda_i, t_k \,|\, N) =  S(\lambda_i, t_k) + g(\lambda_i, t_k \,|\, N) + \epsilon(\lambda_i, t_k) \\
    & \quad \forall i \in \{1, 2, \dots, M\} \quad \forall k \in \{1, 2, \dots, T\}
    \end{split}
\end{equation}
where the data becomes a function of the wavelength $\lambda_i$. The subsets of detector positions defining the data containing the signal of interest and the dataset used as calibrator described by Equation~\ref{Eq:subsets} now becomes:
\begin{equation}
    \begin{split}
    \setPs &= \left\{ \lambda_i \right\}, \quad i \in \{1, 2, \dots, N\}\\
    \setPc &= \left\{\left\{ \lambda_j \,|\, |\lambda_1-\lambda_j|>\Delta\lambda \right\}, \dots, \left\{ \lambda_j \,|\, |\lambda_N-\lambda_j|>\Delta\lambda \right\}\right\}
    \end{split}
\end{equation}
where we used $\Delta\lambda = 0.07\mu$m, or about 3 resolution elements. Note that for each $\lambda_i$ in $\setPs$ we have a separate subset $\setPci$ of wavelengths which will be used to calibrated the timeseries data at $\lambda_i$.
In practice a linear regression model is sufficient to simultaneously model the transit signal and systematics. As the regressors can be co-linear, we use the \emph{ridge regression} estimator proposed by \citep{Hoerl1970}. The strength of the regularization is determined by Generalized cross validation \citep{Golub1979}. The regression model for the transit signal and systematics thus becomes: 
\begin{multline}
    \argmin_{\omega_i, \,\alpha_{i,j}} \left|\left|
    \vphantom{\sum\limits_{j \,\in\,\setPc}}
    \right.\right.
    \underbrace{
    \vphantom{\left(\omega \hat{S}(\lambda_i, t_k) + \sum\limits_{j \,\in\,\setPc} \alpha_{i,j} D(\lambda_j, t_k)\right)}
    D(\lambda_i, t_k)}_{\text{data}} -
    \underbrace{\left(\omega_i \, \hat{S}(\lambda_i, t_k) + \sum\limits_{\lambda_j\, \in\, \setPci} \alpha_{i,j} \, D(\lambda_j, t_k)\right)}_{\text{planet + systematics model}} \left.\left. 
    \vphantom{\sum\limits_{j \,\in\,\setPc}}
    \right|\right|^2 +
    \underbrace{
    \kappa_i \left|\left| (\omega_i, \vec{\alpha}_i) \vphantom{\sum\limits_{j \,\in\,\setPc}}
    \right|\right|^2}_{\text{Regularization}},
    \forall \lambda_i \in \setPs.
    \label{eq:minimization_cascade}
\end{multline}
where $\hat{S}$ is the model for the transit signal $S$ and $\kappa$ the regularization parameter. The parameters $\omega$ and $\alpha$ are the fitted parameters for the transit depth and the contributions of the timeseries data from the calibration subset, respectively. We estimate the regularization parameter for each wavelength using a generalized cross validation. Apart from the data-derived regressors, other linear terms can be added to Equation ~\ref{eq:minimization_cascade} such as a constant value, time $t$ or $\Delta X(t_k)$ the relative cross-dispersion position as shown in Figure~\ref{fig:cascade_tel_move}. For our final solution we added a constant term and the relative cross-dispersion position as adding these terms gave the highest signal-to-noise on the derived transit spectrum.

As we do not need to know the absolute transit depth but only the relative transit depth we use the regression estimate as defined by Equation ~\ref{eq:minimization_cascade} as follows:
\begin{equation}
\dfrac{ D(\lambda_i, t_k)}{\E[D(\lambda_i,t_k) \,|\, D(\lambda_j,t_k)] - \omega_i \, \hat{S}(\lambda_i, t_k)}
\end{equation}
to estimate the relative transit depth as a function of wavelength. The resulting combined regression model for the transit signal and systematics for the white light-curve can be seen in the top panel of Figure~\ref{fig:cascade_lcfit}. The residuals of the data minus the CASCADE regression model are shown in the lower panel of the same Figure. The derived calibrated normalized light curve and transit model is shown in the middle panel of Figure~\ref{fig:cascade_lcfit}.
The final resulting transit spectrum for WASP~117b with CASCADE is shown in Figure~\ref{fig: Final_spectrum}.

\section{Corner plots for retrieval of models used for atmosphere analysis of WASP-117b}

In the following, the full corner plots as used in Sections~\ref{sec: Significance} and \ref{sec: retrieval} are shown for the fitted parameters and for each model used to constrain the atmospheric properties of WASP-117b during transit based on HST/WFC3 data.

 \begin{figure*}[h!]
   \centering
   \textbf{Full retrieval with \ce{H2O} - Nominal spectrum}\par\medskip
   \includegraphics[width=0.95\textwidth]{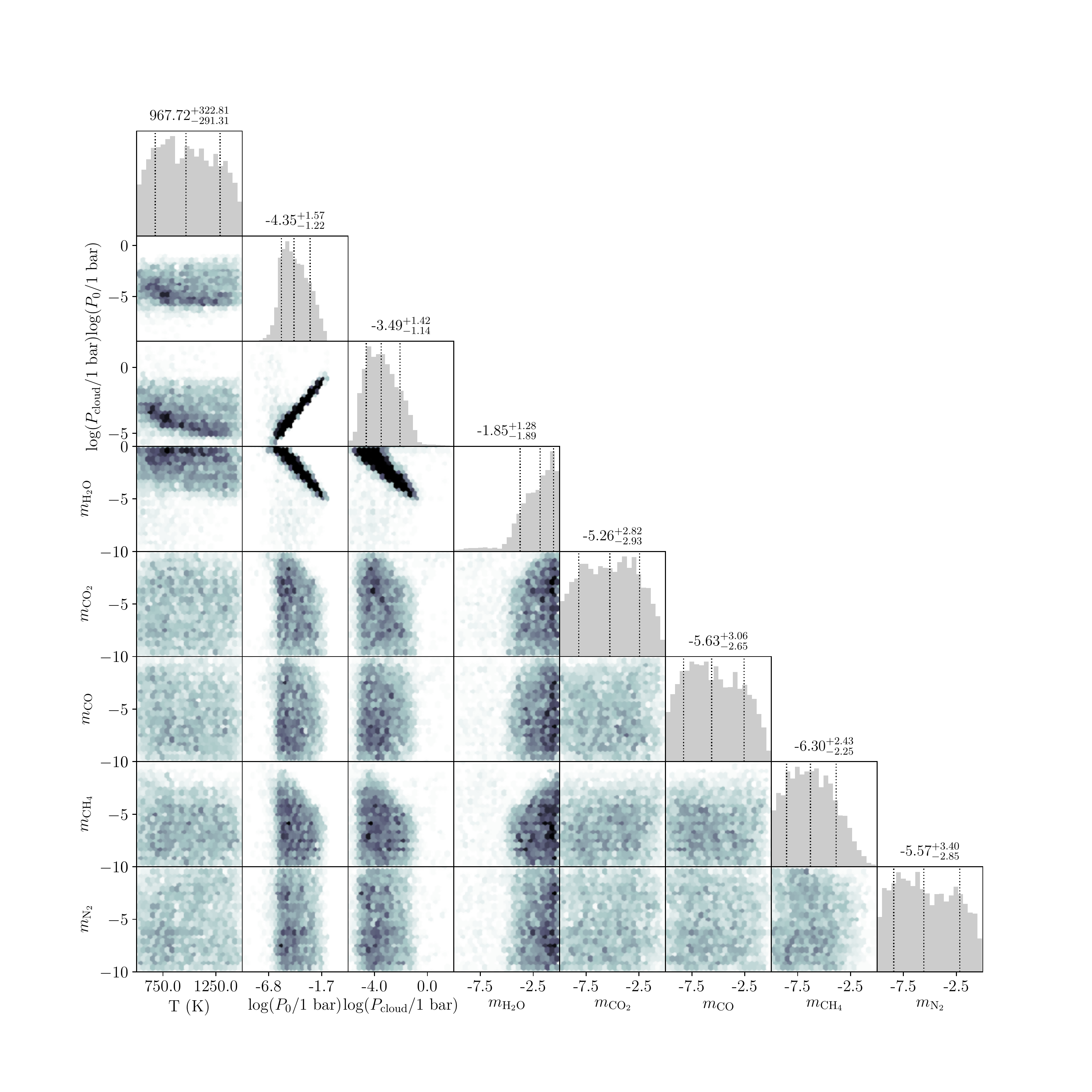}
      \caption{Posterior  probability  distributions of the retrieval on HST/WFC3 dataset for WASP-117b, reduced with the nominal pipeline. Here, the full retrieval including \ce{H2O} was applied (see Section~\ref{sec: Significance}).  Median  parameter  values and  68\%  confidence intervals for the marginalized 1D posterior probability distributions are indicated with horizontal error bars.
              }
         \label{fig: Corner_plot_Full_H2O_nom}
   \end{figure*}
   
   \begin{figure*}[h!]
   \centering
   \textbf{Retrieval without \ce{H2O} - Nominal spectrum }\par\medskip
   \includegraphics[width=0.95\textwidth]{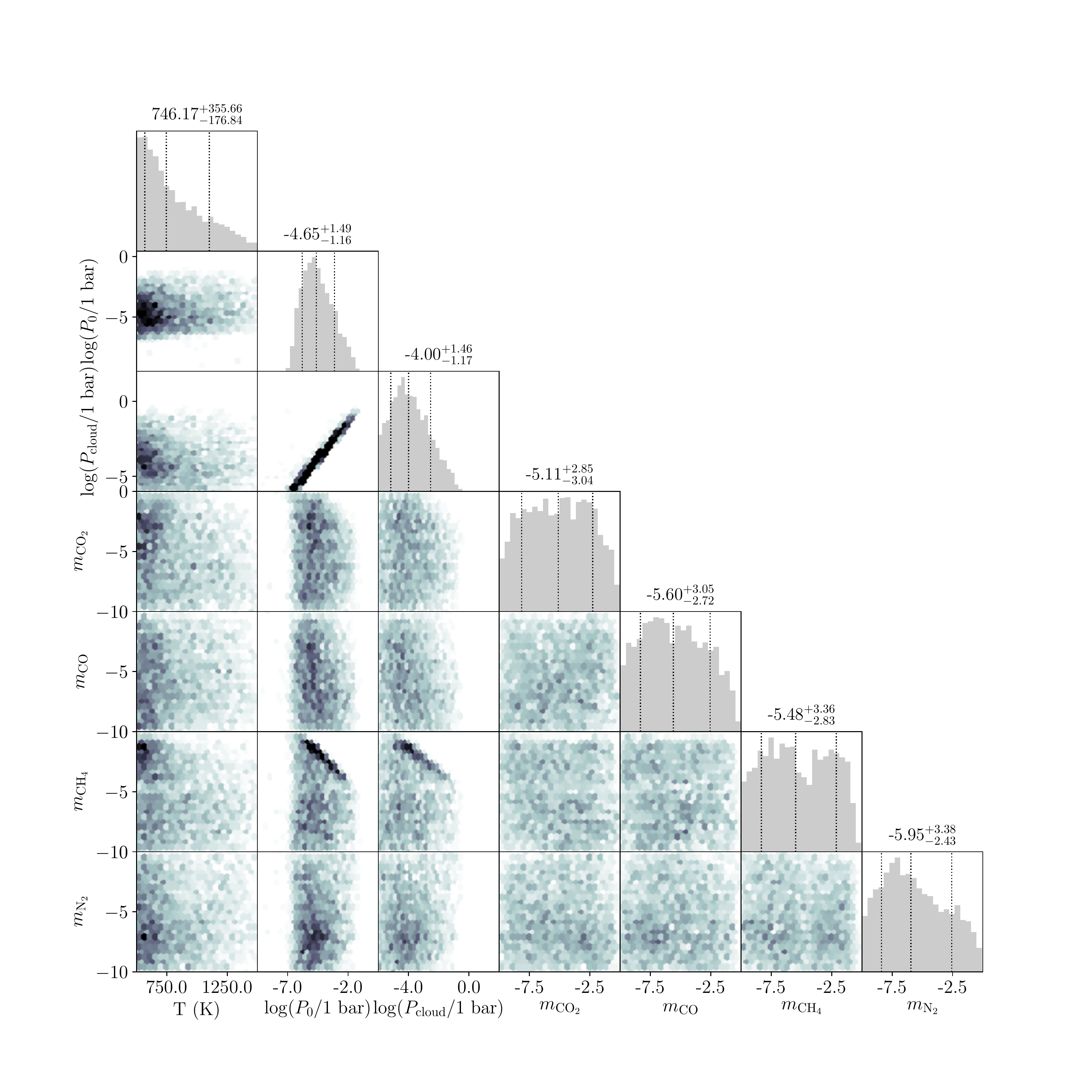}
      \caption{Posterior  probability  distributions of the retrieval on HST/WFC3 dataset for WASP-117b, reduced with the nominal pipeline. Here, the retrieval without \ce{H2O} was applied (see Section~\ref{sec: Significance}).  Median  parameter  values and  68\%  confidence intervals for the marginalized 1D posterior probability distributions are indicated with horizontal error bars.
              }
         \label{fig: Corner_plot_no_H2O_nom}
   \end{figure*}
   
   \begin{figure*}[h!]
   \centering
   \textbf{Full retrieval with \ce{H2O} - CASCADE spectrum}\par\medskip
   \includegraphics[width=0.95\textwidth]{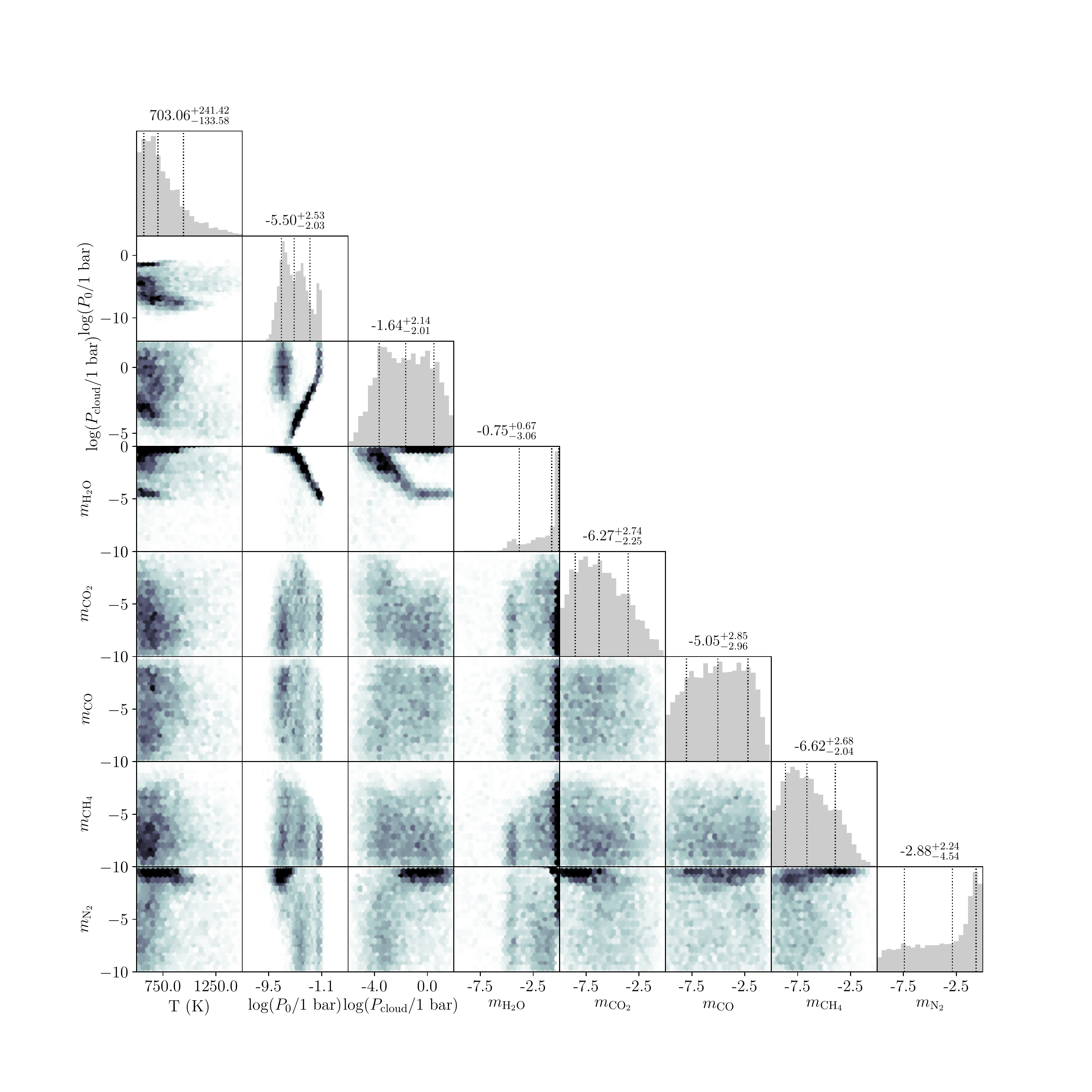}
      \caption{Posterior  probability  distributions of the retrieval on HST/WFC3 dataset for WASP-117b, reduced with the CASCADE pipeline. Here, the full retrieval including \ce{H2O} was applied (see Section~\ref{sec: Significance}).  Median  parameter  values and  68\%  confidence intervals for the marginalized 1D posterior probability distributions are indicated with horizontal error bars.
              }
         \label{fig: Corner_plot_H2O_CASCADE}
   \end{figure*}
   
   \begin{figure*}[h!]
   \centering
   \textbf{Retrieval without \ce{H2O} - CASCADE spectrum}\par\medskip
   \includegraphics[width=0.95\textwidth]{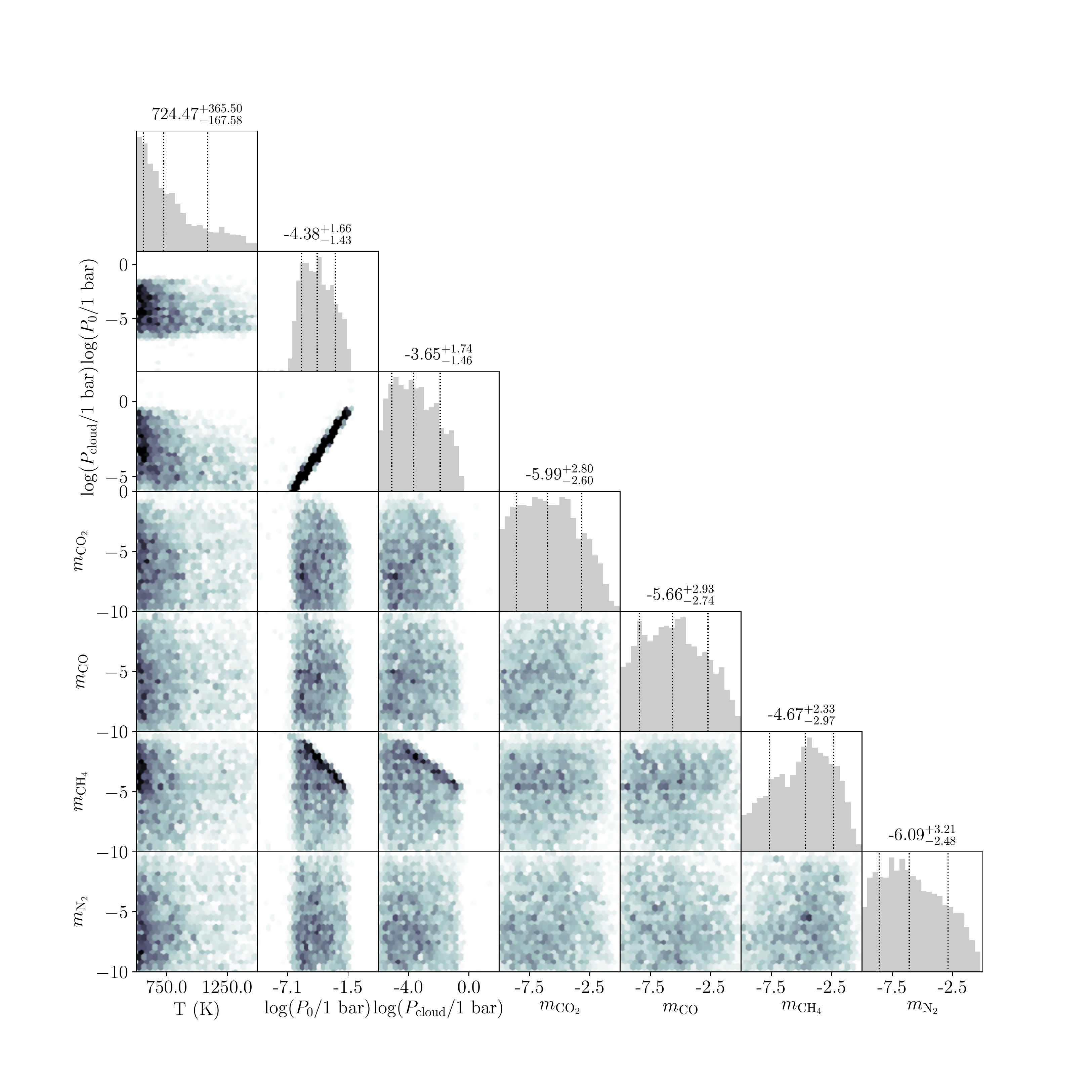}
      \caption{Posterior  probability  distributions of the retrieval on HST/WFC3 dataset for WASP-117b, reduced with the nominal pipeline. Here, the retrieval without \ce{H2O} was applied (see Section~\ref{sec: Significance}).  Median  parameter  values and  68\%  confidence intervals for the marginalized 1D posterior probability distributions are indicated with horizontal error bars.
              }
         \label{fig: Corner_plot_no_H2O_CASCADE}
   \end{figure*}

 \begin{figure*}[h!]
   \centering
   \textbf{Full retrieval with Na/K - Nominal spectrum + VLT/ESPRESSO}\par\medskip
   \includegraphics[width=0.95\textwidth]{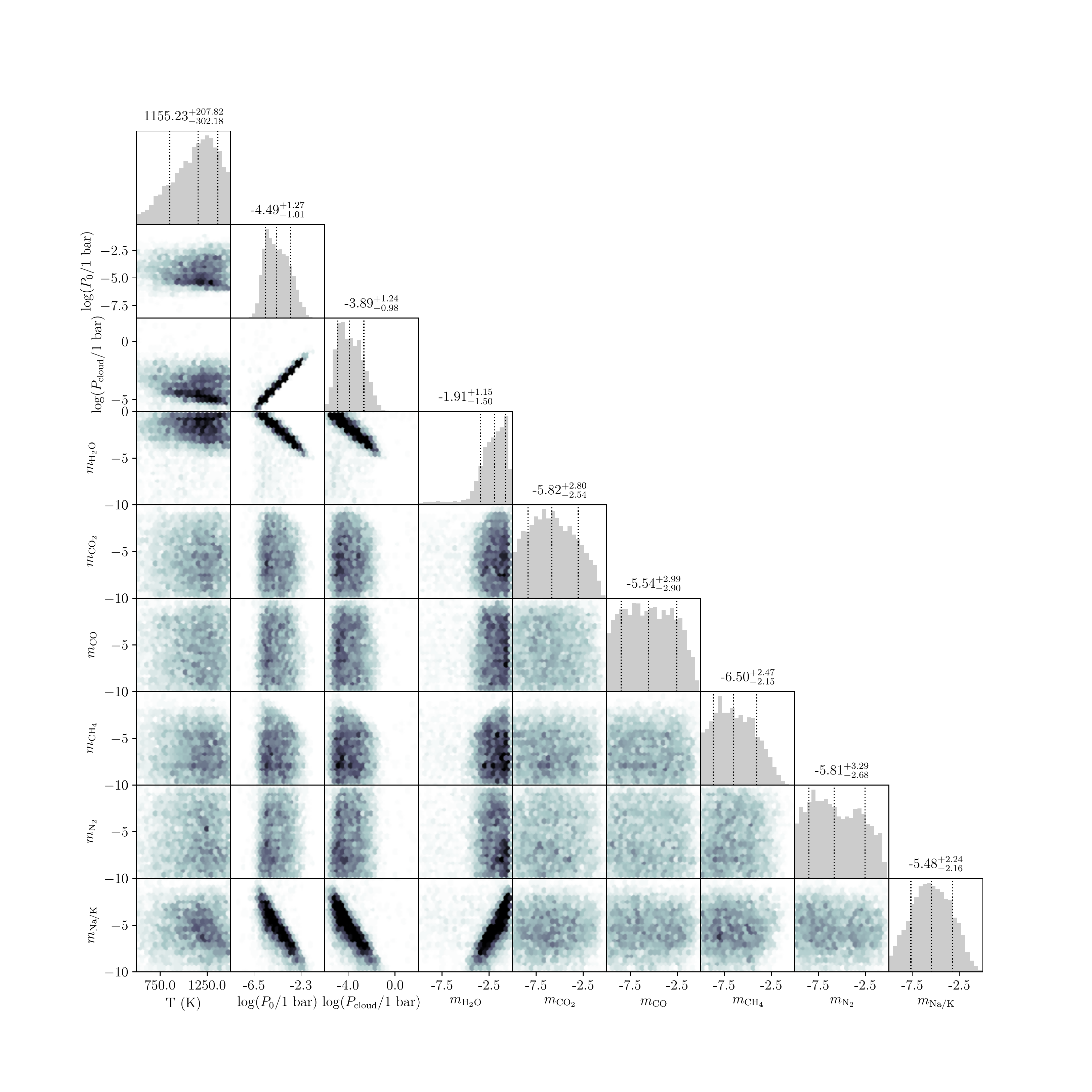}
      \caption{Posterior  probability  distributions of the joint analysis of  the HST and VLT ESPRESSO dataset for WASP-117b, reduced with the nominal pipeline. Here, the full retrieval with Na and K was applied (see Section~\ref{sec: Significance}).  Median  parameter  values and  68\%  confidence intervals for the marginalized 1D posterior probability distributions are indicated with horizontal error bars.
              }
         \label{fig: Corner_plot_VLT_nom}
   \end{figure*}
   
   \begin{figure*}[h!]
   \centering
   \textbf{Retrieval without Na/K - Nominal spectrum + VLT/ESPRESSO}\par\medskip
   \includegraphics[width=0.95\textwidth]{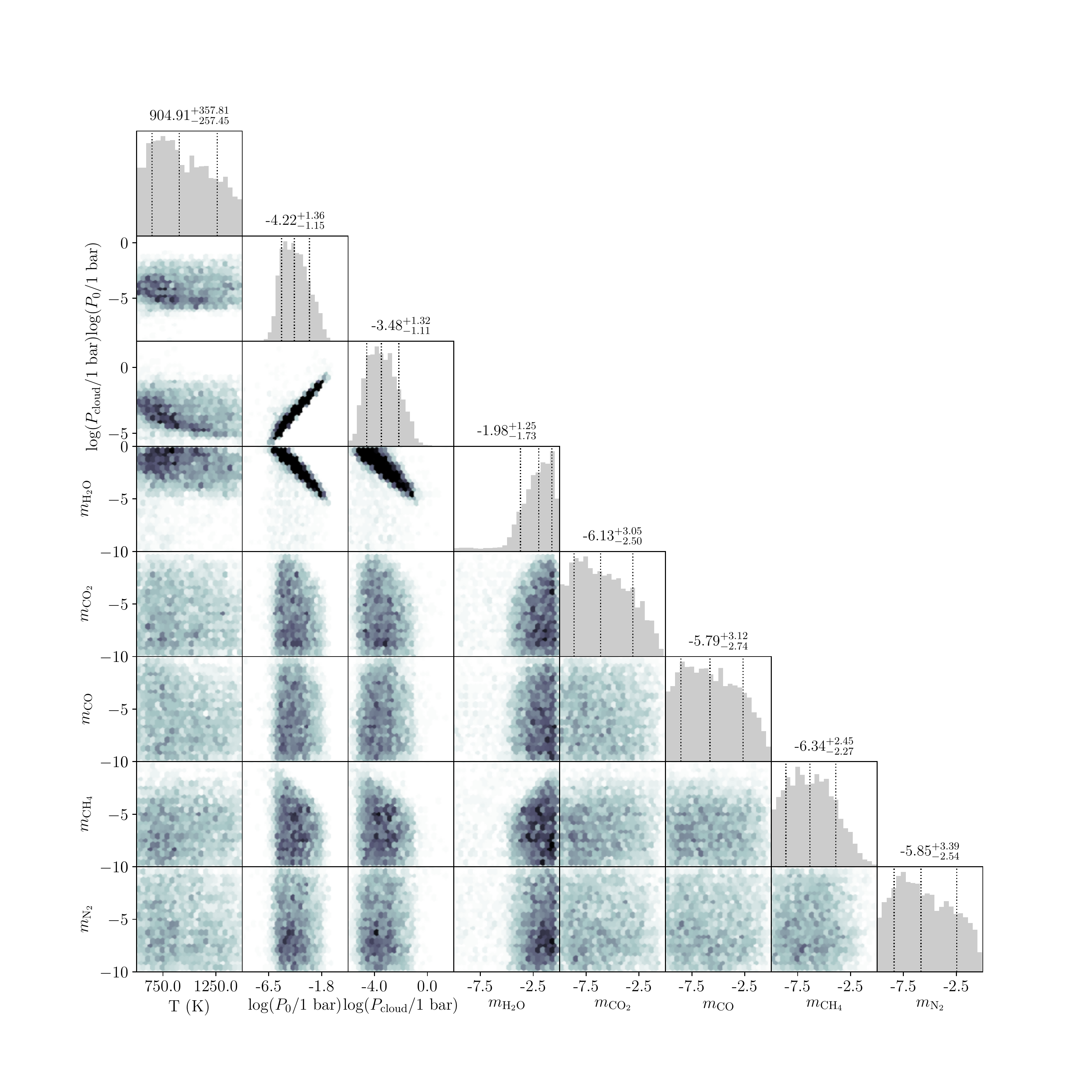}
      \caption{Posterior  probability  distributions of the joint analysis of  the HST and VLT ESPRESSO dataset for WASP-117b, reduced with the nominal pipeline. Here, the retrieval without Na and K was applied (see Section~\ref{sec: Significance}).  Median  parameter  values and  68\%  confidence intervals for the marginalized 1D posterior probability distributions are indicated with horizontal error bars.
              }
         \label{fig: Corner_plot_VLT_no_Na_K}
   \end{figure*}
   
   \begin{figure*}[h!]
   \centering
   \textbf{Full retrieval with Na/K - CASCADE spectrum + VLT/ESPRESSO}\par\medskip
   \includegraphics[width=0.95\textwidth]{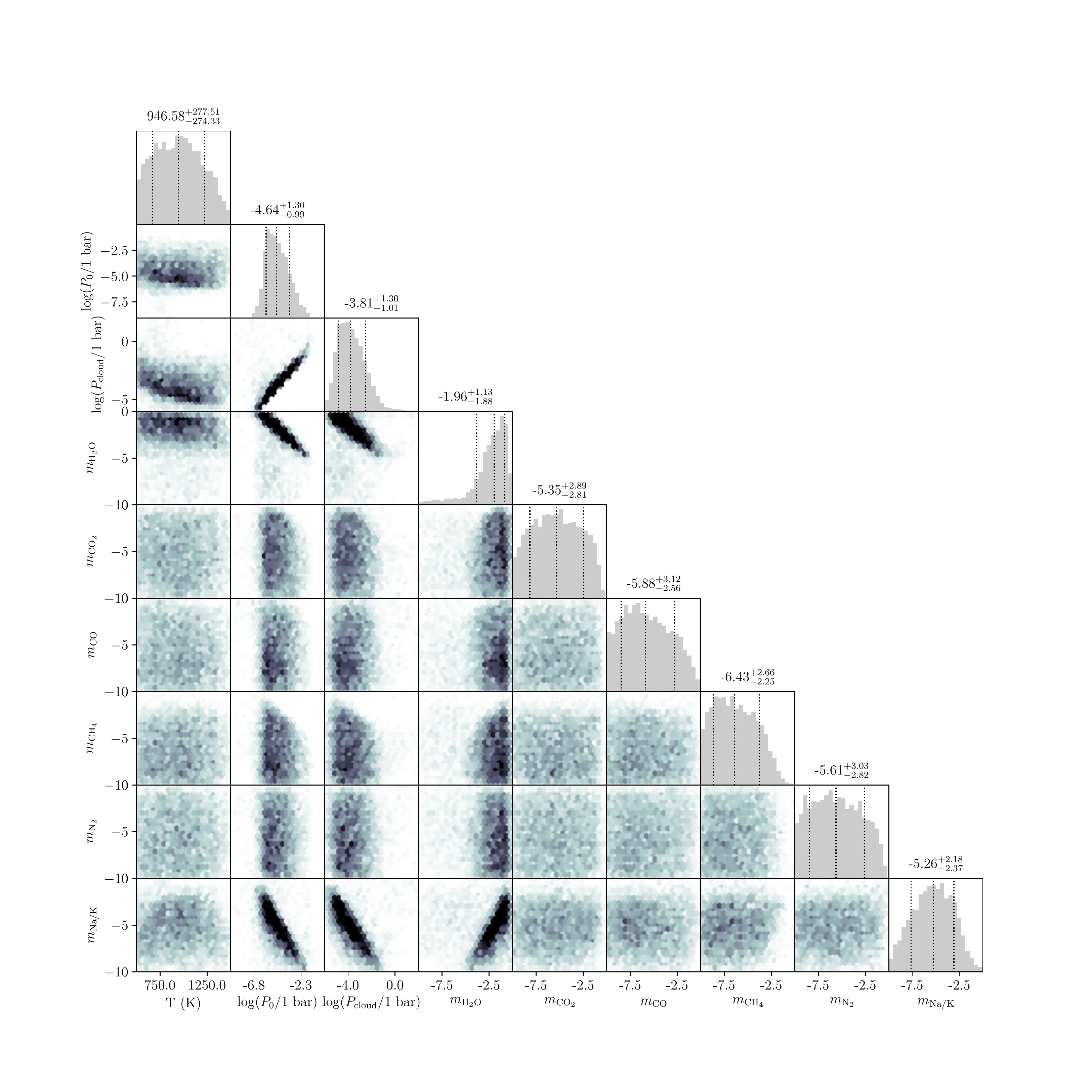}
      \caption{Posterior  probability  distributions of the joint analysis of  the HST and VLT ESPRESSO dataset for WASP-117b, reduced with the CASCADE pipeline. Here, the full retrieval with Na and K was applied (see Section~\ref{sec: Significance}).  Median  parameter  values and  68\%  confidence intervals for the marginalized 1D posterior probability distributions are indicated with horizontal error bars.
              }
         \label{fig: Corner_plot_VLT_CASCADE}
   \end{figure*}
   
   \begin{figure*}[h!]
   \centering
   \textbf{Retrieval without Na/K - CASCADE spectrum + VLT/ESPRESSO}\par\medskip
   \includegraphics[width=0.95\textwidth]{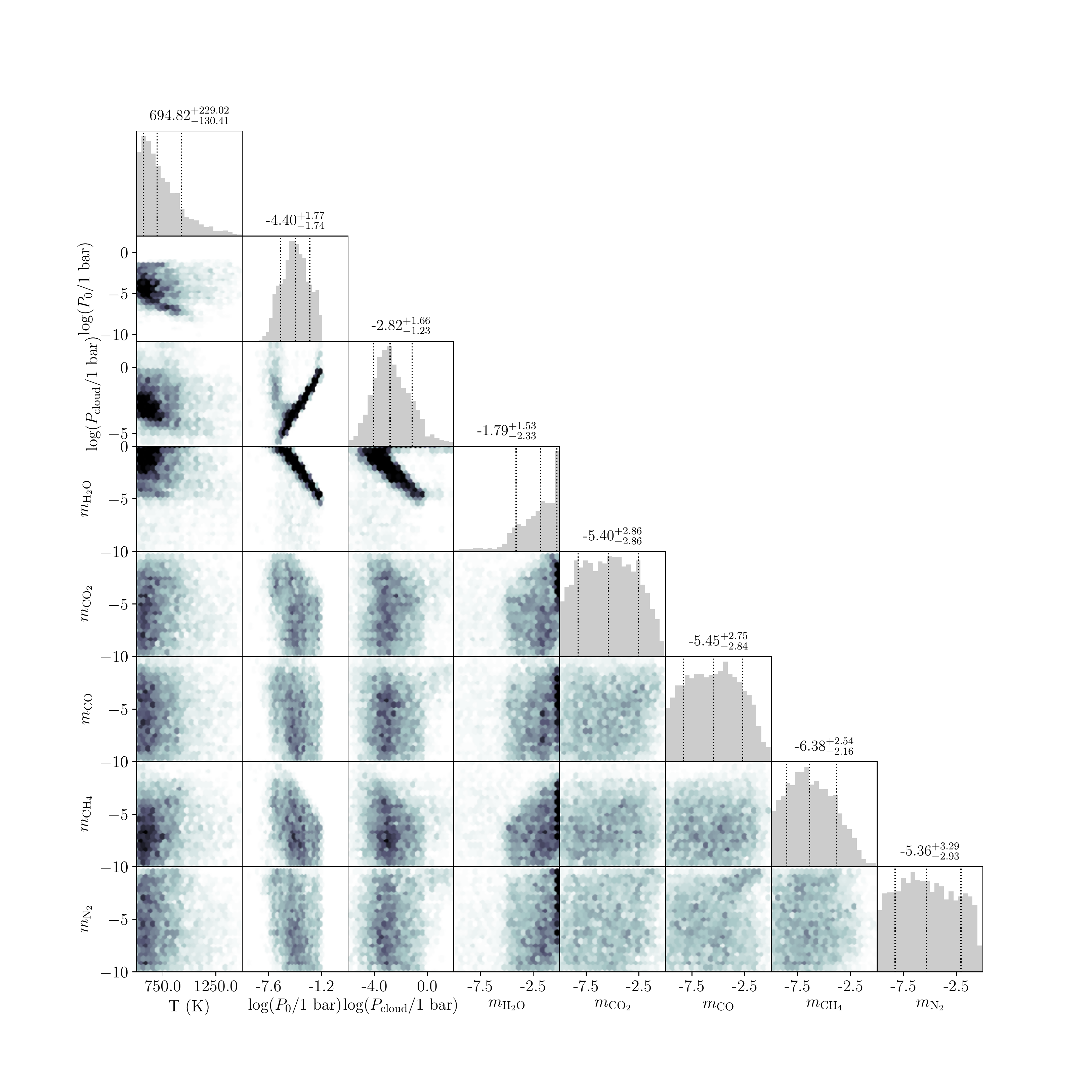}
      \caption{Posterior  probability  distributions of the joint analysis of  the HST and VLT ESPRESSO dataset for WASP-117b, reduced with the CASCADE pipeline. Here, the retrieval without Na and K was applied (see Section~\ref{sec: Significance}).  Median  parameter  values and  68\%  confidence intervals for the marginalized 1D posterior probability distributions are indicated with horizontal error bars.
              }
         \label{fig: Corner_plot_VLT_no_Na_K_CASCADE}
   \end{figure*}

 \begin{figure*}[ht!]
   \centering
   \textbf{\textit{Model 1} - Nominal spectrum}\par\medskip
   \includegraphics[width=0.95\textwidth]{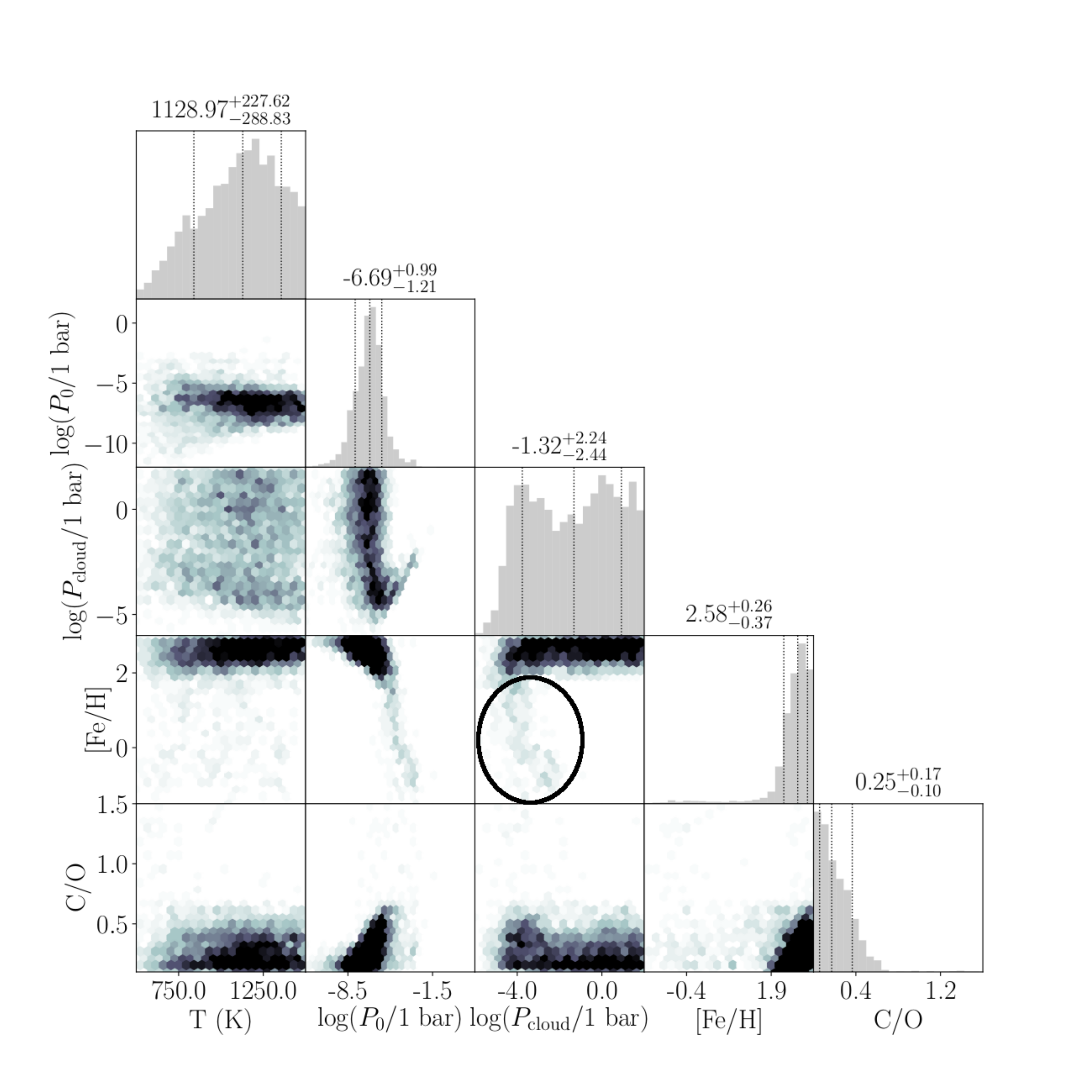}
      \caption{Posterior  probability  distributions  for  the HST dataset, reduced with the nominal pipeline. Here, \textit{Model 1} was applied with unconstrained $\rm{[Fe/H]}$ and condensate clouds. Median  parameter values and  68\%  confidence intervals for the marginalized 1D posterior probability distributions are indicated with horizontal error bars. The black ellipse highlights the subset of cloudy low metallicity solutions, which are explored with \textit{Model 2} and \textit{Model 3} further.
              }
         \label{fig: Corner_plot_Zhou_m1}
   \end{figure*}

\begin{figure*}
   \centering
   \textbf{\textit{Model 1} - CASCADE spectrum}\par\medskip
   \includegraphics[width=0.98\textwidth]{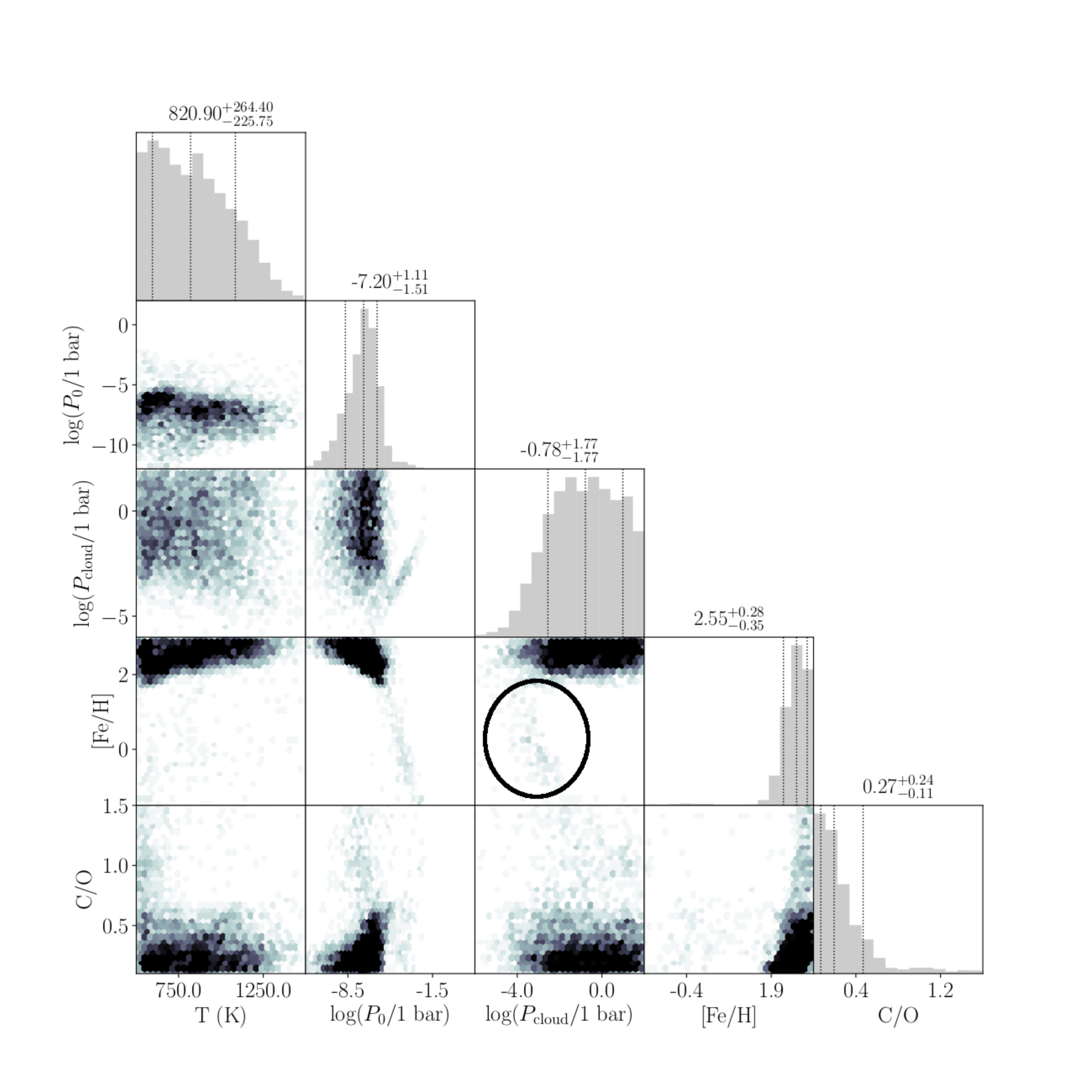}
      \caption{Posterior probability  distributions for the HST data set, reduced with the CASCADE pipeline. Here, \textit{Model 1} was applied with unconstrained $\rm{[Fe/H]}$ and condensate clouds. Median parameter values  and  68\%  confidence intervals for the marginalized 1D posterior probability distributions are indicated with horizontal error bars. The black ellipse highlights the subset of cloudy low metallicity solutions, which are explored with \textit{Model 2} and \textit{Model 3} further.
              }
         \label{fig: Corner_plot_Bouwman_m1}
   \end{figure*}
   
   \begin{figure*}[h!]
   \centering
   \textbf{\textit{Model 2} - Nominal spectrum}\par\medskip
   \includegraphics[width=0.95\textwidth]{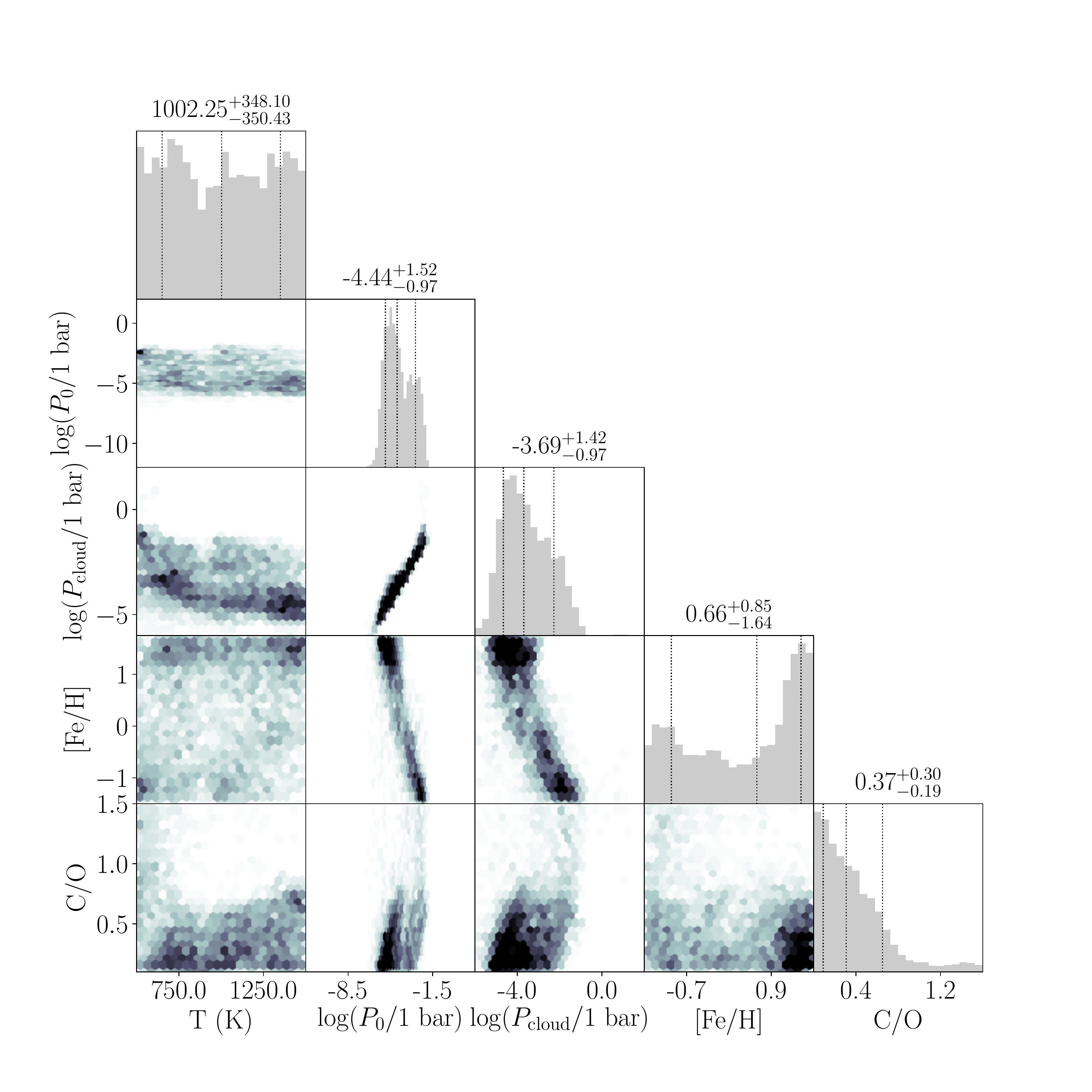}
      \caption{Posterior  probability  distributions for  the HST dataset, reduced with the nominal pipeline. Here, \textit{Model 2} was applied with prior ${\rm [Fe/H]}<1.75$ and condensate clouds.  Median  parameter  values and  68\%  confidence intervals for the marginalized 1D posterior probability distributions are indicated with horizontal error bars.
              }
         \label{fig: Corner_plot_Zhou_m2}
   \end{figure*}

\begin{figure*}
   \centering
   \textbf{\textit{Model 2} - CASCADE spectrum}\par\medskip
   \includegraphics[width=0.98\textwidth]{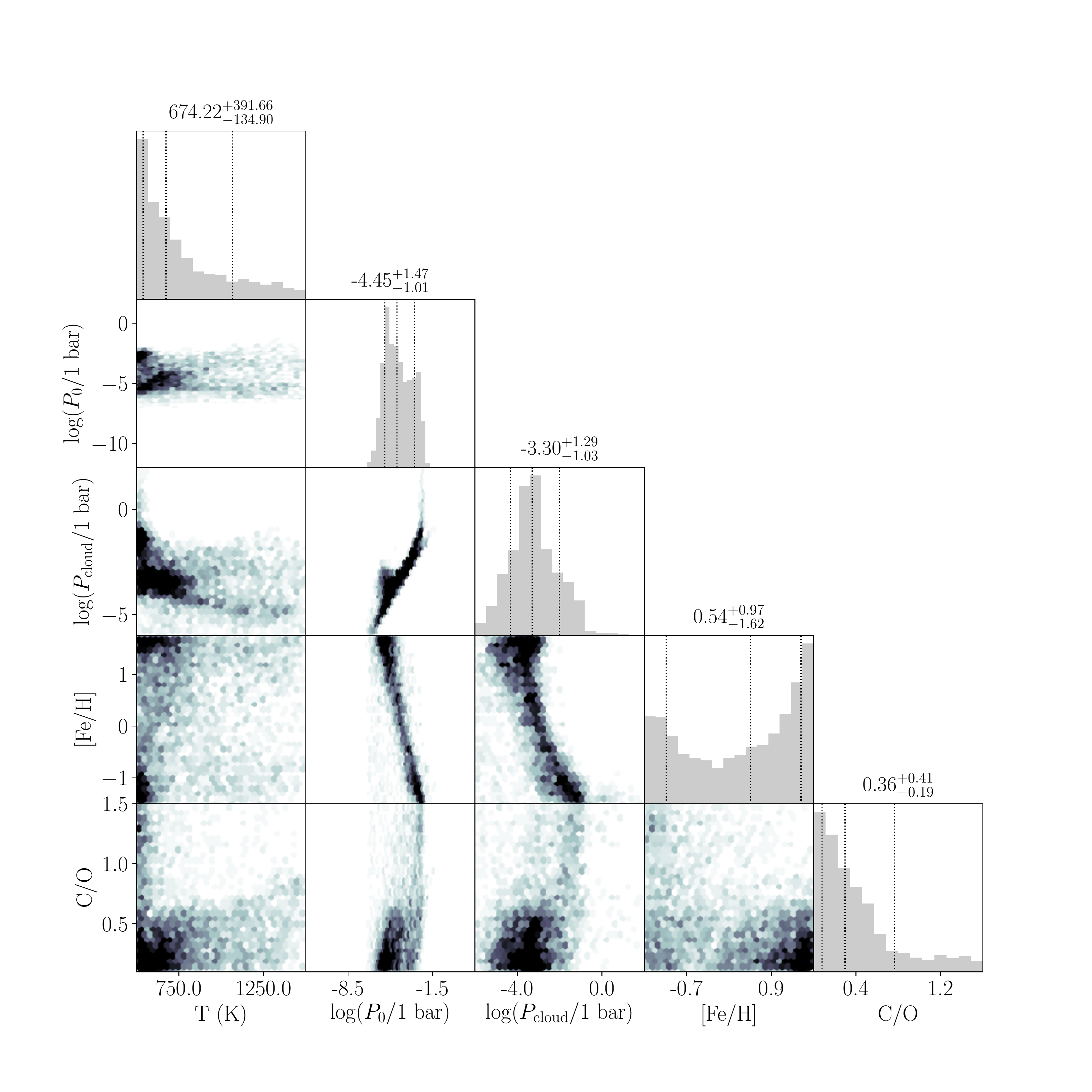}
      \caption{Posterior probability  distributions for the HST dataset, reduced with the CASCADE pipeline. Here, \textit{Model 2} was applied with prior ${\rm [Fe/H]}<1.75$ and condensate clouds. Median parameter values  and  68\%  confidence intervals for the marginalized 1D posterior probability distributions are indicated with horizontal error bars.
              }
         \label{fig: Corner_plot_Bouwman_m2}
   \end{figure*}

  \begin{figure*}[h!]
   \centering
   \textbf{\textit{Model 3} - Nominal spectrum}\par\medskip
   \includegraphics[width=0.95\textwidth]{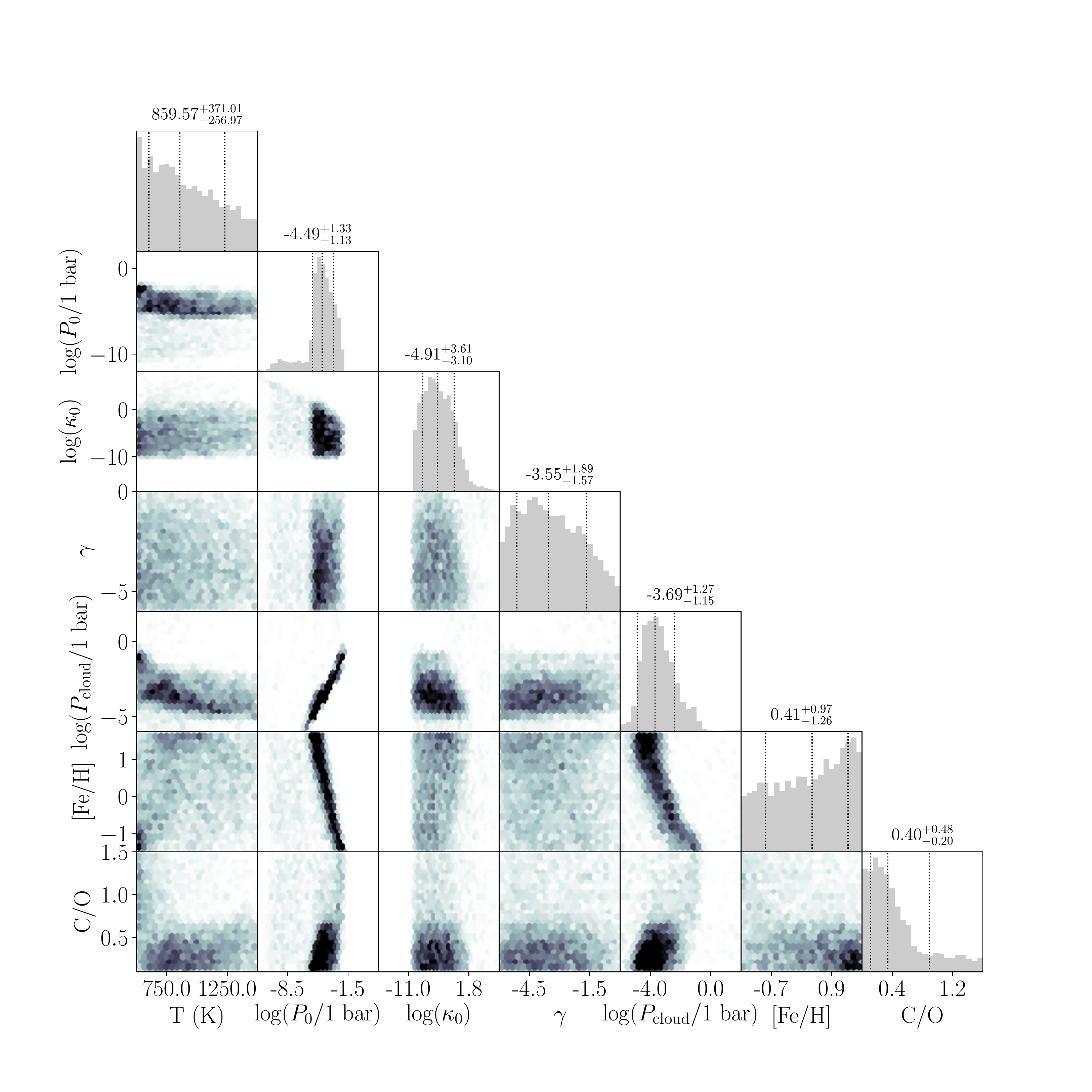}
      \caption{Posterior  probability  distributions for  the HST dataset, reduced with the nominal pipeline. Here, \textit{Model 3} was applied with prior ${\rm [Fe/H]}<1.75$, condensate clouds a haze layer with a prior of $\gamma >-6$ on the scattering slope.  Median  parameter  values and  68\%  confidence intervals for the marginalized 1D posterior probability distributions are indicated with horizontal error bars.
              }
         \label{fig: Corner_plot_Zhou_m3}
   \end{figure*}

\begin{figure*}
   \centering
   \textbf{\textit{Model 3} - CASCADE spectrum}\par\medskip
   \includegraphics[width=0.98\textwidth]{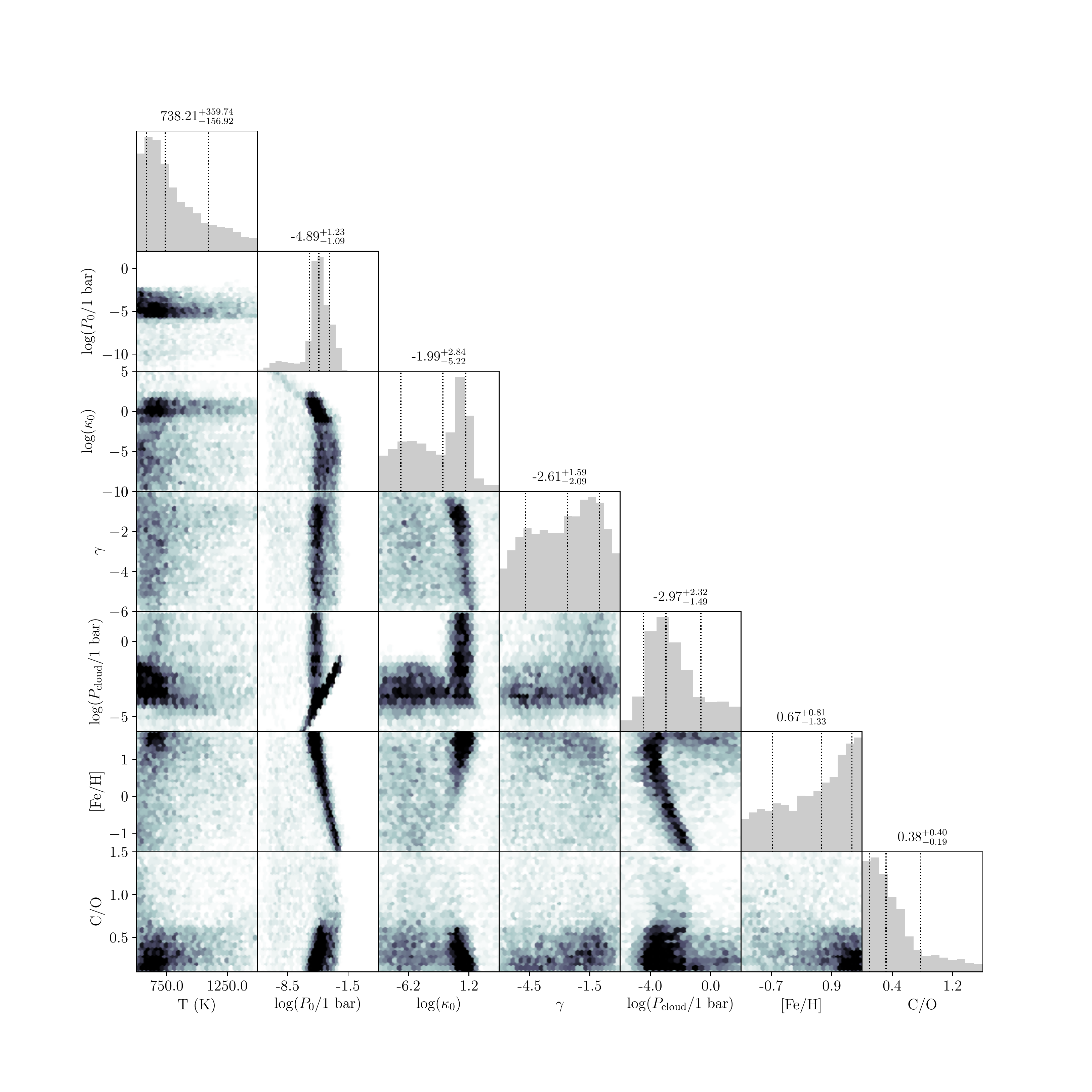}
      \caption{Posterior probability distributions of atmosphere properties of WASP-117b derived from WASP-117b HST observations. Raw data was reduced with the CASCADE pipeline. Here, \textit{Model 3} was applied with prior ${\rm [Fe/H]}<1.75$, condensate clouds a haze layer with a prior of $\gamma >-6$ on the scattering slope. Median parameter values  and  68\%  confidence intervals for the marginalized 1D posterior probability distributions are indicated with horizontal error bars.
              }
         \label{fig: Corner_plot_Bouwman_m3}
   \end{figure*}

   \begin{figure*}
   \centering
   \textbf{H$_2$O \& CH$_4$: \textit{Model 1}}\par
   \textbf{CASCADE} \hspace{7.5 cm} \textbf{Nominal}\par\medskip
   \includegraphics[width=0.45\textwidth]{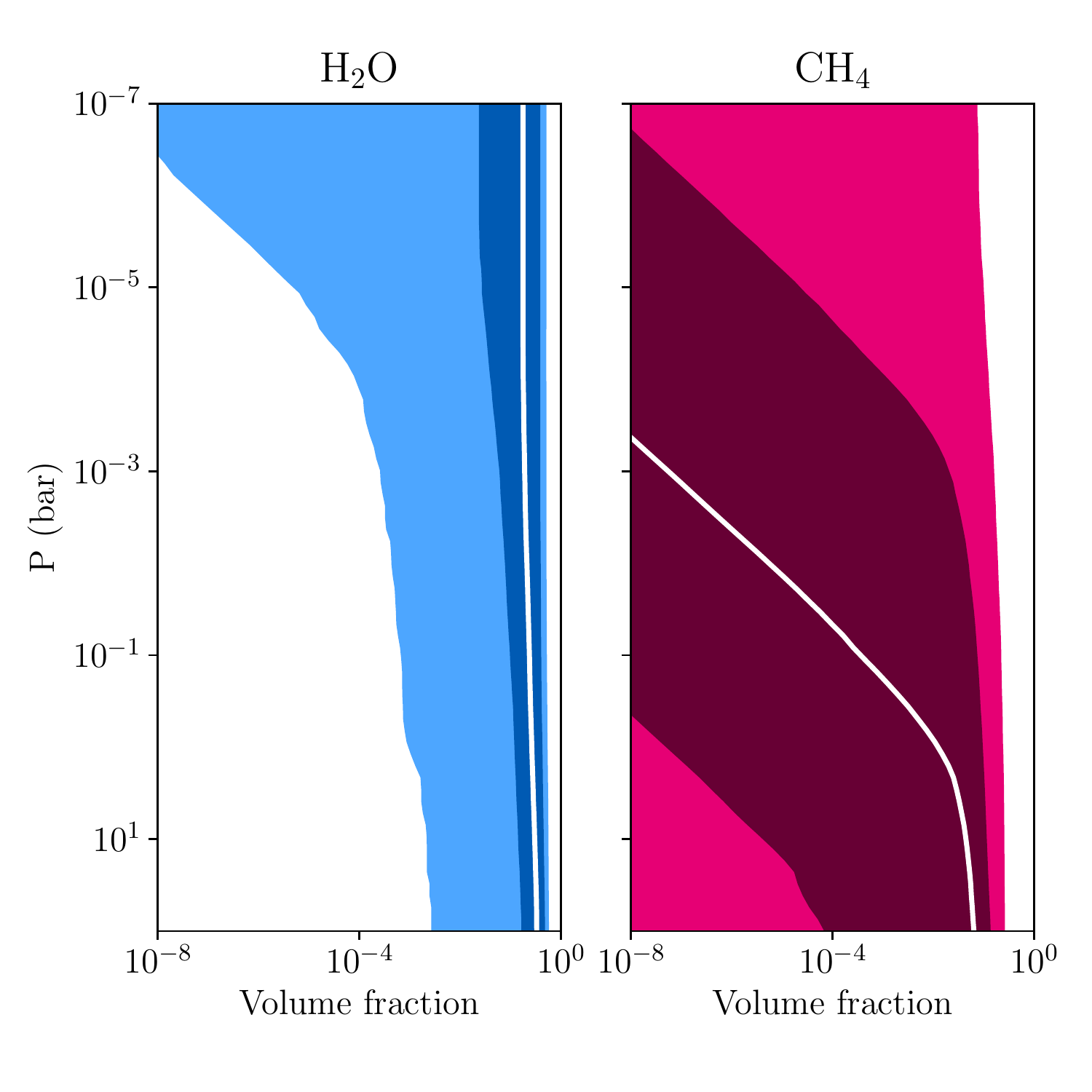}
   \includegraphics[width=0.45\textwidth]{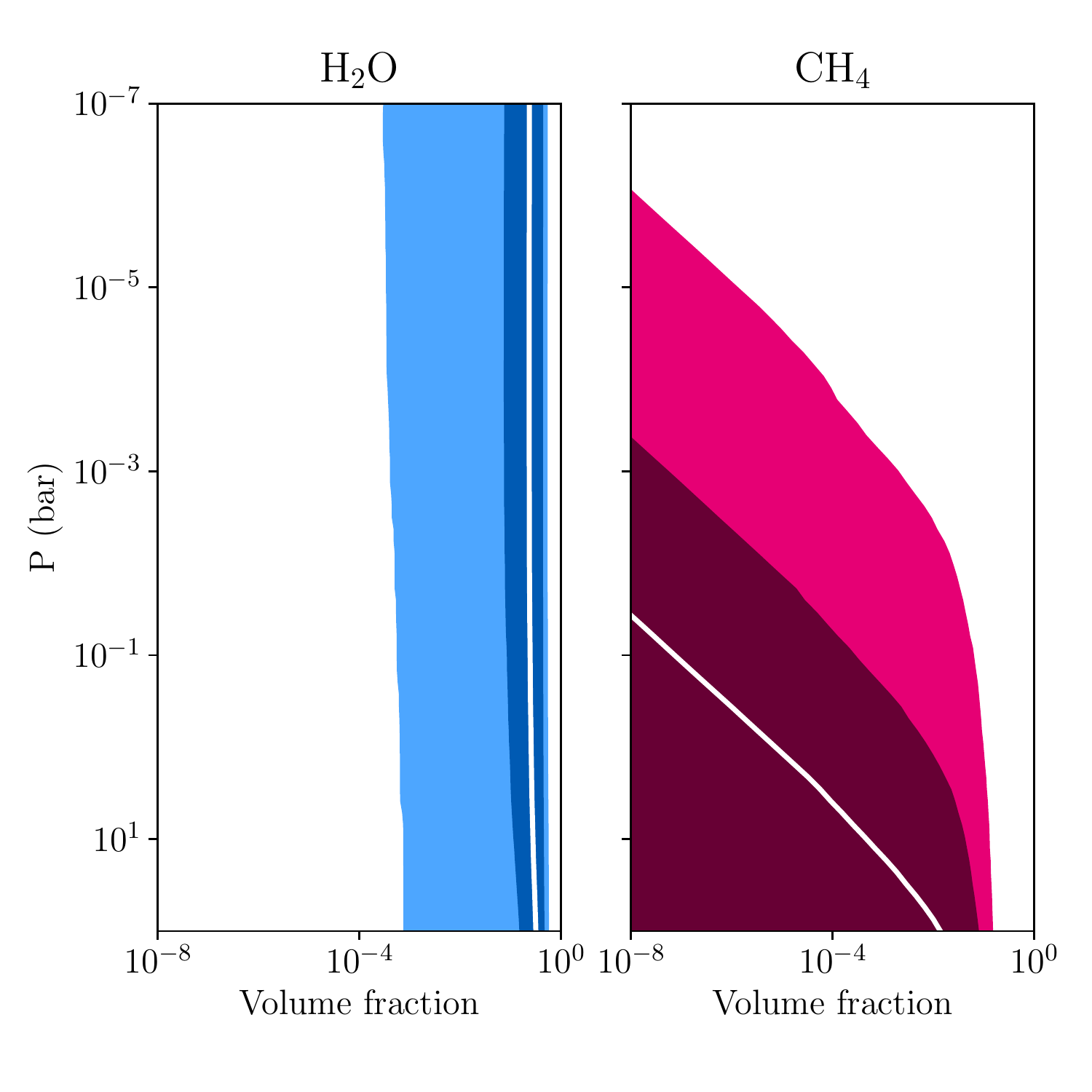}
      \caption{Posterior probability distributions of H$_2$O and CH$_4$ content in the atmosphere of WASP-117b based on HST observations, reduced with the CASCADE (left) and nominal pipeline (right). Here, \textit{Model 1} was applied, with condensate clouds. Median parameter values  and  99.7\%  (light blue and red) and 68\% (dark blue and red) confidence intervals are given .
              }
         \label{fig: Corner_plot_H2O_CH4_m1}
   \end{figure*}

      \begin{figure*}
   \centering
   \textbf{H$_2$O \& CH$_4$: \textit{Model 2}}\par
   \textbf{CASCADE} \hspace{7.5 cm} \textbf{Nominal}\par\medskip
   \includegraphics[width=0.45\textwidth]{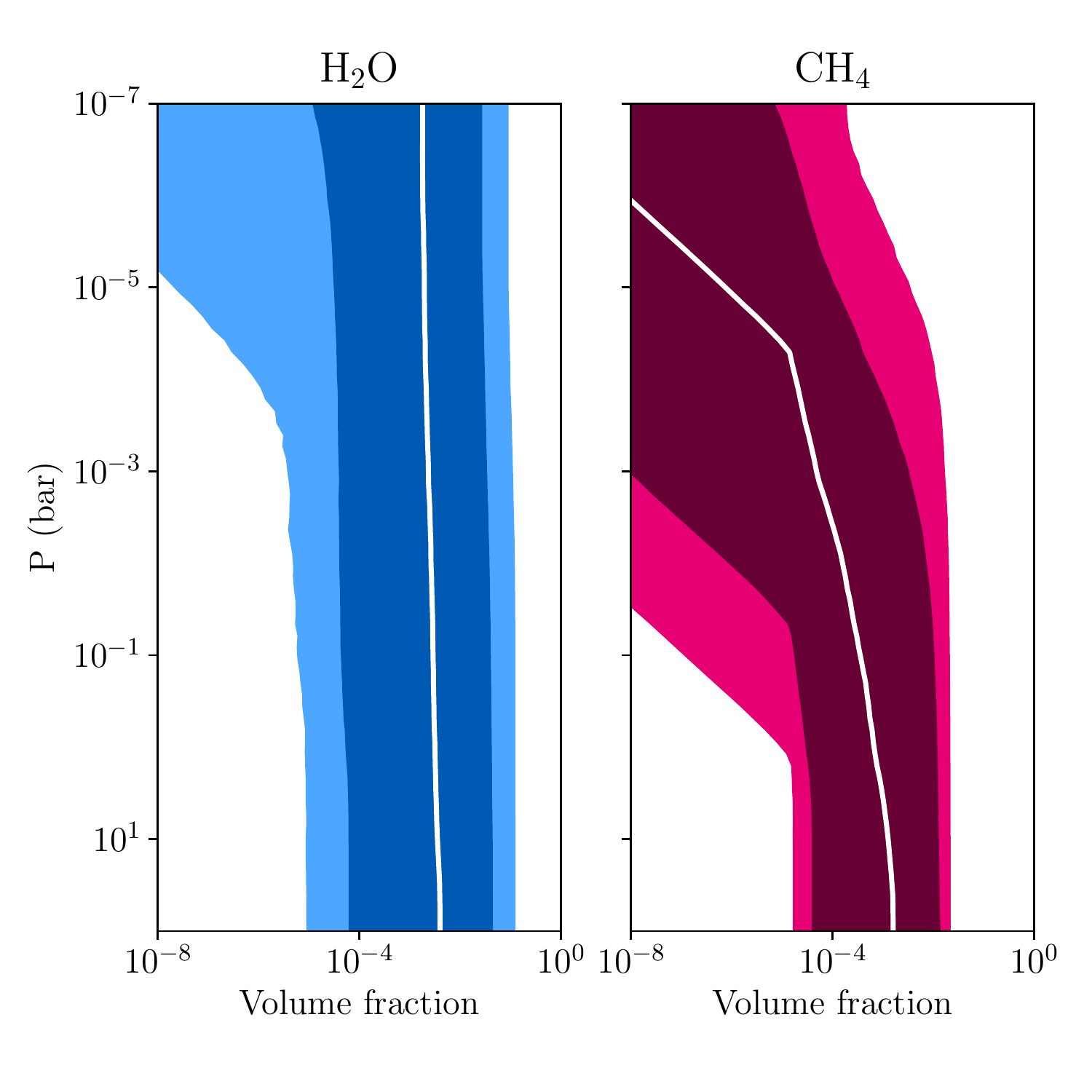}
   \includegraphics[width=0.45\textwidth]{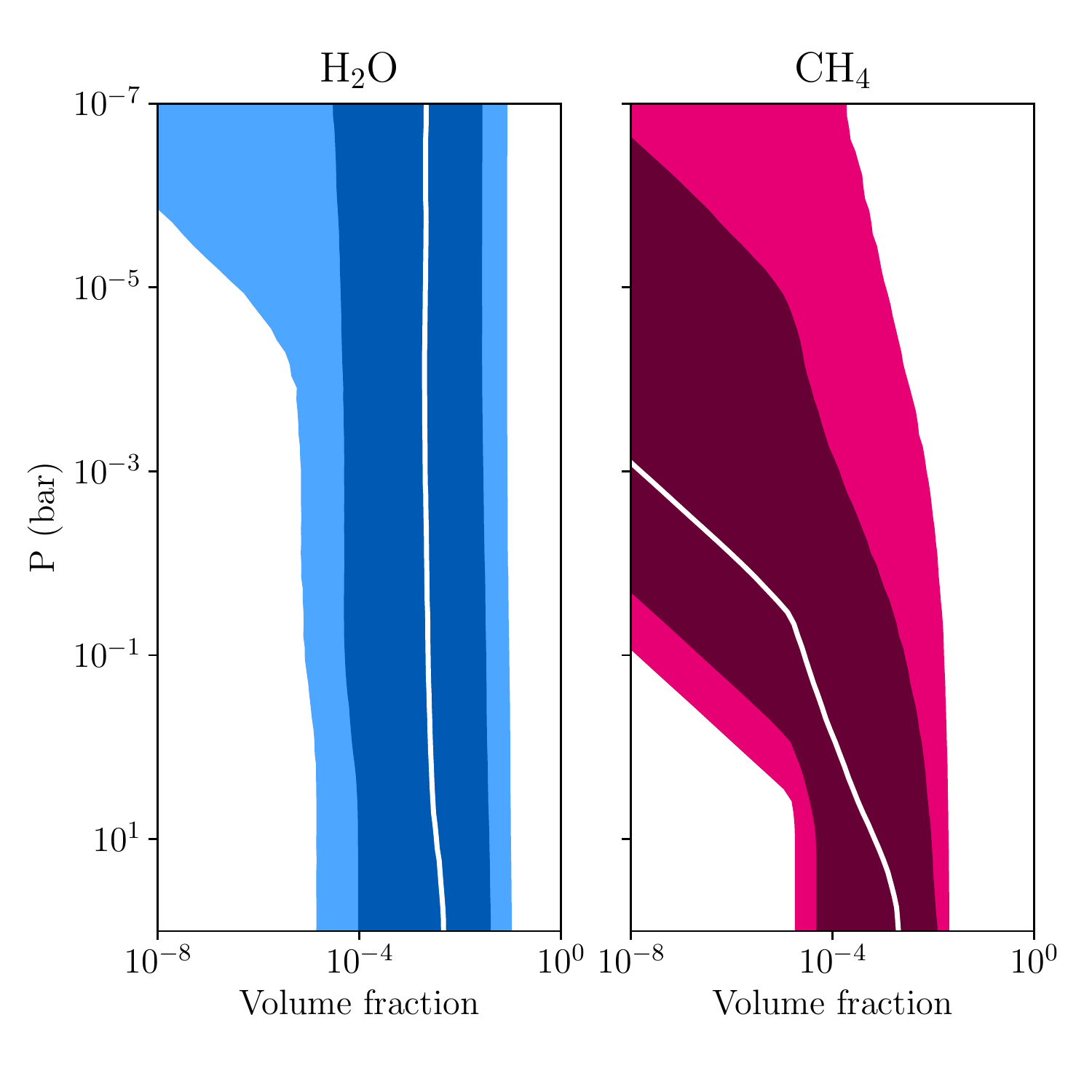}
      \caption{Posterior probability  distributions of H$_2$O and CH$_4$ content in the atmosphere of WASP-117b based on HST observations, reduced with the CASCADE (left) and nominal pipeline (right). Here, \textit{Model 2} was applied, with condensate clouds and ${\rm [Fe/H]}<1.75$. Median parameter values  and  99.7\%  (light blue and red) and 68\% (dark blue and red) confidence intervals are given .
              }
         \label{fig: Corner_plot_H2O_CH4_m2}
   \end{figure*}

      \begin{figure*}
   \centering
   \textbf{H$_2$O \& CH$_4$: \textit{Model 3}}\par
   \textbf{CASCADE} \hspace{7.5 cm} \textbf{Nominal}\par\medskip
   \includegraphics[width=0.45\textwidth]{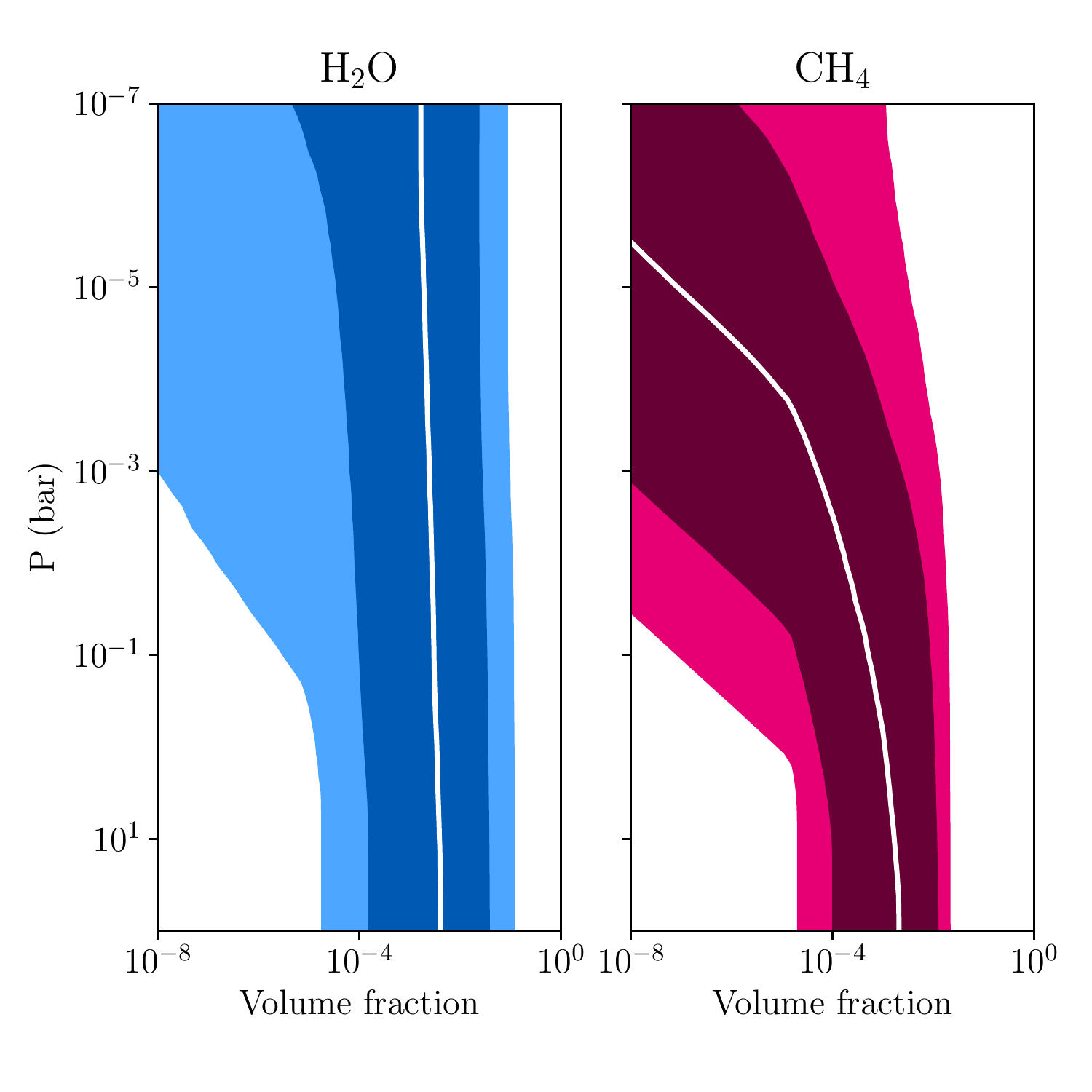}
   \includegraphics[width=0.45\textwidth]{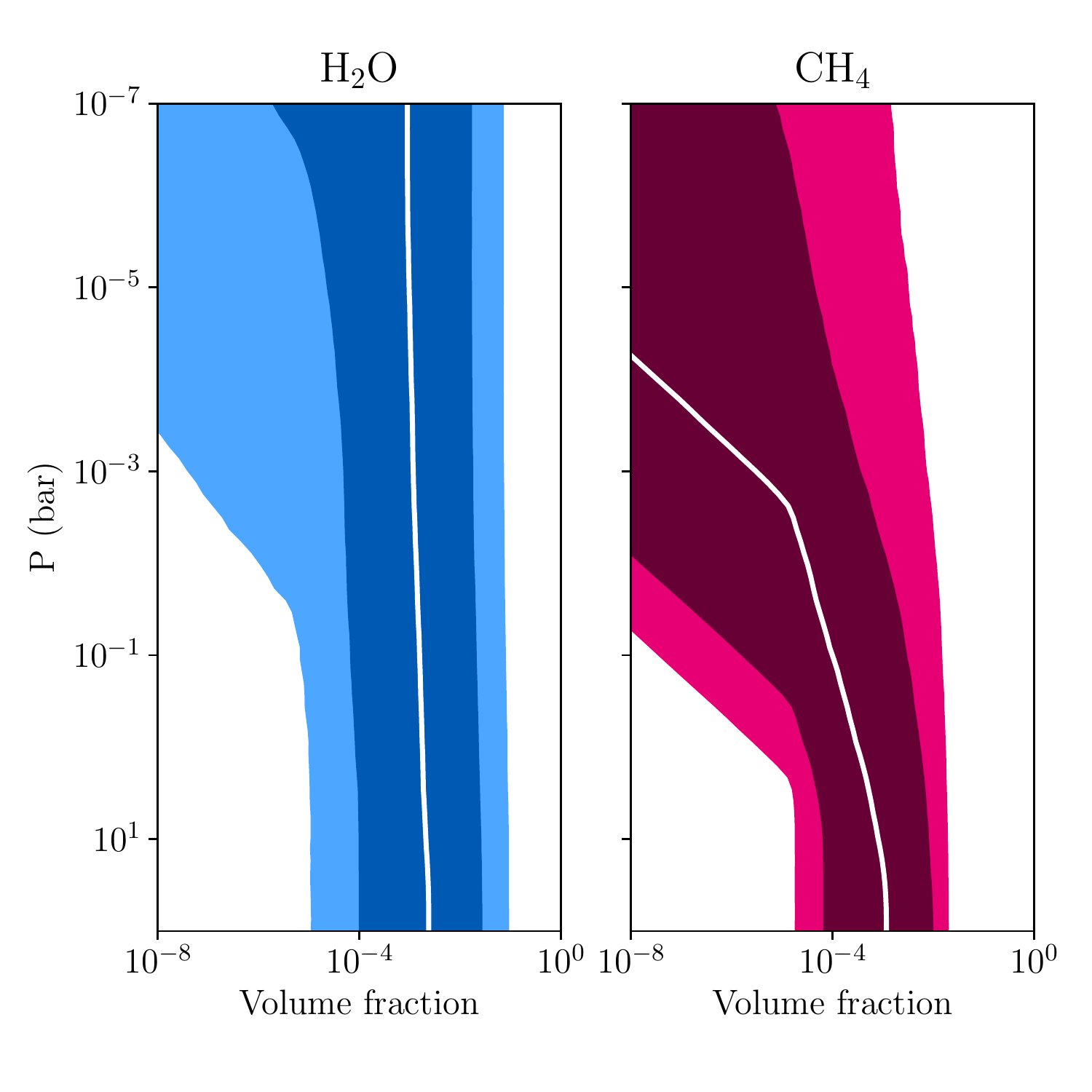}
      \caption{Posterior probability  distributions of H$_2$O and CH$_4$ content in the atmosphere of WASP-117b based on HST observations, reduced with the CASCADE (left) and nominal pipeline (right). Here, \textit{Model 3} was applied, with condensate clouds and ${\rm [Fe/H]}<1.75$ and haze layer with scattering slope $\gamma <3$. Median parameter values  and  99.7\%  (light blue and red) and 68\% (dark blue and red) confidence intervals are given .
              }
         \label{fig: Corner_plot_H2O_CH4_m3}
   \end{figure*}

\clearpage   
\section{Effect of vertical mixing on the methane abundance}   
\label{sec: Vertical mixing}

\begin{figure*}
    \centering
    \includegraphics[width=0.7\textwidth]{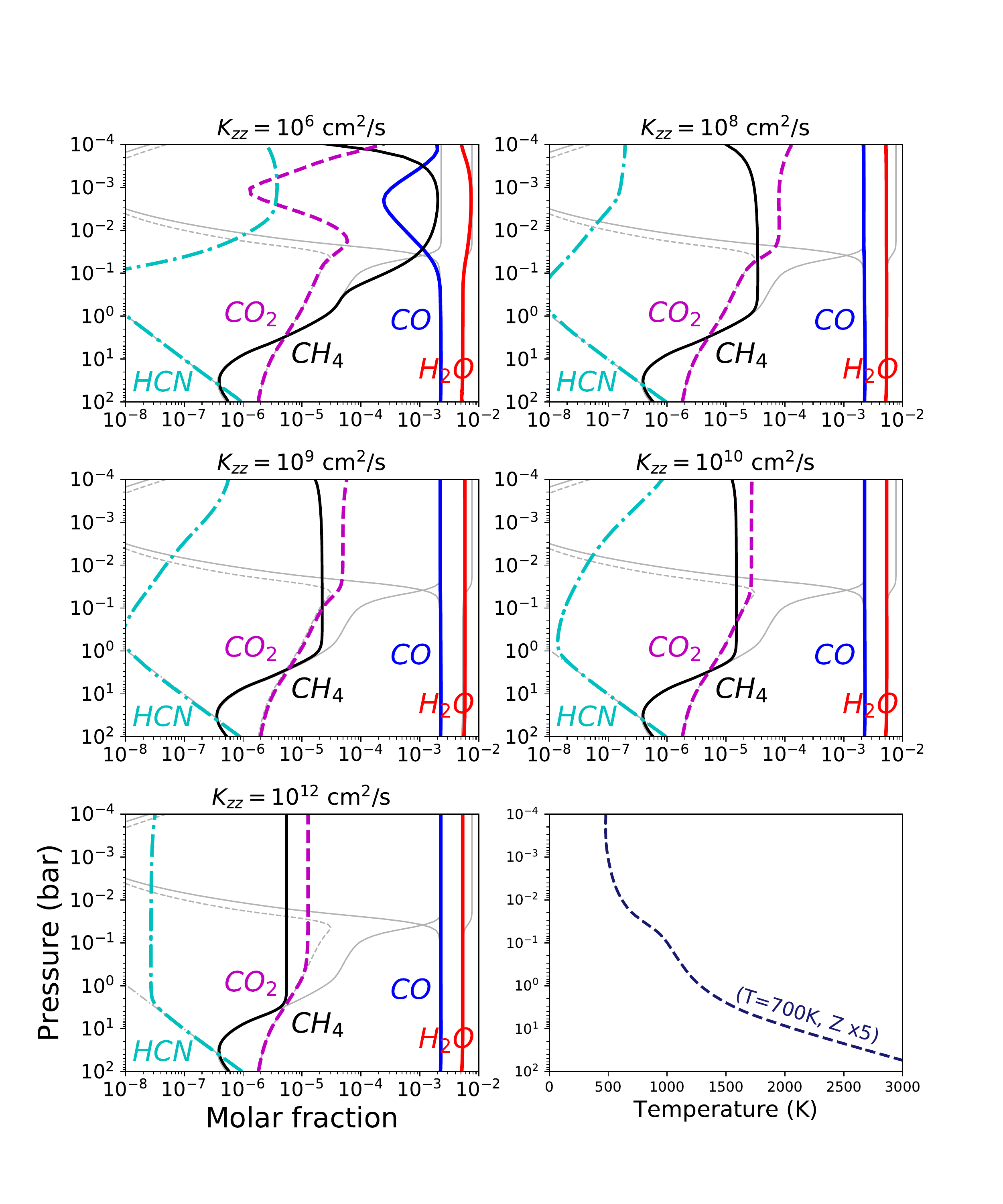}
    \caption{
    Disequilibrium abundances of the dominant molecules in the atmosphere of WASP-117b computed for different vertical eddy coefficients $K_{zz}$ with the chemical kinetics code of \cite{Venot2012}. The gray lines indicate the chemical equilibrium state. 
    The bottom right plot shows the one-dimensional pressure-temperature profile of WASP-117b, calculated with \texttt{petitCODE} \citep{Molliere2015,Molliere2017}. The model setup is based on the warm, slightly enriched scenario for WASP-117b ($T_{\rm eq} = 700$~K, $Z= 5\times$~solar, C/O $= 0.3$).}
    \label{fig: Disequi}
\end{figure*}

To constrain the amount of mixing needed to reproduce the retrieved methane abundance, we performed additional simulations for the cool equilibrium temperature ($T_{\rm eq} = 700$~K) and low metallicity (5$\times$ solar) set-up, with varying eddy diffusion coefficients: $K_{zz}= 10^{6}, 10^{8}, 10^{9}, 10^{10}, 10^{12}$~cm$^2$/s (see Figure~\ref{fig: Disequi}). Here, chemical equilibrium considerations would indicate that the atmosphere of WASP-117b is dominated by \ce{H2O} and \ce{CH4} above $\sim 100$~mbar. This is still the case for eddy diffusion coefficients up to $\sim 10^6$~cm${}^2$/s. For higher values of $K_{zz}$, $10^8$~cm${}^2$/s and above, \ce{CH4} is getting quenched at pressures of $\sim1$~bar and deeper. For these strongly mixed cases, the dominant atmospheric molecules are \ce{H2O} and \ce{CO} throughout the atmosphere. 

\clearpage
\section{HST Transmission Spectra}

\begin{table}
\caption{WASP-117b transmission spectra derived with the nominal pipeline} \label{table: Nominal_spectrum}     
\centering                        
\begin{tabular}{c c c}       
\hline
\hline
Wavelength  & Transit Depth  &  Uncertainty \\ 
\hline
1.137955 &	0.00747861	&	4.312E-05	\\
1.165961 &	0.00768836  &	4.304E-05	\\
1.193967 &	0.00753981  &	4.226E-05	\\
1.221973 &	0.00756615  &	4.110E-05	\\
1.249979 &	0.00750309  &	4.087E-05	\\
1.277985 &	0.00766432  &	4.064E-05\\
1.305991 &	0.00742532  &	4.063E-05	\\
1.333997 &	0.00763581  &	3.978E-05	\\
1.362003 &	0.00753166  &	3.996E-05	\\
1.390009 &	0.00766481  &	4.177E-05	\\
1.418015 &	0.00765191  &	4.198E-05	\\
1.446021 &	0.00761142  &	4.215E-05	\\
1.474027 &	0.00761617  &	4.229E-05	\\
1.502033 &	0.00763603  &	4.401E-05	\\
1.530039 &	0.00759753  &	4.515E-05	\\
1.558045 &	0.00757245  &	4.706E-05	\\
1.586051 &	0.00744342  &	4.838E-05	\\
1.614057 &	0.00758623  &	4.878E-05	\\
1.642063 &	0.00747719  &	5.2362E-05	\\
\hline                      
\end{tabular}
\end{table}

\begin{table}
\caption{WASP-117b transmission spectra derived with the CASCADE pipeline} \label{table: CASCADE_spectrum}     
\centering                        
\begin{tabular}{c c c}       
\hline
\hline
Wavelength  & Transit Depth  &  Uncertainty \\ 
\hline
1.137955	&	0.00762161	&  3.305E-05	\\
1.165961	&	0.00760239	&	3.566E-05	\\
1.193967	&	0.00759120	&	3.623E-05	\\
1.221973	&	0.00751966	&	3.202E-05	\\
1.249979	&	0.00747235	&	3.511E-05	\\
1.277985	&	0.00753574	&	3.619E-05	\\
1.305991	&	0.00747213	&	3.311E-05	\\
1.333997	&	0.00753405	&	3.352E-05	\\
1.362003	&	0.00759060	&	3.677E-05	\\
1.390009	&	0.00759895	&	3.262E-05	\\
1.418015	&	0.00755960	&	4.392E-05	\\
1.446021	&	0.00756905	&	3.370E-05	\\
1.474027	&	0.00760003	&	3.245E-05	\\
1.502033	&	0.00762398	&	3.702E-05	\\
1.530039	&	0.00750796	&	3.711E-05	\\
1.558045	&	0.00751503	&	3.558E-05	\\
1.586051	&	0.00754406	&	3.703E-05	\\
1.614057	&	0.00750870	&	3.696E-05	\\
1.642063	&	0.00752046	&	4.042E-05	\\
\hline                      
\end{tabular}
\end{table}

\end{appendix}

\end{document}